\documentclass[journal=tosc,version=preprint]{iacrj}
\usepackage{amssymb}
\usepackage{amsfonts}
\usepackage{bm}
\usepackage{booktabs}
\usepackage{multirow}
\usepackage{tabularx}
\usepackage{array}
\usepackage{makecell}
\usepackage{threeparttable}
\usepackage{xcolor}
\usetikzlibrary{arrows.meta,backgrounds,positioning}

\newcommand{\bx}{\boldsymbol{x}}
\newcommand{\by}{\boldsymbol{y}}
\newcommand{\bb}{\boldsymbol{b}}
\newcommand{\bu}{\boldsymbol{u}}
\newcommand{\bv}{\boldsymbol{v}}
\newcommand{\bn}{\boldsymbol{n}}

\newcommand{\bz}{\boldsymbol{z}}
\newcommand{\bc}{\boldsymbol{c}}
\newcommand{\bw}{\boldsymbol{w}}
\newcommand{\bo}{\boldsymbol{o}}
\newcommand{\bk}{\boldsymbol{k}}
\newcommand{\bp}{\boldsymbol{p}}
\newcommand{\bq}{\boldsymbol{q}}
\newcommand{\ba}{\boldsymbol{a}}
\newcommand{\be}{\boldsymbol{e}}
\newcommand{\br}{\boldsymbol{r}}

\newcommand{\bgamma}{\boldsymbol{\gamma}}
\newcommand{\bdelta}{\boldsymbol{\delta}}

\newcommand{\bA}{\boldsymbol{A}}
\newcommand{\bS}{\boldsymbol{S}}
\newcommand{\bI}{\boldsymbol{I}}

\newcommand{\bP}{\boldsymbol{P}}
\newcommand{\bQ}{\boldsymbol{Q}}
\newcommand{\bD}{\boldsymbol{D}}
\newcommand{\bM}{\boldsymbol{M}}
\newcommand{\bK}{\boldsymbol{K}}
\newcommand{\bV}{\boldsymbol{V}}
\newcommand{\bO}{\boldsymbol{O}}
\newcommand{\bC}{\boldsymbol{C}}
\newcommand{\bB}{\boldsymbol{B}}
\newcommand{\bW}{\boldsymbol{W}}
\newcommand{\bT}{\boldsymbol{T}}
\newcommand{\bY}{\boldsymbol{Y}}
\newcommand{\bX}{\boldsymbol{X}}

\newcommand{\bL}{\boldsymbol{L}}
\newcommand{\bU}{\boldsymbol{U}}
\newcommand{\bR}{\boldsymbol{R}}
\newcommand{\bZ}{\boldsymbol{Z}}

\newcolumntype{C}{>{\centering\arraybackslash}X}

\allowdisplaybreaks[4]
\title[running={Transformer Block Extraction under Finite Precision}]{Cryptanalytic Extraction of a Single Full Transformer Block under Finite Precision}

\addauthor[inst={1}, email={zhengjq23@mails.tsinghua.edu.cn}, surname={Zheng}]{Jiaqi Zheng}
\addauthor[inst={1}, email={chenzr25@mails.tsinghua.edu.cn}, surname={Chen}]{Zirui Chen}
\addauthor[inst={1}, email={xiaoyangdong@tsinghua.edu.cn}, surname={Dong}, footnote={Corresponding author.}]{Xiaoyang Dong}
\addaffiliation[city={Beijing}, country={China}]{Tsinghua University}

\begin{document}

\maketitle

% === IACR Generative AI Policy (2026-11-10) - Section 2.2 Prominent Designation ===
\noindent\textbf{Generative AI Use Designation:} Substantial AI contribution (see the Disclosure of Generative AI Use at the end of Section~\ref{sect:conclusions}).
\medskip

\addkeywords{Model extraction, Transformer, Multi-head Attention, RMSNorm, SwiGLU, Low-precision arithmetic}

\begin{abstract}
Cryptanalytic model extraction seeks to recover a neural network's
hidden parameters from black-box input--output queries.
Prior Transformer attacks either target isolated layers without
normalization and skip connections, or require all-position outputs,
positional encodings, and extended-precision arithmetic.
Modern Transformer blocks, however, tightly couple dual pre-RMSNorm
layers, residual connections, multi-head softmax Attention, and
biased ReLU or SwiGLU feed-forward networks (FFNs) under last-token
outputs and finite-precision arithmetic.

We present an end-to-end, precision-agnostic extraction attack on a
complete dual pre-RMSNorm residual Transformer block, handling both
biased ReLU and bias-free SwiGLU FFNs using only bounded,
positional-encoding-free queries and strict last-token outputs.
Rather than relying on isolated sublayer access or numerically
fragile pole-finding, our attack decouples the block sequentially:
recovering the aggregate Attention map, extracting the coupled FFN,
reconstructing all effective Attention heads through a shared
backend, and jointly refining the full block.
We evaluate the same pipeline across native FP64, FP32, FP16, and
BF16 implementations up to \(d=256\), \(H=8\) heads, and \(m=1024\)
(over one million parameters).
At this scale, extraction uses \(0.26\text{--}0.47\) vector queries
per parameter and achieves holdout relative \(L_2\) errors of at most
\(2.20\times 10^{-15}\) in FP64, \(5.49\times 10^{-7}\) in FP32,
\(1.07\times 10^{-3}\) in FP16, and \(8.16\times 10^{-3}\) in BF16.
\end{abstract}

\begin{textabstract}
Cryptanalytic model extraction seeks to recover a neural network's hidden parameters from black-box input-output queries. Prior Transformer attacks either target isolated layers without normalization and skip connections, or require all-position outputs, positional encodings, and extended-precision arithmetic. Modern Transformer blocks, however, tightly couple dual pre-RMSNorm layers, residual connections, multi-head softmax Attention, and biased ReLU or SwiGLU feed-forward networks (FFNs) under last-token outputs and finite-precision arithmetic. We present an end-to-end, precision-agnostic extraction attack on a complete dual pre-RMSNorm residual Transformer block, handling both biased ReLU and bias-free SwiGLU FFNs using only bounded, positional-encoding-free queries and strict last-token outputs. Rather than relying on isolated sublayer access or numerically fragile pole-finding, our attack decouples the block sequentially: recovering the aggregate Attention map, extracting the coupled FFN, reconstructing all effective Attention heads through a shared backend, and jointly refining the full block. We evaluate the same pipeline across native FP64, FP32, FP16, and BF16 implementations up to d=256, H=8 heads, and m=1024 (over one million parameters), using 0.26 to 0.47 vector queries per parameter and achieving holdout relative L2 errors of at most 2.20e-15 in FP64, 5.49e-7 in FP32, 1.07e-3 in FP16, and 8.16e-3 in BF16.
\end{textabstract}

\section{Introduction}
\label{sect:introduction}
Cryptanalytic model extraction seeks to recover a neural
network's hidden parameters from chosen inputs and their
outputs. Unlike training a substitute only to reproduce
predictions, parameter recovery uses the structure of the
target computation to identify its constituent components.
Carlini et al.~\cite{carlini2020} demonstrated how changes
in the local behavior of ReLU networks reveal information
about hidden neurons. Extending this approach to
Transformers~\cite{attention} requires handling softmax attention,
normalization, and feed-forward networks whose effects
are combined through residual connections.

\paragraph{Previous work.}
Recent results have established several routes to
Attention extraction.
Bhattamishra et al.~\cite{attention-bhattamishra}
study single-head attention recovery from chosen queries.
Wang et al.~\cite{attention-wang} and
Kim and Kim~\cite{attention-kim} recover canonical multi-head
representations by constructing rational responses from
repeated-token queries and separating their poles.
For gated feed-forward networks, Shi and
Fu~\cite{swigluExtraction} recover isolated bias-free
GLU blocks, including SwiGLU, using curvature information
and the relation between responses at opposite inputs.
These results provide recovery techniques for the
individual components of a Transformer.

More recently, Lin et al.~\cite{transformer-lin}
developed an extraction method for multilayer
Transformers with pre-RMSNorm, residual connections,
known rotary positional encodings, and SwiGLU networks.
Their method separates layers through input scaling
and reconstructs their interactions from response
coefficients. Their observation model returns all
final residual-state vectors, and their experiments
use binary128 arithmetic for target evaluations and
finite differences. This advances recovery of composed
Transformer architectures and motivates studying how
extraction behaves under more restricted observations
and native low-precision execution.

\paragraph{Our setting and challenges.}
We study a complete Transformer block containing
multi-head softmax attention, two pre-RMSNorm layers,
two residual connections, and either a biased ReLU
or a bias-free SwiGLU feed-forward network.
The architecture and normalization constants are known,
while the weights and intermediate states are hidden.
The attacker chooses continuous token vectors subject
to a norm bound and observes only the final token's
output after both sublayers. The model has no positional
encoding. Our recovery target is the effective Attention
and FFN parameterization, which absorbs the normalization
gains and accounts for parameter symmetries.

The main difficulty is that the observed output does
not directly expose either component.
The FFN receives a normalized hidden vector that already
depends on the unknown Attention parameters, while its
output is added to that hidden vector.
Consequently, an extraction method requiring isolated
Attention or FFN responses cannot be applied directly.
It is necessary to construct useful observations of
one component while the other is still unknown or
only approximately recovered.

Finite precision creates a second difficulty.
Small changes in the input can produce output
differences comparable to rounding errors, and these
errors can be amplified during differentiation and
subsequent linear solves.
Recent error analysis documents this issue for
cryptanalytic extraction~\cite{xu2026finite}.
Higher-precision offline computation cannot restore
information already lost in the target's returned
values. We therefore study extraction when the target
itself executes in FP64, FP32, FP16, or BF16, and
evaluate both recovered parameters and agreement
with the complete block.
Table~\ref{tab:summary} compares the architectures,
observations, and numerical settings considered here
and in prior work.

\begin{table}[t]
\centering
\caption{Comparison of model scope, query interfaces,
and evaluated precision in Attention and FFN extraction.}
\label{tab:summary}
\footnotesize
\renewcommand{\arraystretch}{1.15}
\setlength{\tabcolsep}{3pt}

\resizebox{\textwidth}{!}{%
\begin{tabular}{@{}llccccc@{}}
\toprule

\multicolumn{2}{c}{\textbf{Model / Task}}
&
\multicolumn{4}{c}{\textbf{Extraction Setting}}
&
\multicolumn{1}{c}{\textbf{Approach}}
\\
\cmidrule(lr){1-2}
\cmidrule(lr){3-6}
\cmidrule(l){7-7}

\textbf{Architecture}
&
\textbf{Target}
&
\shortstack{\textbf{Oracle}\\\textbf{output}}
&
\shortstack{\textbf{Biased}\\\textbf{FFN}}
&
\shortstack{\textbf{RMSNorm /}\\\textbf{residuals}}
&
\shortstack{\textbf{Reported numerical}\\\textbf{precision}}
&
\textbf{Reference}
\\
\midrule

SHSA
&
Effective params.
&
Scalar
&
--
&
$0/0$
&
Not reported
&
\cite{attention-bhattamishra}
\\

SHSA + Bias-free ReLU
&
Equivalent model
&
Scalar
&
No
&
$0/0$
&
Not reported
&
\cite{attention-bhattamishra}
\\

SHSA + Biased ReLU
&
Equivalent model
&
Scalar
&
Yes
&
$0/0$
&
FP64
&
\cite{attention-wang}
\\
\midrule

MHSA
&
Canonical params.
&
Scalar
&
--
&
$0/0$
&
FP64
&
\cite{attention-wang}
\\

MHSA
&
Canonical params.
&
Scalar
&
--
&
$0/0$
&
\shortstack{180-digit;\\FP64-rounded outputs$^\dagger$}
&
\cite{attention-kim}
\\

MHSA + Bias-free ReLU
&
Equivalent model
&
Scalar
&
No
&
$0/0$
&
Not reported
&
\cite{attention-kim}
\\
\midrule

\shortstack[l]{Isolated Bias-free GLU\\(SwiGLU / GeGLU)}
&
FFN parameters
&
Vector
&
No
&
$0/0$
&
\shortstack{FP64, FP32, FP16,\\BF16, TF32}
&
\cite{swigluExtraction}
\\
\midrule

\shortstack[l]{Multilayer MHSA +\\Bias-free SwiGLU (RoPE)}
&
Equivalent model
&
Vector
&
No
&
$2L/2L$
&
High precision$^\ddagger$
&
\cite{transformer-lin}
\\
\midrule

\textbf{MHSA + Biased ReLU}
&
Effective params.
&
\shortstack{Last-token\\vector}
&
\textbf{Yes}
&
$\mathbf{2/2}$
&
\shortstack{\textbf{FP64, FP32,}\\\textbf{FP16, BF16}}
&
\textbf{This work}
\\

\textbf{MHSA + Bias-free SwiGLU}
&
Effective params.
&
\shortstack{Last-token\\vector}
&
No
&
$\mathbf{2/2}$
&
\shortstack{\textbf{FP64, FP32,}\\\textbf{FP16, BF16}}
&
\textbf{This work}
\\

\bottomrule
\end{tabular}%
}

\begin{minipage}{\textwidth}
\scriptsize
SHSA/MHSA: single-/multi-head softmax Attention;
the RMSNorm/residual column gives their respective counts ($L$: layers);
``--'' means not applicable.
$^\dagger$ Offline arithmetic uses 180 digits;
listed precision formats describe evaluation settings,
not exact recovery of stored weights.
$^\ddagger$ Concurrent work (Nov.\ 2026): targets small synthetic
multilayer models using input-scaling decoupling, known RoPE, and
high-precision residual observations.
\end{minipage}

\end{table}

\paragraph{Our approach and contributions.}
The extraction proceeds from the aggregate effect of
Attention to the FFN, then to the individual heads.
The recovered components are finally refined against
complete-block outputs.
Figure~\ref{fig:extraction-pipeline} illustrates this
sequence. Our contributions are as follows.

\begin{itemize}
    \item \textbf{End-to-end extraction of a complete Transformer block.}
    We present an end-to-end black-box attack that recovers effective
    parameters of a complete dual pre-RMSNorm residual Transformer
    block---supporting both biased ReLU and bias-free SwiGLU FFNs---using
    only bounded, positional-encoding-free queries and strict last-token
    outputs.

    \item \textbf{Modular decoupling across sublayers and FFN families.}
    We decouple the Attention and FFN sublayers without isolated
    component access, matrix inversion of the aggregate Attention map,
    or numerically unstable rational pole-finding.
    Once the FFN is recovered, both ReLU and SwiGLU branches share the
    same multi-head Attention extractor and joint refinement stage.

    \item \textbf{Million-parameter scalability.}
    Whereas prior cryptanalytic evaluations of combined multi-head
    Attention and FFN blocks have been limited to small synthetic
    models~\cite{attention-wang,attention-kim,transformer-lin}, our
    attack scales end to end to \(d=256\), \(H=8\) heads, and FFN
    width \(m=1024\)---up to \(788{,}224\) parameters for biased ReLU
    and \(1{,}049{,}088\) parameters for bias-free SwiGLU---with query
    cost dropping to \(0.26\text{--}0.47\) vector queries per
    parameter at the largest scale.

    \item \textbf{Robustness across native low-precision formats.}
    A single precision-agnostic pipeline succeeds across native FP64,
    FP32, FP16, and BF16 target implementations without format-specific
    branching.
    Across all five evaluated scales in Table~\ref{tab:scaling-summary},
    holdout relative block-output errors remain at most
    \(2.20\times 10^{-15}\) in FP64, \(5.49\times 10^{-7}\) in FP32,
    \(1.07\times 10^{-3}\) in FP16, and \(8.16\times 10^{-3}\) in BF16.
\end{itemize}

The analysis establishes the query identities and
reconstruction conditions underlying these steps.
The implementation uses numerical estimates throughout,
with complete-block refinement correcting errors
remaining after component recovery.
Together, the results demonstrate parameter extraction
and close output agreement for the tested synthetic
blocks under bounded queries and native finite
precision.

\paragraph{Organization.}
Section~\ref{sect:preliminary} defines the model and
observation interface.
Sections~\ref{sect:ReLU} and~\ref{sect:SwiGLU}
describe aggregate-map and FFN recovery for the two
FFN families.
Section~\ref{sect:Attention} presents the shared
Attention recovery and joint refinement.
Section~\ref{sect:experiments} reports the experimental
results, and Section~\ref{sect:conclusions} concludes
with directions for future work.

\section{Preliminaries}
\label{sect:preliminary}
% This section introduces the dual pre-RMSNorm Transformer block and
% its equivalent effective parameterization in row-vector convention
% (Section~\ref{sec:arch}), and formalizes the black-box threat model
% and extraction goal (Section~\ref{sec:adversarial-model}).

Throughout the paper, indices start from \(1\).
Following the Transformer convention~\cite{attention}, bold lowercase
letters (e.g., \(\bx\in\mathbb R^d\)) denote \emph{row} vectors
(\(1\times d\)), with \(i\)-th coordinate \(\bx[i]\) and transpose
column vector \(\bx^\top\); bold uppercase letters (e.g., \(\bA\))
denote matrices, acting on row vectors from the right (\(\bx\bA\));
and calligraphic letters (e.g., \(\mathcal{F},\mathcal{G},\mathcal{N}_j\))
denote vector- or matrix-valued functions.
An \(\ell\)-token sequence of row vectors \(\bx_1,\ldots,\bx_\ell \in\mathbb R^{d}\) is
stacked row-wise into \(\bX\in\mathbb R^{\ell\times d}\), written
compactly as \(\bX=[\bx_1,\ldots,\bx_\ell]\), where \(\bX[i,\cdot]\) is the \(i\)-th row, \(\bX[\cdot,j]\) is the \(j\)-th column, and \(\bX[i,j]\) is the \((i,j)\)-th entry.
Table~\ref{tab:notation} summarizes the primary notation
used across Sections~\ref{sect:preliminary}--\ref{sect:Attention} and
Appendices~\ref{app:relu-scale-refinement} and~\ref{app:relu-paired-continuation}.

\begin{table}[t]
\centering
\caption{Summary of primary notation used across
Sections~\ref{sect:preliminary}--\ref{sect:Attention} and
Appendices~\ref{app:relu-scale-refinement} and~\ref{app:relu-paired-continuation}.
Unless explicitly redefined locally, symbols retain these meanings
throughout; architecture-specific FFN parameter notation is
introduced locally in Sections~\ref{sect:ReLU} and~\ref{sect:SwiGLU}.}
\label{tab:notation}
\footnotesize
\renewcommand{\arraystretch}{1.12}
\begin{tabularx}{\linewidth}{@{}lX@{}}
\toprule
\textbf{Symbol} & \textbf{Meaning} \\
\midrule
\(d, H, r, m, \ell\)
  & Token dimension, head count, per-head width \(r=d/H\), FFN width, and sequence length. \\
\(\epsilon_1,\epsilon_2>0\)
  & Public RMSNorm stability constants (hidden positive gains \(\bgamma_1,\bgamma_2\); \(\bD_j=\operatorname{diag}(\bgamma_j)\)). \\
\(\bX=[\bx_1,\ldots,\bx_\ell]\in\mathcal{X}\)
  & Input sequence of \(\ell\) row tokens \(\bx_i\in\mathbb R^d\) (\(\bX\in\mathbb R^{\ell\times d}\)) in permitted query domain \(\mathcal{X}\). \\
\(\mathcal{N}_j:\mathbb R^d\to\mathcal{B}_d\)
  & Gain-free RMSNorm onto \(\mathcal{B}_d=\{\bu\in\mathbb R^d:\|\bu\|_2<\sqrt{d}\}\); \(\bu_i=\mathcal{N}_1(\bx_i)\), \(\bU=[\bu_1,\ldots,\bu_\ell]\). \\
\(\bS_h,\bC_h\in\mathbb R^{d\times d}\)
  & Effective score and value matrices of head \(h\) (rank \(\leq r\)); aggregate value map \(\bM=\sum_{h=1}^{H}\bC_h\). \\
\(\operatorname{Att}_{\bS,\bC}(\bX),\ \mathcal{A}(\bU)\)
  & Effective Attention output on physical input \(\bX\) and normalized input \(\bU\) (\(\mathcal{A}(\bU)=\operatorname{Att}_{\bS,\bC}(\mathcal{N}_1^{-1}(\bU))\)). \\
\(\bz(\bX),\ \bw=\mathcal{N}_2(\bz)\)
  & First residual \(\bz(\bX)=\bx_\ell+\operatorname{Att}_{\bS,\bC}(\bX)\) and gain-free normalized FFN input. \\
\(\mathcal{G}(\bw),\ \mathcal{F}(\bX)\)
  & Effective FFN map and complete-block last-token output \(\mathcal{F}(\bX)=\bz(\bX)+\mathcal{G}(\mathcal{N}_2(\bz(\bX)))\). \\
\(\Psi_{\mathcal{G}}(\bz),\ \mathcal{P}_{\mathcal{G}}(\bz)\)
  & Residual-to-output map \(\Psi_{\mathcal{G}}(\bz)=\bz+\mathcal{G}(\mathcal{N}_2(\bz))\) and paired row map \(\mathcal{P}_{\mathcal{G}}(\bz)=[\Psi_{\mathcal{G}}(\bz),\,\Psi_{\mathcal{G}}(-\bz)]\). \\
\(\cdot^\dagger\)
  & Moore--Penrose pseudoinverse. \\
\bottomrule
\end{tabularx}
\end{table}

\subsection{Single-Block Transformer and Effective Parameterization}
\label{sec:arch}

\begin{figure}[t]
\centering
\resizebox{\linewidth}{!}{%
\begin{tikzpicture}[
    x=1cm, y=1cm,
    font=\fontsize{8.5}{10}\selectfont,
    line width=0.65pt,
    flow/.style={
        -{Stealth[length=2.0mm,width=1.35mm]}
    },
    block/.style={
        draw=black!75,
        rounded corners=1.5pt,
        align=center,
        inner sep=4pt
    },
    norm/.style={
        block,
        fill=blue!6,
        minimum width=1.45cm,
        minimum height=0.85cm
    },
    head/.style={
        block,
        fill=blue!10,
        minimum width=2.25cm,
        minimum height=0.68cm
    },
    proj/.style={
        block,
        fill=blue!10,
        minimum width=1.50cm,
        minimum height=0.90cm
    },
    add/.style={
        circle,
        draw=black!85,
        fill=white,
        minimum size=4.6mm,
        inner sep=0pt,
        font=\normalsize
    },
    smalllabel/.style={
        font=\fontsize{7.5}{9}\selectfont
    },
    every node/.style={text=black}
]

% Input and first normalization.
\node[align=center] (input) at (0.35,3.0)
    {\(\bX\)};
\node[smalllabel] at (0.35,2.60)
    {input sequence};
\coordinate (inputtap) at (1.05,3.0);

\node[norm] (normone) at (2.25,3.0)
    {RMSNorm\\ \(\mathcal{N}_1(\cdot)\bD_1\)};

% Multi-head Attention.
\node[head] (headone) at (5.65,3.65)
    {Head \(1\)\\ \(\bQ_1,\bK_1,\bV_1\)};
\node[head] (headlast) at (5.65,2.35)
    {Head \(H\)\\ \(\bQ_H,\bK_H,\bV_H\)};
\node at (5.65,3.0) {\(\vdots\)};

\node[proj] (outproj) at (8.00,3.0)
    {Output proj.\\ \(\cdot\,\bO\)};

\node[
    font=\fontsize{8.5}{10}\selectfont\bfseries
] at (6.475,4.40) {Multi-head Attention};

\begin{scope}[on background layer]
    \draw[
        draw=black!45,
        fill=blue!2,
        rounded corners=2pt
    ] (4.05,1.75) rectangle (8.90,4.65);
\end{scope}

\draw (input.east) -- (inputtap);
\draw[flow] (inputtap) -- (normone.west);
\draw[flow] (normone.east) -- (4.25,3.0)
    node[midway,above=2pt,smalllabel] {\(\bU\bD_1\)};

\fill (4.25,3.0) circle (1.0pt);
\draw (4.25,2.35) -- (4.25,3.65);
\draw[flow] (4.25,3.65) -- (headone.west);
\draw[flow] (4.25,2.35) -- (headlast.west);

\draw[flow]
    (headone.east) -- (8.00,3.65) -- (outproj.north);
\draw[flow]
    (headlast.east) -- (8.00,2.35) -- (outproj.south);

% First residual connection.
\node[add] (addone) at (9.80,3.0) {\(+\)};
\draw[flow] (outproj.east) -- (addone.west);

\fill (inputtap) circle (1.0pt);
\draw[flow]
    (inputtap) -- (1.05,5.20) -- (9.80,5.20)
    node[midway,above=3pt] {last token \(\bx_\ell\)}
    -- (addone.north);

% First residual feeds the second sublayer.
\coordinate (ztap) at (0.95,-0.25);
\draw[flow]
    (addone.east) -- (11.55,3.0) -- (11.55,1.10)
    -- (0.95,1.10) -- (ztap);

\node[fill=white,inner sep=2pt] at (6.25,1.10)
    {\(\bz=\bx_\ell+\operatorname{Att}_{\bS,\bC}(\bX)\)};

\fill (ztap) circle (1.0pt);
\node[left=3pt] at (ztap) {\(\bz\)};

% Second normalization and FFN.
\node[norm] (normtwo) at (2.45,-0.25)
    {RMSNorm\\ \(\mathcal{N}_2(\cdot)\bD_2\)};

\node[
    block,
    fill=green!7,
    minimum width=3.8cm,
    minimum height=1.05cm
] (ffn) at (6.425,-0.25)
    {\(\operatorname{FFN}\)\\ ReLU or SwiGLU};

% Second residual connection and output.
\node[add] (addtwo) at (9.80,-0.25) {\(+\)};
\node[align=center] (output) at (11.20,-0.25)
    {\(\mathcal{F}(\bX)\)};
\node[smalllabel] at (11.20,-0.75)
    {last-token output};

\draw[flow] (ztap) -- (normtwo.west);
\draw[flow] (normtwo.east) -- (ffn.west)
    node[midway,above=2pt,smalllabel] {\(\bw\bD_2\)};
\draw[flow] (ffn.east) -- (addtwo.west)
    node[midway,above=2pt,smalllabel] {\(\mathcal{G}(\bw)\)};
\draw[flow] (addtwo.east) -- (output.west);

\draw[flow]
    (ztap) -- (0.95,-1.65) -- (9.80,-1.65)
    node[midway,below=3pt] {\(\bz\)}
    -- (addtwo.south);

\end{tikzpicture}%
}
\caption{Last-token computation of the dual pre-RMSNorm residual
Transformer block in~\eqref{eq:block}.
The upper sublayer applies \(\operatorname{RMSNorm}_1(\bx_i)=\mathcal{N}_1(\bx_i)\bD_1\),
evaluates \(H\) Attention heads projected by \(\bO\) (equivalently
\(\operatorname{Att}_{\bS,\bC}(\bX)\) in~\eqref{eq:attention}), and
adds the last input token \(\bx_\ell\) to form \(\bz\).
The lower sublayer applies \(\operatorname{RMSNorm}_2(\bz)=\bw\bD_2\)
with \(\bw=\mathcal{N}_2(\bz)\) and the FFN (biased ReLU or bias-free SwiGLU,
yielding \(\mathcal{G}(\bw)=\operatorname{FFN}(\bw\bD_2)\)), followed by the
second residual addition to output \(\mathcal{F}(\bX)\).}
\label{fig:transformer-block}
\end{figure}

Figure~\ref{fig:transformer-block} illustrates the complete dual
pre-RMSNorm~\cite{rmsnorm} residual~\cite{residualconnections}
Transformer block considered in this paper.
In autoregressive generation with a KV cache, the block processes an
\(\ell\)-token input sequence \(\bX=[\bx_1,\ldots,\bx_\ell]\in\mathbb R^{\ell\times d}\)
and outputs the updated row representation at the final token
position, which serves as the query
token~\cite{huggingface_cache,attention-bhattamishra,attention-kim,attention-wang}.
The block has \(H\) Attention heads of per-head width \(r=d/H\) and
FFN hidden width \(m\).

\paragraph{RMSNorm.}
The block uses two pre-RMSNorm layers indexed by \(j\in\{1,2\}\),
each with a public stability constant \(\epsilon_j>0\) and a hidden
positive gain vector \(\bgamma_j\in(0,\infty)^d\).
Writing \(\bD_j=\operatorname{diag}(\bgamma_j)\in\mathbb R^{d\times d}\),
the \(j\)-th RMSNorm layer acts on a row token \(\bx\in\mathbb R^d\) as
\begin{equation}
    \operatorname{RMSNorm}_j(\bx)
    =
    \mathcal{N}_j(\bx)\bD_j,
    \qquad
    \mathcal{N}_j(\bx)
    :=
    \frac{\bx}{\sqrt{\epsilon_j+\|\bx\|_2^2/d}}
    \in\mathcal{B}_d,
    \label{eq:rms}
\end{equation}
where \(\mathcal{N}_j\) is the gain-free normalization map onto the open ball
\(\mathcal{B}_d=\{\bu\in\mathbb R^d:\|\bu\|_2<\sqrt{d}\}\).
For an input sequence \(\bX=[\bx_1,\ldots,\bx_\ell]\), we write
\(\bu_i=\mathcal{N}_1(\bx_i)\) and \(\bU=[\bu_1,\ldots,\bu_\ell]\in\mathbb R^{\ell\times d}\),
so applying \(\operatorname{RMSNorm}_1\) row-wise gives \(\bU\bD_1\).

\paragraph{Multi-head Attention and effective head matrices.}
For each head \(h\in\{1,\ldots,H\}\), let
\(\bQ_h,\bK_h,\bV_h\in\mathbb R^{d\times r}\) be the query, key, and
value projection matrices, and partition the Attention output
projection row-wise as
\(\bO=[\bO_1^\top,\ldots,\bO_H^\top]^\top\in\mathbb R^{d\times d}\)
with \(\bO_h\in\mathbb R^{r\times d}\).
On the normalized tokens \(\bU\bD_1\), head \(h\) forms the
last-token query \(\bu_\ell\bD_1\bQ_h\), keys \(\bU\bD_1\bK_h\), and
values \(\bU\bD_1\bV_h\).
The scaled dot-product score between the last token and token \(i\) is
\((\bu_\ell\bD_1\bQ_h)(\bu_i\bD_1\bK_h)^\top/\sqrt{r}=\bu_\ell\bS_h\bu_i^\top\),
and projecting the softmax-weighted values by \(\bO_h\) multiplies
\(\bu_i\) on the right by \(\bC_h=\bD_1\bV_h\bO_h\).
Defining the effective score and value matrices
\begin{equation}
    \bS_h
    :=
    \frac{\bD_1\bQ_h\bK_h^\top\bD_1}{\sqrt{r}},
    \qquad
    \bC_h
    :=
    \bD_1\bV_h\bO_h,
   % \qquad
    %\bM
   % :=
   % \sum_{h=1}^{H}\bC_h,
    \label{eq:effective}
\end{equation}
where each \(\bS_h,\bC_h\in\mathbb R^{d\times d}\) has rank at most
\(r\), the Attention weights and last-token Attention output are
\begin{equation}
    \alpha_{h,i}(\bX)
    =
    \frac{\exp(\bu_\ell\bS_h\bu_i^\top)}
         {\sum_{k=1}^{\ell}\exp(\bu_\ell\bS_h\bu_k^\top)},
    \qquad i=1,\ldots,\ell,
    \label{eq:attention-weights}
\end{equation}
\begin{equation}
    \operatorname{Att}_{\bS,\bC}(\bX)
    =
    \sum_{h=1}^{H}
    \left(\sum_{i=1}^{\ell}\alpha_{h,i}(\bX)\bu_i\right)\bC_h
    =
    \sum_{h=1}^{H}
    \operatorname{softmax}\!\bigl(\bu_\ell\bS_h\bU^\top\bigr)\bU\bC_h.
    \label{eq:attention}
\end{equation}

\paragraph{Complete block and effective FFN.}
Absorbing the second RMSNorm gain \(\bD_2\) into the FFN via
\(\mathcal{G}(\bw):=\operatorname{FFN}(\bw\bD_2)\), the complete block
\(\mathcal{F}:\mathbb R^{\ell\times d}\to\mathbb R^d\) computes the last-token
row output
\begin{equation}
    \bz(\bX)
    =
    \bx_\ell+\operatorname{Att}_{\bS,\bC}(\bX),
    \qquad
    \mathcal{F}(\bX)
    =
    \bz(\bX)+\mathcal{G}\bigl(\mathcal{N}_2(\bz(\bX))\bigr),
    \label{eq:block}
\end{equation}
where \(\bw=\mathcal{N}_2(\bz)\) is the gain-free normalized FFN input.
We consider two standard FFN architectures---biased ReLU~\cite{relu} and
bias-free SwiGLU~\cite{swiglu}---whose matrix parameterizations for
\(\mathcal{G}\) after absorbing \(\bD_2\) are introduced at the start
of Sections~\ref{sect:ReLU} and~\ref{sect:SwiGLU}, respectively.

The raw projection matrices and normalization gains are not uniquely
identifiable from black-box queries: for any invertible
\(\bP_h,\bR_h\in\mathbb R^{r\times r}\), replacing
\((\bQ_h,\bK_h)\mapsto(\bQ_h\bP_h,\bK_h\bP_h^{-\top})\) or
\((\bV_h,\bO_h)\mapsto(\bV_h\bR_h,\bR_h^{-1}\bO_h)\) leaves
\((\bS_h,\bC_h)\) unchanged, and \(\bgamma_1,\bgamma_2\) fold into
adjacent linear maps.
The effective parameters \((\{\bS_h,\bC_h\}_{h=1}^{H},\mathcal{G})\) eliminate
these unidentifiable internal factorizations while preserving the
exact block map~\eqref{eq:block} for every input \(\bX\).

\subsection{Threat Model and Extraction Goal}
\label{sec:adversarial-model}

Let \(\mathcal{X}\) denote the permitted input domain (incorporating
any public bounds on sequence length and token norms).
The adversary has adaptive black-box query access to the target block
\(\mathcal{F}\) in~\eqref{eq:block}: submitting any input sequence
\(\bX\in\mathcal{X}\) returns only the last-token output vector
\(\mathcal{F}(\bX)\in\mathbb{R}^d\).

\paragraph{Extraction goal.}
From adaptive queries \(\bX\in\mathcal{X}\) and their responses
\(\mathcal{F}(\bX)\), the adversary aims to recover the effective Attention
and FFN parameters of \(\mathcal{F}\):
\begin{itemize}
    \item \emph{Multi-head Attention:} the rank-\(\leq r\) matrix
    pairs \(\{(\bS_h,\bC_h)\}_{h=1}^{H}\), up to a common permutation
    of the \(H\) heads;
    \item \emph{Feed-forward network \(\mathcal{G}\):} the identifiable
    effective FFN parameters after absorbing \(\bD_2\), detailed for
    biased ReLU in Section~\ref{sect:ReLU} and for bias-free SwiGLU in
    Section~\ref{sect:SwiGLU}.
\end{itemize}
By Section~\ref{sec:arch}, recovering these effective parameters
yields a functionally equivalent block agreeing with \(\mathcal{F}(\bX)\) for
every \(\bX\in\mathcal{X}\), without attempting to recover the
unidentifiable raw \(\bQ/\bK/\bV/\bO\) factorizations or separate
RMSNorm gains \(\bgamma_1,\bgamma_2\).

\paragraph{Adversary assumptions.}
\begin{itemize}
    \item \textbf{Known architecture.}
    The adversary knows the block structure, FFN type (biased ReLU or
    bias-free SwiGLU), dimensions \((d,H,r,m)\), and public
    normalization constants \(\epsilon_1,\epsilon_2\).
    All weights, biases, and normalization gains are secret.

    \item \textbf{Bounded chosen inputs without positional encodings.}
    The adversary may adaptively query arbitrary real-valued token
    sequences \(\bX\in\mathcal{X}\) respecting any declared length
    and token-norm bounds, without requiring positional encodings or
    unbounded input scaling.

    \item \textbf{Strict last-token complete-block outputs.}
    Each query returns only the final-token output vector
    \(\mathcal{F}(\bX)\in\mathbb R^d\) after both residual connections.
    Attention weights, intermediate residuals \(\bz(\bX)\), isolated
    FFN responses, earlier token outputs, and model gradients are not
    exposed.

    \item \textbf{Exact and finite-precision arithmetic.}
    Our theoretical identifiability and convergence analyses assume
    exact real-valued oracle responses \(\mathcal{F}(\bX)\).
    Our experiments (Section~\ref{sect:experiments}) evaluate the
    same precision-agnostic pipeline against native \textsc{FP64},
    \textsc{FP32}, \textsc{FP16}, and \textsc{BF16} target
    implementations, measuring both aligned effective parameter
    recovery error and end-to-end holdout functional agreement.
\end{itemize}

With the effective parameterization and oracle model fixed,
Figure~\ref{fig:extraction-pipeline} summarizes the data-flow
dependencies across Sections~\ref{sect:ReLU}--\ref{sect:Attention},
showing how the architecture-specific Stage~1--2 branches converge
into the shared Stage~3--4 Attention recovery and joint refinement
pipeline.

\begin{figure}[t]
\centering
\resizebox{\linewidth}{!}{%
\begin{tikzpicture}[
    x=1cm, y=1cm,
    font=\fontsize{8.0}{9.5}\selectfont,
    line width=0.65pt,
    flow/.style={-{Stealth[length=2.0mm,width=1.35mm]}},
    qarrow/.style={-{Stealth[length=1.8mm,width=1.25mm]}, dashed, draw=black!65},
    stage/.style={
        draw=black!75,
        rounded corners=1.8pt,
        align=center,
        inner sep=4pt
    },
    smalllabel/.style={font=\fontsize{7.2}{8.5}\selectfont},
    every node/.style={text=black}
]

% Top oracle bar.
\node[
    stage,
    fill=gray!10,
    draw=black!70,
    minimum width=17.55cm,
    minimum height=0.72cm
] (oracle) at (8.825, 5.45)
    {\textbf{Black-Box Last-Token Oracle} \(\mathcal{F}(\bX)\in\mathbb{R}^d\)
     \quad (\emph{native FP64/FP32/FP16/BF16, bounded inputs})};

% Stage 1 and Stage 2: ReLU branch (top row).
\node[
    stage,
    fill=green!10,
    minimum width=3.10cm,
    minimum height=1.20cm
] (s1r) at (1.85, 3.15)
    {\textbf{Stage 1} (\S\ref{sec:relu-aggregate})\\
     Aggregate map \(\bM\)\\
     {\scriptsize repeated-prefix radial fit}};

\node[
    stage,
    fill=green!10,
    minimum width=3.50cm,
    minimum height=1.20cm
] (s2r) at (6.15, 3.15)
    {\textbf{Stage 2} (\S\ref{sec:relu-ffn-recovery})\\
     Biased ReLU \(\mathcal{G}\)\\
     {\scriptsize bridge, \(2m\) proxy, folding, VarPro}};

% Stage 1 and Stage 2: SwiGLU branch (bottom row).
\node[
    stage,
    fill=orange!10,
    minimum width=3.10cm,
    minimum height=1.20cm
] (s1s) at (1.85, 0.85)
    {\textbf{Stage 1} (\S\ref{sec:swiglu-aggregate})\\
     Aggregate map \(\bM\)\\
     {\scriptsize origin / Richardson pilot}};

\node[
    stage,
    fill=orange!10,
    minimum width=3.50cm,
    minimum height=1.20cm
] (s2s) at (6.15, 0.85)
    {\textbf{Stage 2} (\S\ref{sec:swiglu-ffn-recovery})\\
     Bias-free SwiGLU \(\mathcal{G}\)\\
     {\scriptsize bridge, factor SVD, flag, GN}};

% Stage 3: Shared multi-head Attention recovery.
\node[
    stage,
    fill=blue!10,
    minimum width=3.30cm,
    minimum height=3.40cm
] (s3) at (11.45, 2.00)
    {\textbf{Stage 3} (\S\ref{sec:att-observations}--\S\ref{sec:att-cur})\\[2pt]
     \textbf{Shared Attention}\\[4pt]
     {\scriptsize 1.\ Paired \(\pm\bX\) inversion (\(\bz\))}\\[2pt]
     {\scriptsize 2.\ Macro probes \& JD}\\[2pt]
     {\scriptsize 3.\ 2-token stitch (\(\bC_h\))}\\[2pt]
     {\scriptsize 4.\ \(\operatorname{artanh}\) + CUR (\(\bS_h\))}};

% Stage 4a and 4b: Fixed-rank tangent-space refinement.
\node[
    stage,
    fill=blue!14,
    minimum width=3.00cm,
    minimum height=1.20cm
] (s4a) at (15.85, 3.05)
    {\textbf{Stage 4a} (\S\ref{sec:att-refinement})\\
     Attention GN\\
     {\scriptsize rank-\(\le r\) manifold}};

\node[
    stage,
    fill=blue!14,
    minimum width=3.00cm,
    minimum height=1.20cm
] (s4b) at (15.85, 0.95)
    {\textbf{Stage 4b} (\S\ref{sec:att-refinement})\\
     Joint Block GN\\
     {\scriptsize \((\bS_h,\bC_h,\mathcal{G})\) output}};

% Background frames (non-overlapping with boxes or labels).
\begin{scope}[on background layer]
    \draw[draw=green!50!black, fill=green!3, rounded corners=2pt]
        (0.05, 2.30) rectangle (8.15, 4.35);
    \node[anchor=west, smalllabel, text=green!40!black]
        at (0.18, 4.15) {\textbf{Biased ReLU branch (Section~\ref{sect:ReLU})}};

    \draw[draw=orange!65!black, fill=orange!3, rounded corners=2pt]
        (0.05, 0.00) rectangle (8.15, 2.05);
    \node[anchor=west, smalllabel, text=orange!55!black]
        at (0.18, 1.85) {\textbf{Bias-free SwiGLU branch (Section~\ref{sect:SwiGLU})}};

    \draw[draw=blue!55!black, fill=blue!3, rounded corners=2pt]
        (9.55, 0.00) rectangle (17.60, 4.35);
    \node[anchor=west, smalllabel, text=blue!45!black]
        at (9.68, 4.15) {\textbf{Shared Attention \& Refinement (Section~\ref{sect:Attention})}};
\end{scope}

% Horizontal data-flow arrows.
\draw[flow] (s1r.east) -- (s2r.west)
    node[midway, above=2pt, smalllabel] {\(\bM\)};
\draw[flow] (s1s.east) -- (s2s.west)
    node[midway, above=2pt, smalllabel] {\(\bM\)};

% Merge Stage 2 outputs into shared Stage 3.
\draw (s2r.east) -- (8.35, 3.15) -- (8.35, 0.85) -- (s2s.east);
\fill (8.35, 2.00) circle (1.0pt);
\draw[flow] (8.35, 2.00) -- (s3.west)
    node[pos=0.45, above=3pt, smalllabel] {\(\bM,\mathcal{G}\)};

% Stage 3 to Stage 4a to Stage 4b.
\draw[flow] (s3.east |- s4a.west) -- (s4a.west)
    node[midway, above=2pt, smalllabel] {\(\bS_h,\bC_h\)};
\draw[flow] (s4a.south) -- (s4b.north)
    node[midway, right=3pt, smalllabel] {\emph{0 new \(Q\)}};

% Dashed adaptive query arrows from oracle.
\draw[qarrow] (6.15, 5.09) -- (6.15, 4.35)
    node[midway, left=4pt, smalllabel] {adaptive queries};
\draw[qarrow] (11.45, 5.09) -- (11.45, 4.35);
\draw[qarrow] (15.85, 5.09) -- (15.85, 4.35)
    node[midway, left=4pt, smalllabel] {Stage 4a only};

\end{tikzpicture}%
}
\caption{Overview of the four-stage precision-agnostic extraction
pipeline.
Sections~\ref{sect:ReLU} and~\ref{sect:SwiGLU} recover the aggregate
Attention value map \(\bM\) (Stage~1) and coupled FFN \(\mathcal{G}\) (Stage~2)
for biased ReLU and bias-free SwiGLU blocks, respectively.
Both branches then feed the same shared multi-head Attention extractor
(Stage~3) and fixed-rank manifold refinement (Stage~4) in
Section~\ref{sect:Attention}, where Stage~4b reuses Stage~4a responses
with zero additional oracle queries.}
\label{fig:extraction-pipeline}
\end{figure}

\section{Recovering the Aggregate Map and Biased ReLU FFN}
\label{sect:ReLU}
We begin with the biased ReLU~\cite{relu} branch of the
block~\eqref{eq:block}.
Given raw parameters \(\bA_1\in\mathbb R^{d\times m}\),
\(\bb_1\in\mathbb R^m\), \(\bA_2\in\mathbb R^{m\times d}\), and
\(\bb_2\in\mathbb R^d\), the raw ReLU FFN acts on a row vector
\(\bx\in\mathbb R^d\) as
\(\operatorname{FFN}(\bx)=\operatorname{ReLU}(\bx\bA_1+\bb_1)\bA_2+\bb_2\).
Absorbing the second RMSNorm gain \(\bD_2\) into the first linear map
(replacing \(\bA_1\) by \(\bD_2\bA_1\)) leaves \(\bb_1,\bA_2,\bb_2\)
unchanged and yields the effective FFN
\(\mathcal{G}(\bw):=\operatorname{FFN}(\bw\bD_2)\) given by
\begin{equation}
    \mathcal{G}(\bw)
    =
    \operatorname{ReLU}(\bw\bA_1+\bb_1)\bA_2+\bb_2,
    \qquad
    \operatorname{ReLU}(s)=\max\{s,0\}.
    \label{eq:relu}
\end{equation}
Because \(\mathcal{N}_2\) maps every residual into the open ball
\(\mathcal{B}_d=\{\bw\in\mathbb R^d:\|\bw\|_2<\sqrt{d}\}\), any neuron
whose activation hyperplane does not intersect the reachable subset of
\(\mathcal{B}_d\) remains identically inactive or identically active on
all allowable queries and cannot be uniquely separated from black-box
outputs.
For non-degenerate networks whose \(m\) activation hyperplanes are
distinct, have nonzero output vectors, intersect the reachable
interior of \(\mathcal{B}_d\), and satisfy the binary
sign-identifiability condition of~\eqref{eq:relu-sign-gamma}, the
identifiable FFN parameters are
\((\bA_1,\bb_1,\bA_2,\bb_2)\) up to neuron permutation and
positive per-neuron rescaling of \((\bA_1,\bb_1)\)
compensated in \(\bA_2\).

\subsection{Recovering the Aggregate Value Map}
\label{sec:relu-aggregate}

%For the biased ReLU block, we estimate \(\bM=\sum_{h=1}^{H}\bC_h\) from complete-block outputs. 
For each unit direction \(\bv\), we vary the number of copies of one input token, keeping the final token zero. %The resulting outputs let us estimate \(\bk=\bv\bM\) as the coefficient of a linear term.

\paragraph{Queries ending in a zero token.}
For a fixed \(r_{\mathrm{in}}>0\), set \(\bp=r_{\mathrm{in}}\bv\).
For \(1\leq q<\ell\), use the sequence
\(\bX_q=[\underbrace{\bp,\ldots,\bp}_{q},
\underbrace{\boldsymbol0,\ldots,\boldsymbol0}_{\ell-q}]\),
provided that \(\bX_q\in\mathcal X\).
Write \(\mathcal N_1(\bp)=\tau\bv\), where
\(\tau=\|\mathcal N_1(\bp)\|_2\).
For \(q<i\leq\ell\), the normalized token is
\(\bu_i=\mathcal N_1(\boldsymbol0)=\boldsymbol0\); in particular,
\(\bu_\ell=\boldsymbol0\).
Substituting $\boldsymbol{u}_\ell = \mathbf{0}$ into the attention score formula in Eq. \eqref{eq:attention-weights} gives: 
$\boldsymbol{u}_\ell \boldsymbol{S}_h \boldsymbol{u}_i^\top = \mathbf{0} \cdot \boldsymbol{S}_h \boldsymbol{u}_i^\top = 0 \quad \text{for all } h, i$. 
Then, $\alpha_{h,i}(\boldsymbol{X}_q) = \frac{1}{\ell}, \quad i = 1, \dots, \ell$. 
According to Eq. \eqref{eq:attention}, we have 
$$\operatorname{Att}_{\boldsymbol{S},\boldsymbol{C}}(\boldsymbol{X}_q) = \sum_{h=1}^H \left(\sum_{i=1}^\ell \frac{1}{\ell}\boldsymbol{u}_i\right)\boldsymbol{C}_h = \sum_{h=1}^H \left(\frac{q}{\ell}\mathcal{N}_1(\boldsymbol{p})\right)\boldsymbol{C}_h=\frac{q\tau}{\ell}\boldsymbol{v}\sum_{h=1}^H \boldsymbol{C}_h = t\boldsymbol{v}\boldsymbol{M} = t\boldsymbol{k},$$
where $\boldsymbol{M}= \sum_{h=1}^H \boldsymbol{C}_h$ (the so-called Aggregate Value Map), $\boldsymbol{v}\boldsymbol{M} = \boldsymbol{k}$, and  \(t=q\tau/\ell\). 
Then, the first equation of Eq. \eqref{eq:block} reduces to:
$\boldsymbol{z}(\boldsymbol{X}_q) = \boldsymbol{x}_\ell + \operatorname{Att}_{\boldsymbol{S},\boldsymbol{C}}(\boldsymbol{X}_q) = \mathbf{0} + t\boldsymbol{k} = t\boldsymbol{k}.$
Then, applying $\mathcal{N}_2$ to $\boldsymbol{z} = t\boldsymbol{k}$ gives:
$$\mathcal{N}_2(t\boldsymbol{k}) = \frac{t\boldsymbol{k}}{\sqrt{\epsilon_2 + \Vert{}t\boldsymbol{k}\Vert{}_2^2 / d}} = \frac{t}{\sqrt{\epsilon_2 + \frac{\Vert{}\boldsymbol{k}\Vert{}_2^2}{d}\,t^2}}\,\boldsymbol{k}.$$
Defining the unknown scalar $\rho := \frac{\Vert{}\boldsymbol{k}\Vert{}_2^2}{d}$ and the scalar function $\varphi_\rho(t) := \frac{t}{\sqrt{\epsilon_2 + \rho t^2}}$, we obtain $\mathcal{N}_2(t\boldsymbol{k}) = \varphi_\rho(t)\boldsymbol{k}$. 
Then, according to Eq. \eqref{eq:block} and \eqref{eq:relu}, we have 
\begin{equation}
    \boldsymbol{y}(t) := \mathcal{F}(\boldsymbol{X}_q) = t\boldsymbol{k} + \mathcal{G}\!\left(\varphi_\rho(t)\boldsymbol{k}\right), \quad \varphi_\rho(t) = \frac{t}{\sqrt{\epsilon_2 + \rho t^2}}, \quad \rho = \frac{\Vert{}\boldsymbol{k}\Vert{}_2^2}{d}.
    \label{eq:relu-v11-response}
\end{equation}
The sample locations \(t\) are known from the inputs.
The vector \(\bk\) and the scalar \(\rho\) are unknown.

At a sampled value \(t\), the FFN input is
\(\bw(t)=\varphi_\rho(t)\bk\).
Suppose that each ReLU unit has the same activation state
at all sampled inputs. Let \(\bD\in\mathbb R^{m\times m}\)
be the diagonal matrix whose \(j\)-th entry is \(1\) if
unit \(j\) is active and \(0\) otherwise.
At these inputs, multiplying on the right by \(\bD\) keeps
the active preactivations and sets the rest to zero.
Substituting into~\eqref{eq:relu} gives
\begin{equation}
    \begin{aligned}
        \mathcal G(\bw(t))
        &=\bigl(\bw(t)\bA_1+\bb_1\bigr)\bD\bA_2+\bb_2=\varphi_\rho(t)\,\bk\bA_1\bD\bA_2
          +\bb_1\bD\bA_2+\bb_2.
    \end{aligned}
    \label{eq:relu-v11-ffn-fixed-pattern}
\end{equation}
Then, Eq. \eqref{eq:relu-v11-response} becomes
\begin{equation}
    \begin{aligned}
        \by(t)&=t\bk+\varphi_\rho(t)\bw_*+\bc_*, ~\mbox{with} ~ \bw_*=\bk\bA_1\bD\bA_2, 
        \bc_*=\bb_1\bD\bA_2+\bb_2.        
    \end{aligned}
    \label{eq:relu-v11-fixed-pattern}
\end{equation}
\paragraph{A first estimate of \(\bv\bM\).}
%The unknown vector \(\bk=\bv\bM\) is the coefficient of \(t\) in the term \(t\bk\) in~\eqref{eq:relu-v11-response}.
Let \(0<t_1<\cdots<t_n\), with \(n\geq3\), be the known
values of \(t\) used in the queries, and let 
\(\by(t_i)\in\mathbb R^d\) be the output. %denote the row output returned at \(t_i\).

Since \(t>0\), 
\(\varphi_\rho(t)=1/\sqrt{\rho+\epsilon_2/t^2}\). 
For \(\rho>0\), this is close to \(1/\sqrt{\rho}\)
when \(\epsilon_2/t^2 \ll \rho\).
The FFN inputs \(\varphi_\rho(t_i)\bk\) are then close
to one another. We approximate their FFN outputs by one
constant vector and fit each returned output by
\(\bc+t_i\ba\), with both \(\bc,\ba\in\mathbb R^d\) unknown.
Then we use the fitted coefficient of \(t_i\) as our first
estimate of \(\bk\):
\begin{equation}
    (\bc_{\mathrm{aff}},\bk_{\mathrm{aff}})
    =\arg\min_{\bc,\ba\in\mathbb R^d}
    \sum_{i=1}^{n}
    \bigl\|\by(t_i)-\bc-t_i\ba\bigr\|_2^2.
    \label{eq:relu-v11-first-fit}
\end{equation}

\paragraph{Updating the estimate of \(\bv\bM\) (Iterative Scale Refinement).}
The first fit gives the initial vector
\(\bk^{(0)}=\bk_{\mathrm{aff}}\).
Since the true vector satisfies \(\rho=\|\bk\|_2^2/d\),
we use the current estimate \(\bk^{(s)}\) to choose
a trial value for \(\rho\):
\begin{equation}
    \xi_s=\max\!\left(\frac{\|\bk^{(s)}\|_2^2}{d},10^{-18}\right).
    \label{eq:relu-v11-trial-scale}
\end{equation}
Once \(\xi_s\) is fixed, we can compute
\(\varphi_{\xi_s}(t_i)\) for every sample.
We then fit the same returned outputs using
\(\bc+t_i\ba+\varphi_{\xi_s}(t_i)\bw\), where
\(\bc,\ba,\bw\in\mathbb R^d\) are unknown. The update is
\begin{equation}
    (\bc^{(s+1)},\bk^{(s+1)},\bw^{(s+1)})
    =
    \arg\min_{\bc,\ba,\bw\in\mathbb R^d}
    \sum_{i=1}^{n}
    \bigl\|\by(t_i)-\bc-t_i\ba-\varphi_{\xi_s}(t_i)\bw\bigr\|_2^2.
    \label{eq:relu-v11-vector-update}
\end{equation}
Thus the coefficient of \(t_i\) gives the new estimate
\(\bk^{(s+1)}\). We use this vector to compute
\(\xi_{s+1}\) in the next update.
We refer to this update loop as \emph{iterative scale refinement};
all iterations reuse the same \(n\) outputs offline without
additional oracle queries.

\begin{proposition}[Local convergence]
\label{prop:relu-scale-refinement}
Suppose the outputs \(\by(t_1),\ldots,\by(t_n)\) are exact, share one
ReLU activation pattern~\eqref{eq:relu-v11-fixed-pattern} with
\(\rho>10^{-18}\), and the three vectors \((1,\ldots,1)\),
\((t_1,\ldots,t_n)\), and \((\varphi_\rho(t_1),\ldots,\varphi_\rho(t_n))\)
are linearly independent.
For a trial value \(\xi>0\), let \(\bk(\xi)\) be the fitted coefficient
of \(t_i\) in~\eqref{eq:relu-v11-vector-update} and set
\(\Phi(\xi):=\|\bk(\xi)\|_2^2/d\).
Then:
\begin{enumerate}
    \item \(\rho\) is a fixed point of~\eqref{eq:relu-v11-trial-scale}--\eqref{eq:relu-v11-vector-update}, with \(\Phi(\rho)=\rho\) and \(\bk(\rho)=\bk\).
    \item If \(|\Phi'(\rho)|<1\), then for any \(\eta\in(|\Phi'(\rho)|,1)\)
    there is a neighborhood of \(\rho\) such that, whenever \(\xi_0\)
    starts in it, the iterates satisfy
    \(|\xi_s-\rho|\leq \eta^s|\xi_0-\rho|\) and \(\bk^{(s)}\to\bk\).
    In particular, reaching \(|\xi_S-\rho|\leq\delta\) takes at most
    \begin{equation}
        S=\left\lceil
            \frac{\log(|\xi_0-\rho|/\delta)}{\log(1/\eta)}
          \right\rceil
        \label{eq:relu-v11-iterations}
    \end{equation}
    iterations.
\end{enumerate}
\end{proposition}

The proof of Proposition~\ref{prop:relu-scale-refinement} is deferred to
Appendix~\ref{app:relu-scale-refinement}, where we also show that for
fixed \(\bk,\bw_*\) and sample locations \(t_1,\ldots,t_n\), the
contraction factor scales as
\(|\Phi'(\rho)|=O\!\left((\epsilon_2/(\rho t_1^2))^2\right)\) as
\(\epsilon_2\to 0\).
In our benchmarks, \(|\Phi'(\rho)|\) is typically very small and the
iteration converges within a few steps; our implementation runs a fixed
\(S=6\) iterations per ray.

\paragraph{Constructing \(\bM\).}
For each basis vector \(\bv_j\), we apply the procedure to both
\(\bv_j\) and \(-\bv_j\).
Before combining their estimates, we negate the result for \(-\bv_j\)
so that both recover the same vector \(\bv_j\bM\).
Let \(\bk_j\) be the resulting estimate.
For an orthonormal basis of row vectors, let \(\bV\in\mathbb R^{d\times d}\)
have rows \(\bv_1,\ldots,\bv_d\) and let \(\bK\in\mathbb R^{d\times d}\)
have rows \(\bk_1,\ldots,\bk_d\).
Since the exact products form the rows of \(\bV\bM\), we assemble \(\bM\) by
\begin{equation}
    \bM=\bV^\top\bK.
    \label{eq:relu-v11-M}
\end{equation}

\subsection{Recovering the ReLU Feed-Forward Network under Finite Precision}
\label{sec:relu-ffn-recovery}

We use the aggregate map from Section~\ref{sec:relu-aggregate}
to obtain observations of the biased ReLU FFN \(\mathcal G\)
in~\eqref{eq:relu}. We then fit a wider ReLU network and
use its activation hyperplanes to initialize a network
of the target width \(m\).

\paragraph{Recovery target.}
Expanding~\eqref{eq:relu} across neurons gives
\begin{equation}
    \mathcal G(\bw)=\bb_2+\sum_{j=1}^m
      \operatorname{ReLU}\!\bigl(\bw\bA_1[\cdot,j]+\bb_1[j]\bigr)\,
      \bA_2[j,\cdot].
    \label{eq:relu-ffn-sum}
\end{equation}
By positive homogeneity of ReLU, dividing \(\bA_1[\cdot,j]\)
and \(\bb_1[j]\) by \(\|\bA_1[\cdot,j]\|_2\) while
multiplying \(\bA_2[j,\cdot]\) by the same norm leaves \(\mathcal G\)
unchanged.
We assume each neuron has nonzero \(\bA_1[\cdot,j]\) and
\(\bA_2[j,\cdot]\), distinct hyperplanes, and an activation hyperplane
intersecting the reachable interior of \(\mathcal{B}_d\).
Without loss of generality, we normalize each neuron so that
\(\|\bA_1[\cdot,j]\|_2=1\) (with \(|\bb_1[j]|<\sqrt{d}\)).
Our target is \(\bb_2\) and the normalized triples
\((\bA_1[\cdot,j],\bb_1[j],\bA_2[j,\cdot])\), up to permutation.
The equation \(\bw\bA_1[\cdot,j]+\bb_1[j]=0\) gives
the activation hyperplane, and the joint sign of
\((\bA_1[\cdot,j],\bb_1[j])\) specifies which side is active.

\paragraph{Obtaining FFN observations.}
Submit a one-token query \(\bX=[\bx]\in\mathcal X\) (\(\ell=1\)).
Since there is only one token, \(\bu_1=\mathcal N_1(\bx)\), and every head
in~\eqref{eq:attention-weights} assigns weight \(\alpha_{h,1}([\bx])=1\).
Substituting into~\eqref{eq:attention} eliminates all score matrices
\(\bS_h\) and collapses the Attention output to the aggregate map \(\bM\):
\begin{equation}
    \operatorname{Att}_{\bS,\bC}([\bx])
    =\sum_{h=1}^{H}\alpha_{h,1}([\bx])\,\bu_1\bC_h
    =\mathcal N_1(\bx)\sum_{h=1}^{H}\bC_h
    =\mathcal N_1(\bx)\bM.
\end{equation}
    
Because \(\bM\) was recovered in Section~\ref{sec:relu-aggregate} and
\(\mathcal N_1,\mathcal N_2\) are public, we can compute both the first
residual and the FFN input in~\eqref{eq:block} directly from \(\bx\):
\begin{equation}
    \bz(\bx)=\bx+\mathcal N_1(\bx)\bM,\qquad
    \bw(\bx)=\mathcal N_2(\bz(\bx)).
    \label{eq:relu-singleton-residual}
\end{equation}
Moreover,~\eqref{eq:block} gives
\(\mathcal F([\bx])=\bz(\bx)+\mathcal G(\bw(\bx))\).
Subtracting the known vector \(\bz(\bx)\) from the oracle output
\(\mathcal F([\bx])\) therefore isolates the FFN output:
\begin{equation}
    \mathcal F([\bx])-\bz(\bx)=\mathcal G(\bw(\bx)).
    \label{eq:relu-forward-bridge}
\end{equation}
Hence each one-token query \([\bx]\) directly yields an FFN
input--output pair \(\bigl(\bw(\bx),\mathcal G(\bw(\bx))\bigr)\),
without knowing the individual heads \((\bS_h,\bC_h)\).

\paragraph{Fitting a wider ReLU network.}
Prior ReLU extraction attacks~\cite{carlini2020,ReluAbad} locate
critical points using tiny-step finite differences along straight lines
in the input space.
In our block, however, tiny-step finite differences are drowned out by
rounding noise under low precision (FP16/BF16), and the FFN input
\(\bw(\bx)\) passes through two nonlinear RMSNorm layers, preventing
direct binary search along lines in \(\bw\)-space.
We therefore recover \(\mathcal G\) by global fitting on macroscopic
pairs \(\bigl(\bw(\bx_i),\mathcal G(\bw(\bx_i))\bigr)\) obtained
via~\eqref{eq:relu-singleton-residual}--\eqref{eq:relu-forward-bridge}
for sampled tokens \(\bx_1,\ldots,\bx_N\).
Because training a width-\(m\) ReLU network from random initialization
often falls into poor local minima and misses target hyperplanes, we
first fit a wider network of width \(K=2m\) with parameters
\(\bW_1\in\mathbb R^{d\times K}\), \(\bc_1\in\mathbb R^K\),
\(\bW_2\in\mathbb R^{K\times d}\), and \(\bc_2\in\mathbb R^d\):
\begin{equation}
    \mathcal G_K(\bw)
    =\bc_2
     +\sum_{j=1}^{K}
       \operatorname{ReLU}\!
       \bigl(\bw\bW_1[\cdot,j]+\bc_1[j]\bigr)\,\bW_2[j,\cdot].
    \label{eq:relu-wide-network}
\end{equation}
The column \(\bW_1[\cdot,j]\) and bias \(\bc_1[j]\)
determine where unit \(j\) changes its activation state, and
the row \(\bW_2[j,\cdot]\) determines its contribution
to the network output.
The extra width allows the optimizer to capture all target hyperplanes
(possibly split across multiple units), which the next steps merge and
prune back to width \(m\).

To sample both sides of every activation hyperplane across
\(\mathcal B_d\), we collect two groups of random one-token queries:
an outer group placing \(\bw(\bx)\) near the boundary of
\(\mathcal B_d\), and an inner group distributed across its interior.
Each of the two input groups is divided into a fitting
subset and a selection subset. The fitting subsets
are used to update the parameters and the selection
subsets are used to compare the resulting networks.
Within either set, a sample belonging to a group
of size \(n_g\) receives squared-error weight
\(\sigma_i^2=1/(2n_g)\). Thus each group contributes
half of the total weight.
Let \(\bP=[\bW_1^\top\ \ \bc_1^\top]\in\mathbb R^{K\times(d+1)}\)
contain the hidden parameters, and let
\(\bT\in\mathbb R^{(K+1)\times d}\) contain the output
parameters, with rows \(\bW_2[1,\cdot],\ldots,\bW_2[K,\cdot],\bc_2\).
The fitting objective compares predicted and
observed outputs:
\begin{equation}
    \mathcal L(\bP,\bT)
    =\frac1d\sum_{i=1}^{N}
      \sigma_i^2
      \bigl\|\mathcal G_K(\bw(\bx_i))-\mathcal G(\bw(\bx_i))\bigr\|_2^2.
    \label{eq:relu-weighted-loss}
\end{equation}

The two parameter groups play different roles in
this objective. Once \(\bP\) is fixed, every hidden
activation is known, and the predicted outputs
depend linearly on \(\bT\).
To express this linear problem, let \(\bX\) and
\(\bY\) have rows \([\bw(\bx_i)\ \ 1]\) and
\(\mathcal G(\bw(\bx_i))\), respectively, and define
\begin{equation}
    \begin{aligned}
        \bD_\sigma
        &=\operatorname{diag}(\sigma_1,\ldots,\sigma_N),\\
        \boldsymbol{H}(\bP)
        &=\bD_\sigma
          [\operatorname{ReLU}(\bX\bP^\top)
            \ \ \boldsymbol{1}_N^\top],\\
        \bY_w&=\bD_\sigma\bY,
    \end{aligned}
    \label{eq:relu-wide-design}
\end{equation}
where \(\boldsymbol{1}_N=(1,\ldots,1)\in\mathbb R^{N}\) is the row
vector of ones accounting for the output bias.
Consequently,
\begin{equation}
    \mathcal L(\bP,\bT)
    =\frac{1}{d}
      \|\boldsymbol{H}(\bP)\bT-\bY_w\|_F^2.
\end{equation}
A minimizing output layer is
\begin{equation}
    \bT=\boldsymbol{H}(\bP)^{\dagger}\bY_w,
    \label{eq:relu-wide-output-fit}
\end{equation}
where \(\dagger\) denotes the Moore--Penrose
pseudoinverse~\cite{penrose1955}.

Each fit starts from randomly initialized hidden weights and
biases together with small random output weights (or, at small
widths, an initial least-squares solve~\eqref{eq:relu-wide-output-fit}).
Full-batch Adam~\cite{kingma2015adam} with a width-scaled
learning-rate schedule then updates both layers to
reduce~\eqref{eq:relu-weighted-loss}.
After these updates, the output layer is solved in closed form
via~\eqref{eq:relu-wide-output-fit} with the current hidden
parameters fixed.
Joint Strong--Wolfe L-BFGS refinement~\cite{liu1989lbfgs}
further adjusts both layers.
Across the initializations and saved iterates,
the network with the lowest selection loss is retained.
Each retained unit with \(\bW_1[\cdot,j]\neq\boldsymbol0\)
provides a candidate hyperplane
\(\bw\bW_1[\cdot,j]+\bc_1[j]=0\)
for the next step.

\paragraph{Combining units on the same hyperplane.}
The wider network may represent one activation boundary
using several hidden units. Before selecting the target
hyperplanes, we therefore combine the contributions of
units that share the same boundary.

First, normalize each fitted unit with
\(\bW_1[\cdot,j]\neq\boldsymbol0\) by positive homogeneity,
dividing \(\bW_1[\cdot,j]\) and \(\bc_1[j]\) by
\(\|\bW_1[\cdot,j]\|_2\) and multiplying \(\bW_2[j,\cdot]\)
by \(\|\bW_1[\cdot,j]\|_2\).
This preserves the unit's output and gives its normal
vector unit length (\(\|\bW_1[\cdot,j]\|_2=1\)).
Units with zero input weights contribute only constants,
which can be included in \(\bc_2\).

Consider a group \(\mathcal C\) whose units share
one hyperplane. Choose a reference equation
\begin{equation}
    f(\bw)=\bw\bn^\top+\beta=0,
    \qquad \|\bn\|_2=1.
    \label{eq:relu-group-reference}
\end{equation}
Each normalized unit in this group has input
\(s_j f(\bw)\), where \(s_j\in\{-1,1\}\).
The sign specifies which side of the common
hyperplane activates that unit.
Units with opposite active sides can be compared
using
\(\operatorname{ReLU}(-f)=\operatorname{ReLU}(f)-f\).
Consequently, the group's total contribution is
\begin{equation}
    \begin{aligned}
    \sum_{j\in\mathcal C}
        \operatorname{ReLU}(s_j f(\bw))\,\bW_2[j,\cdot]
    =
        \operatorname{ReLU}(f(\bw))\,\bo_{\mathcal C}
        -f(\bw)
          \sum_{\substack{j\in\mathcal C\\s_j=-1}}
              \bW_2[j,\cdot]
    \end{aligned}
    \label{eq:relu-folding-identity}
\end{equation}
where \(\bo_{\mathcal C}
    =\sum_{j\in\mathcal C}\bW_2[j,\cdot]\).
The first term contains the change in derivative
at the hyperplane. The second term is affine in
\(\bw\) and records the contribution of units
active on the opposite side.

Let \(\boldsymbol J_+\) and \(\boldsymbol J_-\)
be the Jacobians of this group's contribution
on \(f(\bw)>0\) and \(f(\bw)<0\), respectively
(acting on row perturbations from the right).
The affine term has the same derivative on both
sides, so
\begin{equation}
    \boldsymbol J_+-\boldsymbol J_-
    =\bn^\top\bo_{\mathcal C}.
    \label{eq:relu-group-jump}
\end{equation}
Whereas prior ReLU extraction exploits the derivative jump across a
single neuron's boundary~\cite{carlini2020} and its
outer-product Jacobian form~\cite{ReluAbad},
\eqref{eq:relu-group-jump} extends this identity to a co-hyperplane
group: its net jump vector \(\bo_{\mathcal C}\) is the sum of the
group's normalized output vectors, even across units active on
opposite sides.

\begin{proposition}[Hyperplanes visible in an exact fit]
\label{prop:relu-folding}
Suppose that the target activation hyperplanes are
distinct and their normalized output vectors are
nonzero. If \(\mathcal G_K=\mathcal G\) on an open set
\(\mathcal W\subseteq\mathbb R^d\), then every target
hyperplane intersecting \(\mathcal W\) occurs among
the fitted units. The sum of the normalized fitted
output vectors on that hyperplane equals the
normalized output vector of the target unit.
On any fitted hyperplane intersecting \(\mathcal W\)
that is absent from the target network, this sum
is zero.
\end{proposition}

\begin{proof}
Fix a hyperplane \(\mathcal H\) of either network
that intersects \(\mathcal W\).
Choose a point of \(\mathcal H\cap\mathcal W\)
that lies on no other distinct hyperplane of either
network. Such a point exists because the finitely
many other hyperplanes cannot cover a relatively
open subset of \(\mathcal H\).
A sufficiently small ball around this point lies
in \(\mathcal W\) and meets no other hyperplane.

On this ball, all units whose hyperplanes differ
from \(\mathcal H\) have fixed activation states.
Their contributions are affine and hence have
no Jacobian change across \(\mathcal H\).
Since the two networks agree throughout the ball,
their Jacobian changes across \(\mathcal H\) agree.
Choose a common unit normal \(\bn\) for
\(\mathcal H\). By~\eqref{eq:relu-group-jump},
left-multiplying each Jacobian change by \(\bn\)
gives the corresponding summed normalized output
vector, since \(\bn\bn^\top=1\).
For a target hyperplane, this vector is nonzero,
so a corresponding fitted group must exist and
have the same sum. For a hyperplane absent from
the target, the target has no Jacobian change,
so the fitted sum must be zero.
\end{proof}

In numerical fits, units representing the same boundary
may have slightly different normals and biases.
We therefore group nearby hyperplanes and rank the
resulting groups for selection.

\paragraph{Ranking the fitted hyperplanes.}
A unit can have large output weights while remaining
active at every fitting input. Its sampled contribution
is then affine and does not reveal an activation
boundary. We therefore rank a candidate by the part
of its sampled contribution that cannot be explained
by an affine function. For a group with reference
equation \(f(\bw)=\bw\bn^\top+\beta=0\) and summed
output vector \(\bo_{\mathcal C}\), define
\begin{equation}
    r_{\mathcal C}^2
    =
    \min_{\bv\in\mathbb R^d,\ c\in\mathbb R}
    \sum_{i=1}^{N}\sigma_i^2
    \bigl[
        \operatorname{ReLU}(f(\bw(\bx_i)))
        -\bw(\bx_i)\bv^\top-c
    \bigr]^2.
    \label{eq:relu-affine-score-fit}
\end{equation}
Here \(r_{\mathcal C}\) is the residual after fitting
the candidate's ReLU values with an affine function.
Multiplying by the size of its output vector gives
the score
\begin{equation}
    \eta_{\mathcal C}
    =r_{\mathcal C}\|\bo_{\mathcal C}\|_2.
    \label{eq:relu-kink-energy}
\end{equation}
This accounts for both the change expressed in the
sampled ReLU values and its contribution to the
vector output.

The affine fit is a linear least-squares problem.
Let \(h_i=\sigma_i\operatorname{ReLU}(f(\bw(\bx_i)))\)
and let \(\boldsymbol h=(h_1,\ldots,h_N)\in\mathbb R^{N}\).
Using \(\bX\) and \(\bD_\sigma\) from
\eqref{eq:relu-wide-design}, let
\(\bP_{\mathrm{aff}}\) project orthogonally onto
the column space of \(\bD_\sigma\bX\).
This space contains the weighted sample values
of all affine scalar functions. Hence
\begin{equation}
    \begin{aligned}
        r_{\mathcal C}
        &=\|(\bI_N-\bP_{\mathrm{aff}})
              \boldsymbol h^\top\|_2,\\
        \eta_{\mathcal C}
        &=\|(\bI_N-\bP_{\mathrm{aff}})
              \boldsymbol h^\top\bo_{\mathcal C}\|_F.
    \end{aligned}
    \label{eq:relu-score-projection}
\end{equation}
If all fitting inputs lie on one side of the
hyperplane, the ReLU values are affine on these
inputs and the score is zero. The score is also
unchanged when the reference equation is reversed:
\(\operatorname{ReLU}(-f)=\operatorname{ReLU}(f)-f\),
and the projection removes the affine difference
\(f(\bw)\). For the same reason, the affine term
in~\eqref{eq:relu-folding-identity} does not affect
the score.

For numerical grouping, first compute the score
of each normalized unit with a nonzero output
vector. Compare two units using their normals
and biases, allowing a simultaneous sign change:
\begin{equation}
    \delta_{jk}
    =
    \min_{s\in\{-1,1\}}
    \left(
        \|\bW_1[\cdot,j]-s\bW_1[\cdot,k]\|_2^2
        +|\bc_1[j]-s\bc_1[k]|^2
    \right)^{1/2}.
    \label{eq:relu-hyperplane-distance}
\end{equation}
For each chosen threshold \(\tau>0\), connect pairs
with \(\delta_{jk}\leq\tau\); units linked through
these connections form a group. Within each group,
align the normals and biases to the highest-scoring
member, then average them using the individual
scores as weights. Normalize the averaged normal
to unit length and scale the bias by the same factor.
If all scores are zero, use the reference member;
if the averaged normal is zero, discard the group.
Sum the normalized output vectors without applying
the orientation sign changes, as justified by
\eqref{eq:relu-folding-identity}.
Recompute each group's score using its representative
hyperplane and summed output vector, and retain the
\(m\) highest-scoring groups as a candidate set.
Thresholds yielding fewer than \(m\) groups are
discarded. The next step determines the active
sides of the selected hyperplanes.

\paragraph{Determining the active sides.}
An activation hyperplane determines where a neuron
switches, but does not determine which side is active.
For each selected hyperplane, choose a reference
equation
\(f_j(\bw)=\bw\bA_1[\cdot,j]+\bb_1[j]=0\),
with \(\|\bA_1[\cdot,j]\|_2=1\).
To derive the recovery equations, suppose that these
are the exact target hyperplanes, in some order.
The target unit then uses either
\(\operatorname{ReLU}(f_j(\bw))\) or
\(\operatorname{ReLU}(-f_j(\bw))\).
Introduce \(z_j\in\{0,1\}\), where \(z_j=0\)
keeps the reference direction and \(z_j=1\)
reverses it. The identity
\begin{equation}
    \operatorname{ReLU}((1-2z_j)f_j(\bw))
    =
    \operatorname{ReLU}(f_j(\bw))-z_j f_j(\bw)
    \label{eq:relu-sign-identity}
\end{equation}
shows that reversing a unit changes its contribution
by an affine function. Consequently, the target FFN
can be written as
\begin{equation}
    \begin{aligned}
        \mathcal G(\bw)
        &=\sum_{j=1}^{m}\operatorname{ReLU}(f_j(\bw))\,\bA_2[j,\cdot]
          +\bw\bA+\bc,\\
        \bA
        &=-\sum_{j=1}^{m}
          z_j\,\bA_1[\cdot,j]\,\bA_2[j,\cdot],\qquad
        \bb_2
        =\bc+\sum_{j=1}^{m}
          z_j\,\bb_1[j]\,\bA_2[j,\cdot].
    \end{aligned}
    \label{eq:relu-affine-sign-expansion}
\end{equation}

The first line of~\eqref{eq:relu-affine-sign-expansion}
is linear in
\(\bA_2[1,\cdot],\ldots,\bA_2[m,\cdot],\bA,\bc\).
These coefficients are fitted to the FFN observations
by weighted least squares, using the reference ReLU
values, the input coordinates, and a constant as
features. With exact observations and a
full-column-rank design, the coefficients are
uniquely determined.
The fitted matrix \(\bA\in\mathbb R^{d\times d}\)
then identifies which units must be reversed through
\begin{equation}
    \begin{aligned}
        \boldsymbol\Gamma\bz^\top
        &=-\operatorname{vec}(\bA),\\
        \boldsymbol\Gamma[\cdot,j]
        &=\operatorname{vec}\!
          \bigl(\bA_1[\cdot,j]\,\bA_2[j,\cdot]\bigr),\qquad
        \bz=(z_1,\ldots,z_m),
    \end{aligned}
    \label{eq:relu-sign-gamma}
\end{equation}
where
\(\boldsymbol\Gamma\in\mathbb R^{d^2\times m}\)
and \(\operatorname{vec}\) stacks matrix entries
in a fixed order.
If \(\boldsymbol\Gamma\) is injective on
\(\{0,1\}^m\), the binary solution is unique.
In particular, full column rank of
\(\boldsymbol\Gamma\) allows the exact solution
to be obtained by a linear solve over the reals.
The recovered directions are
\((1-2z_j)(\bA_1[\cdot,j],\bb_1[j])\), and the last
line of~\eqref{eq:relu-affine-sign-expansion}
recovers the output bias \(\bb_2\).

For numerical fits, the reference directions
inherited from the wider network give one candidate.
A second candidate is obtained by solving
\eqref{eq:relu-sign-gamma} over the reals and
rounding the solution to binary entries.
This candidate is considered only when both
linear systems pass full-rank and conditioning
checks and the solution is sufficiently close
to a binary vector.
For each candidate, fix the resulting directions
and refit the output weights and bias on the
fitting set.
Among candidates with full-column-rank output
designs, retain the one with the lowest selection
loss. If none has full column rank, compare the
available candidates by selection loss.
The selected network initializes the final
width-\(m\) refinement.

\paragraph{Refining the width-\(m\) network.}
The selected hyperplanes and active sides provide
an initial network with \(m\) hidden units.
Its parameters are refined using the FFN observations.
Let \(\bP\in\mathbb R^{m\times(d+1)}\) contain
the hidden weights and biases, and let
\(\bT\in\mathbb R^{(m+1)\times d}\) contain
the output weights and bias.
For each trial \(\bP\), the output parameters are
recomputed by weighted least squares.
Substituting this solution into the fitting loss
leaves an objective in \(\bP\) alone:
\begin{equation}
    \begin{aligned}
        \bT(\bP)
        &=\boldsymbol H(\bP)^\dagger\bY_w,\\
        \bR(\bP)
        &=\boldsymbol H(\bP)\bT(\bP)-\bY_w,\\
        \mathcal L(\bP)
        &=\min_{\bT}\mathcal L(\bP,\bT)
          =\frac1d\|\bR(\bP)\|_F^2.
    \end{aligned}
    \label{eq:relu-varpro-loss}
\end{equation}
Here \(\boldsymbol H(\bP)\) and \(\bY_w\) are
defined as in~\eqref{eq:relu-wide-design}, now
with \(m\) hidden units.
This is the variable-projection method of
Golub and Pereyra~\cite{golub1973differentiation}:
the nonlinear search adjusts the hidden parameters,
while each trial uses a least-squares output layer.

The refinement first applies L-BFGS
\cite{liu1989lbfgs} to
\(\mathcal L(\bP)\) on the full fitting set,
followed by damped Gauss--Newton steps on a
subset of that set.
Within a neighborhood where the sampled activation
states remain fixed and the output design has
full column rank, \(\bR(\bP)\) is smooth.
The Gauss--Newton Jacobian accounts for the
dependence of \(\bT(\bP)\) on the hidden parameters.
Each trial step recomputes the output layer;
the Gauss--Newton stage accepts a step only when
the subset fitting loss decreases and the output
design remains full column rank.
After each refinement stage, the output layer
is refitted on the full fitting set.
The updated network is retained only if its
selection loss is lower than that of the previous
candidate; otherwise, the previous candidate is kept.
The accepted width-\(m\) network defines the
recovered FFN \(\mathcal G\).

The resulting aggregate map \(\bM\) and FFN \(\mathcal G\) are used
to construct Attention observations through the shared
procedure in Section~\ref{sec:att-observations}.

\section{Recovering the Aggregate Map and Bias-Free SwiGLU FFN}
\label{sect:SwiGLU}
We now turn to the bias-free SwiGLU~\cite{swiglu} branch of the
block~\eqref{eq:block}.
Given raw weight matrices \(\bA_g,\bA_u\in\mathbb R^{d\times m}\) and
\(\bA\in\mathbb R^{m\times d}\), the raw SwiGLU FFN acts on a row
vector \(\bx\in\mathbb R^d\) as
\(\operatorname{FFN}(\bx)=(\operatorname{SiLU}(\bx\bA_g)\odot(\bx\bA_u))\bA\),
where \(\odot\) denotes coordinate-wise multiplication.
Absorbing the second RMSNorm gain \(\bD_2\) into the first two
projections (replacing \(\bA_g,\bA_u\) by \(\bD_2\bA_g,\bD_2\bA_u\)),
the effective FFN \(\mathcal G(\bw):=\operatorname{FFN}(\bw\bD_2)\)
expands across neurons as
\begin{equation}
    \mathcal{G}(\bw)
    =
    \sum_{j=1}^{m}
    \operatorname{SiLU}\!\bigl(\bw\bA_g[\cdot,j]\bigr)
    \bigl(\bw\bA_u[\cdot,j]\bigr)\,\bA[j,\cdot],
    \qquad
    \operatorname{SiLU}(s)=\frac{s}{1+\exp(-s)}.
    \label{eq:swiglu}
\end{equation}
The identifiable FFN parameters are the neuron pairs
\(\bigl(\bA_g[\cdot,j],\,\bA_u[\cdot,j]\bA[j,\cdot]\bigr)\) up to
neuron permutation.
For bias-free SwiGLU, the singleton derivative at the origin
provides a direct way to identify the aggregate value map \(\bM\).
The recovered aggregate map and FFN then feed into the shared
Attention reconstruction in Section~\ref{sect:Attention}.

\subsection{Recovering the Aggregate Value Map}
\label{sec:swiglu-aggregate}
We consider the bias-free SwiGLU block~\eqref{eq:block}
with \(\mathcal G\) given by~\eqref{eq:swiglu}.
Our first objective is to recover
\(\bM=\sum_{h=1}^{H}\bC_h\in\mathbb R^{d\times d}\)
without first identifying the individual heads or the FFN.
We construct approximate vector--matrix products
\(\ba_i\approx\bu_i\bM\) at known normalized inputs
\(\bu_i=\mathcal N_1(\bx_i)\), and estimate \(\bM\) by normalized linear
least squares.
\paragraph{Local structure of the bias-free SwiGLU FFN.}
Using the identity
\(\operatorname{SiLU}(s)=s/2+(s/2)\tanh(s/2)\)~\cite{swigluExtraction},
the FFN in~\eqref{eq:swiglu} splits into even and odd parts,
\(\mathcal G=\mathcal G_{\mathrm{even}}+\mathcal G_{\mathrm{odd}}\), where
\begin{equation}
    \begin{aligned}
        \mathcal G_{\mathrm{even}}(\bw)
        &=\frac12\sum_{j=1}^{m}
          \bigl(\bw\bA_g[\cdot,j]\bigr)\bigl(\bw\bA_u[\cdot,j]\bigr)\,
          \bA[j,\cdot],\\
        \mathcal G_{\mathrm{odd}}(\bw)
        &=\frac12\sum_{j=1}^{m}
          \bigl(\bw\bA_g[\cdot,j]\bigr)
          \tanh\!\left(\frac{\bw\bA_g[\cdot,j]}{2}\right)
          \bigl(\bw\bA_u[\cdot,j]\bigr)\,\bA[j,\cdot].
    \end{aligned}
    \label{eq:swiglu-aggregate-parity}
\end{equation}
Each term of \(\mathcal G_{\mathrm{even}}\) is quadratic in \(\bw\);
in particular,
\(\mathcal G_{\mathrm{even}}(-\bw)=\mathcal G_{\mathrm{even}}(\bw)\).
Near zero, \(\tanh(s/2)=s/2+O(s^3)\).
Substituting this expansion into \(\mathcal G_{\mathrm{odd}}\) gives
\begin{equation}
    \mathcal G_{\mathrm{odd}}(\bw)
    =\frac14\sum_{j=1}^{m}
      \bigl(\bw\bA_g[\cdot,j]\bigr)^2\bigl(\bw\bA_u[\cdot,j]\bigr)\,
      \bA[j,\cdot]
     +O(\|\bw\|_2^5),
    \qquad \bw\to\boldsymbol0.
    \label{eq:swiglu-aggregate-odd-leading}
\end{equation}
The displayed term is cubic in \(\bw\).
Thus \(\mathcal G\) has no constant or linear terms:
its even part is quadratic, while its odd part is
\(O(\|\bw\|_2^3)\) near zero.

\paragraph{Opposite one-token queries.}
For a one-token query \([\bx]\in\mathcal X\), let \(\bu=\mathcal N_1(\bx)\).
As in~\eqref{eq:relu-singleton-residual}, every head assigns attention
weight \(\alpha_{h,1}([\bx])=1\), so
\(\operatorname{Att}_{\bS,\bC}([\bx])=\bu\bM\) and the first residual
is \(\bz(\bx)=\bx+\bu\bM\).
Since both \(\mathcal N_1\) and \(\mathcal N_2\) are odd
(\(\mathcal N_j(-\bv)=-\mathcal N_j(\bv)\)), negating the token gives
\(\mathcal N_1(-\bx)=-\bu\), \(\bz(-\bx)=-\bz(\bx)\), and
\(\mathcal N_2(\bz(-\bx))=-\mathcal N_2(\bz(\bx))\).
Setting \(\bw=\mathcal N_2(\bx+\bu\bM)\) and splitting
\(\mathcal G=\mathcal G_{\mathrm{even}}+\mathcal G_{\mathrm{odd}}\)
in~\eqref{eq:block} yields
\begin{equation}
    \begin{aligned}
        \mathcal F([\bx])
        &=(\bx+\bu\bM)+\mathcal G_{\mathrm{even}}(\bw)+\mathcal G_{\mathrm{odd}}(\bw),\\
        \mathcal F([-\bx])
        &=-(\bx+\bu\bM)+\mathcal G_{\mathrm{even}}(-\bw)+\mathcal G_{\mathrm{odd}}(-\bw)\\
        &=-(\bx+\bu\bM)+\mathcal G_{\mathrm{even}}(\bw)-\mathcal G_{\mathrm{odd}}(\bw).
    \end{aligned}
\end{equation}
Taking half the difference cancels \(\mathcal G_{\mathrm{even}}(\bw)\),
and subtracting \(\bx\) isolates \(\bu\bM\) up to the odd FFN term:
\begin{equation}
    \begin{aligned}
        \ba_{\mathrm{odd}}(\bx)
        &:=\frac{\mathcal F([\bx])-\mathcal F([-\bx])}{2}-\bx\\
        &=\bu\bM+
          \mathcal G_{\mathrm{odd}}\bigl(\mathcal N_2(\bx+\bu\bM)\bigr).
    \end{aligned}
    \label{eq:swiglu-singleton}
\end{equation}
The first term is the vector--matrix product we seek and the second is
the residual odd FFN distortion.
Near zero, \(\mathcal N_1(\bx)=\bx/\sqrt{\epsilon_1+\|\bx\|_2^2/d}\)
gives \(\bx=\sqrt{\epsilon_1}\bu+O(\|\bu\|_2^3)\), so
\(\bx+\bu\bM=O(\|\bu\|_2)\) and hence
\(\mathcal N_2(\bx+\bu\bM)=O(\|\bu\|_2)\).
Since \(\mathcal G_{\mathrm{odd}}(\bw)=O(\|\bw\|_2^3)\) by
\eqref{eq:swiglu-aggregate-odd-leading}, we obtain
\begin{equation}
    \ba_{\mathrm{odd}}(\bx)
    =\bu\bM+O(\|\bu\|_2^3),
    \qquad \bu\to\boldsymbol0.
    \label{eq:swiglu-singleton-product-approximation}
\end{equation}
Thus small opposite one-token pairs yield cubic-accuracy approximate
vector--matrix products for linear reconstruction of \(\bM\).

\paragraph{Underflow fallback via zero-padded extrapolation.}
In narrow-exponent formats such as FP16, reducing \(\|\bx\|_2\) to
make~\eqref{eq:swiglu-singleton-product-approximation} negligible can
cause the input vector to underflow to zero before entering the model.
To handle this without branching on the target's numerical format, we
first run an 8-query test on four small random opposite pairs
\([\pm\bx_i]\) (RMS radius in \([10^{-15},3\times 10^{-15}]\)) and check
whether all raw half-differences \(\tfrac12(\mathcal F([\bx_i])-\mathcal F([-\bx_i]))\)
(evaluated \emph{before} subtracting \(\bx_i\)) are finite and nonzero.
When this test succeeds (as occurs in FP64, FP32, and BF16), we use the
one-token estimator~\eqref{eq:swiglu-singleton}.
When it underflows (as occurs automatically in FP16), we keep the
nonzero token \(\bx\) strictly above the FP16 subnormal threshold
(RMS radius in \([6.2\times 10^{-5},\,8.0\times 10^{-5}]\)) and switch
automatically to zero-padded multi-copy sequences whose uniform
Attention weights attenuate the first residual further without
causing input underflow.

For integers \(1\leq k<\ell\), define
\begin{equation}
    \bX_{k,\ell}(\bx)
    =\bigl[
        \underbrace{\bx,\ldots,\bx}_{k\ \mathrm{tokens}},
        \underbrace{\boldsymbol0,\ldots,\boldsymbol0}
                   _{\ell-k\ \mathrm{tokens}}
     \bigr],
    \qquad
    \alpha=\frac{k}{\ell}.
    \label{eq:swiglu-zero-padded-input}
\end{equation}
Let \(\bu=\mathcal N_1(\bx)\) and \(\ba=\bu\bM\).
For \(\bX_{k,\ell}(\bx)\in\mathcal X\), the final token is
\(\bx_\ell=\boldsymbol0\), so \(\bu_\ell=\mathcal N_1(\boldsymbol0)=\boldsymbol0\).
Substituting \(\bu_\ell=\boldsymbol0\) into~\eqref{eq:attention-weights}
makes every attention score \(\bu_\ell\bS_h\bu_i^\top=0\), giving uniform
weights \(\alpha_{h,i}(\bX_{k,\ell}(\bx))=1/\ell\) for \(i=1,\ldots,\ell\).
By~\eqref{eq:attention}, the Attention output is therefore
\begin{equation}
    \operatorname{Att}_{\bS,\bC}\bigl(\bX_{k,\ell}(\bx)\bigr)
    =\sum_{h=1}^{H}\left(\frac{k}{\ell}\bu\right)\bC_h
    =\frac{k}{\ell}\bu\sum_{h=1}^{H}\bC_h
    =\alpha\bu\bM
    =\alpha\ba.
\end{equation}
Since \(\bx_\ell=\boldsymbol0\), the first residual in~\eqref{eq:block}
reduces to \(\bz(\bX_{k,\ell}(\bx))=\boldsymbol0+\alpha\ba=\alpha\ba\),
so the complete-block output is
\begin{equation}
    \mathcal F(\bX_{k,\ell}(\bx))
    =\alpha\ba+\mathcal G\bigl(\mathcal N_2(\alpha\ba)\bigr).
    \label{eq:swiglu-zero-padded-response}
\end{equation}
For the negated sequence \(\bX_{k,\ell}(-\bx)\), oddness of
\(\mathcal N_1\) and \(\mathcal N_2\) replaces \(\alpha\ba\) by
\(-\alpha\ba\) and \(\mathcal N_2(\alpha\ba)\) by
\(-\mathcal N_2(\alpha\ba)\), giving
\begin{equation}
    \mathcal F(\bX_{k,\ell}(-\bx))
    =-\alpha\ba
     +\mathcal G_{\mathrm{even}}\bigl(\mathcal N_2(\alpha\ba)\bigr)
     -\mathcal G_{\mathrm{odd}}\bigl(\mathcal N_2(\alpha\ba)\bigr).
\end{equation}
Taking half the difference between \(\bX_{k,\ell}(\bx)\) and
\(\bX_{k,\ell}(-\bx)\) again cancels \(\mathcal G_{\mathrm{even}}\):
\begin{equation}
    \by_\alpha
    =\frac{
        \mathcal F(\bX_{k,\ell}(\bx))
        -\mathcal F(\bX_{k,\ell}(-\bx))
      }{2}
    =\alpha\ba+
      \mathcal G_{\mathrm{odd}}\bigl(\mathcal N_2(\alpha\ba)\bigr),
    \qquad \alpha=\frac{k}{\ell}.
    \label{eq:swiglu-aggregate-odd-identity}
\end{equation}
Within the analyticity radius \(\alpha\|\ba\|_2/\sqrt{d}<\sqrt{\epsilon_2}\)
of \(\mathcal N_2(\alpha\ba)\), the smooth odd map
\(\mathcal G_{\mathrm{odd}}\circ \mathcal N_2\) expands in odd powers
starting at cubic:
\begin{equation}
    \frac{\by_\alpha}{\alpha}
    =\ba+\alpha^2\bc_3(\ba)+\alpha^4\bc_5(\ba)
     +O\bigl(\alpha^6\|\ba\|_2^7\bigr),
    \qquad \alpha\|\ba\|_2/\sqrt{d}<\sqrt{\epsilon_2},
\end{equation}
where \(\bc_3(\ba)\) and \(\bc_5(\ba)\) are homogeneous of degrees three
and five, respectively.

Fixing a sequence length \(\ell\) (the implementation uses \(\ell=256\))
and querying \(k\in\{1,2,3\}\) copies of \(\pm\bx\) with
\((3/\ell)\|\ba\|_2/\sqrt{d}<\sqrt{\epsilon_2}\) yields three
rescaled odd responses \(\br_k:=(\ell/k)\by_{k/\ell}\) with ratios
\(\alpha_k=k/\ell\):
\begin{equation}
    \br_k
    =\ba+\frac{k^2}{\ell^2}\bc_3(\ba)+\frac{k^4}{\ell^4}\bc_5(\ba)
     +O\bigl(\ell^{-6}\|\ba\|_2^7\bigr),
    \qquad k\in\{1,2,3\}.
\end{equation}
Solving the \(3\times 3\) Vandermonde weights in \((1,k^2,k^4)\) gives
three-point Richardson extrapolation~\cite{richardson1927}:
\begin{equation}
    \ba_{\mathrm{ext}}(\bx)
    :=\frac{3}{2}\br_1-\frac{3}{5}\br_2+\frac{1}{10}\br_3
    =\ba+O\bigl(\ell^{-6}\|\ba\|_2^7\bigr).
    \label{eq:swiglu-aggregate-extrapolation}
\end{equation}
Indeed, \(\frac32-\frac35+\frac1{10}=1\),
\(\frac32(1)-\frac35(4)+\frac1{10}(9)=0\), and
\(\frac32(1)-\frac35(16)+\frac1{10}(81)=0\), canceling both the cubic
and quintic distortion simultaneously using six complete-block queries
per direction.

\paragraph{Recovering \(\bM\) from the observations.}
Both tiers provide approximate products
\(\ba_i\approx\bu_i\bM\).
For input vectors \(\bx_1,\ldots,\bx_n\), let
\(\bu_i=\mathcal N_1(\bx_i)\neq\boldsymbol0\) and
\(s_i=\|\bu_i\|_2\).
Let \(\ba_i\) denote the target produced by
the active tier.
Normalize each row vector by the same \(s_i\) and stack
them as rows:
\begin{equation}
    \bU=
    \begin{pmatrix}
        \bu_1/s_1\\
        \vdots\\
        \bu_n/s_n
    \end{pmatrix},
    \qquad
    \bY=
    \begin{pmatrix}
        \ba_1/s_1\\
        \vdots\\
        \ba_n/s_n
    \end{pmatrix}
    \in\mathbb R^{n\times d}.
    \label{eq:swiglu-aggregate-linear-design}
\end{equation}
When \(\bU\) has column rank \(d\), we recover \(\bM\)
by linear least squares:
\begin{equation}
    \bM
    =\arg\min_{\bT\in\mathbb R^{d\times d}}
      \|\bU\bT-\bY\|_F^2
    =(\bU^\top\bU)^{-1}\bU^\top\bY.
    \label{eq:swiglu-aggregate-least-squares}
\end{equation}
In particular, when the normalized probe directions \(\bu_i/s_i\)
comprise \(\nu\geq 1\) orthonormal bases of \(\mathbb R^d\), we have
\(\bU^\top\bU=\nu\bI\) and~\eqref{eq:swiglu-aggregate-least-squares}
reduces directly to \(\bM=\nu^{-1}\bU^\top\bY\).
The recovered matrix \(\bM\) is then used to recover the FFN.

\subsection{Recovering the SwiGLU Feed-Forward Network under Finite Precision}
\label{sec:swiglu-ffn-recovery}

We now use the aggregate map \(\bM\) from
Section~\ref{sec:swiglu-aggregate} to obtain observations of the
FFN \(\mathcal G\) in~\eqref{eq:swiglu}.
For neuron \(j\), write \(\bP_j=\bA_u[\cdot,j]\,\bA[j,\cdot]\in\mathbb R^{d\times d}\).
The function depends on \(\bA_u[\cdot,j]\) and \(\bA[j,\cdot]\) through
\(\bP_j\), since \((\bw\bA_u[\cdot,j])\bA[j,\cdot]=\bw\bP_j\).
The recovery target is the collection of pairs \((\bA_g[\cdot,j],\bP_j)\)
up to a permutation of the neurons.

\paragraph{Obtaining FFN observations from the block.}
As in Section~\ref{sec:relu-ffn-recovery}, querying a one-token input
\([\bx]\in\mathcal X\) and subtracting the first residual via \(\bM\)
gives an FFN input--output pair \(\bigl(\bw(\bx),\mathcal G(\bw(\bx))\bigr)\) by
\eqref{eq:relu-singleton-residual}--\eqref{eq:relu-forward-bridge}.
When a query requires an FFN input with normalized RMS radius close to a
prescribed \(\rho\in(0,1)\), we fix a unit direction \(\bv\), evaluate
\(r_{\bv}(t)=\|\bw(t\sqrt d\,\bv)\|_2/\sqrt d\) over a grid
of scales \(t\), refine any bracketing interval by bisection, and query
the block at the resulting token \(\bx=t\sqrt d\,\bv\).

Because \(\mathcal N_1\) and \(\mathcal N_2\) are odd, opposite queries \([\pm\bx]\)
produce \(\bz(-\bx)=-\bz(\bx)\) and \(\bw(-\bx)=-\bw(\bx)\).
After subtracting the two residuals separately, the half-difference and
half-sum of the FFN outputs separate the odd and even parts of \(\mathcal G\).
Following Shi and Fu~\cite{swigluExtraction}, let
\(\phi(s)=\tfrac{s}{2}\tanh(s/2)\); then
\begin{equation}
    \begin{aligned}
        \mathcal G_{\mathrm{odd}}(\bw)
        &=\sum_{j=1}^{m}\phi\bigl(\bw\bA_g[\cdot,j]\bigr)\,\bw\bP_j,\\
        \mathcal G_{\mathrm{even}}(\bw)
        &=\frac12\sum_{j=1}^{m}\bigl(\bw\bA_g[\cdot,j]\bigr)\,\bw\bP_j.
    \end{aligned}
    \label{eq:swiglu-ffn-odd-even}
\end{equation}
Since \(\phi\) is even, replacing \(\bA_g[\cdot,j]\) by \(-\bA_g[\cdot,j]\) leaves
\(\mathcal G_{\mathrm{odd}}\) unchanged while reversing neuron \(j\)'s sign in
\(\mathcal G_{\mathrm{even}}\). We therefore use the even responses later to
resolve gate-vector signs.

\paragraph{Hessians of the odd part.}
The Hessians of the odd output contain matrices whose
row spaces give the gate directions. To see where
these matrices come from, fix an input \(\bq\) and an
output coordinate \(i\). By~\eqref{eq:swiglu-ffn-odd-even},
the contribution of neuron \(j\) to this coordinate is
\begin{equation}
    \phi\bigl(\bq\bA_g[\cdot,j]\bigr)\,(\bq\bP_j)[i]
    =\bA[j,i]\,\phi\bigl(\bq\bA_g[\cdot,j]\bigr)\bigl(\bq\bA_u[\cdot,j]\bigr).
    \label{eq:swiglu-ffn-neuron-odd}
\end{equation}
Here we have expanded \(\bP_j=\bA_u[\cdot,j]\,\bA[j,\cdot]\)
to show the two factors that depend on \(\bq\).
Differentiating the gate factor twice produces
\(\bA_g[\cdot,j]\bA_g[\cdot,j]^\top\). Differentiating the gate and the
linear factor once each produces the two cross terms.
Define
\begin{equation}
    \bA_j=\bA_g[\cdot,j]\bA_g[\cdot,j]^\top,\qquad
    \bB_j=\bA_g[\cdot,j]\bA_u[\cdot,j]^\top+\bA_u[\cdot,j]\bA_g[\cdot,j]^\top.
    \label{eq:swiglu-ffn-generators}
\end{equation}
Applying the product rule, as in the Hessian decomposition
of Shi and Fu~\cite{swigluExtraction},
gives the Hessian of~\eqref{eq:swiglu-ffn-neuron-odd}
with respect to \(\bq\):
\begin{equation}
    \bA[j,i]\Bigl[
        \bigl(\bq\bA_u[\cdot,j]\bigr)\phi''\bigl(\bq\bA_g[\cdot,j]\bigr)\bA_j
        +\phi'\bigl(\bq\bA_g[\cdot,j]\bigr)\bB_j
    \Bigr].
    \label{eq:swiglu-ffn-neuron-hessian}
\end{equation}
Varying \(\bq\) or \(i\) changes the coefficients of
\(\bA_j\) and \(\bB_j\). The matrices themselves are
fixed. For a nonzero gate vector, \(\bA_j\) has rank
one and its column space is the line spanned by \(\bA_g[\cdot,j]\).

We use Hessians averaged around each chosen input \(\bw\).
Let \(\boldsymbol{\zeta}\sim
\mathcal N(\boldsymbol{0},\sigma^2\bI)\), where
\(\sigma>0\) sets the scale of the perturbations. Define
\begin{equation}
    \boldsymbol{H}_{i,\sigma}(\bw)
    =\mathbb E_{\boldsymbol{\zeta}}\!
       \left[
       \left.\nabla_{\bq}^{2}\mathcal G_{\mathrm{odd}}(\bq)[i]
       \right|_{\bq=\bw+\boldsymbol{\zeta}}
       \right].
    \label{eq:swiglu-ffn-smoothed-hessian}
\end{equation}
The average in~\eqref{eq:swiglu-ffn-smoothed-hessian}
acts on the scalar coefficients in
\eqref{eq:swiglu-ffn-neuron-hessian}. It therefore preserves
the matrix space shared by the Hessians. Summing over
the \(m\) neurons gives
\begin{equation}
    \boldsymbol{H}_{i,\sigma}(\bw)\in\mathcal E,\qquad
    \mathcal E=\operatorname{span}
       \{\bA_j,\bB_j:1\leq j\leq m\}.
    \label{eq:swiglu-ffn-hessian-span}
\end{equation}

\paragraph{Estimating the subspace from FFN observations.}
Shi and Fu~\cite{swigluExtraction} estimate Hessians by second-order
finite differences of isolated FFN responses.
In our block, however, second-order finite differences amplify
low-precision (FP16/BF16) rounding noise by the inverse square of the
step size, and the nonlinear RMSNorm maps prevent direct coordinate
perturbations in \(\bw\)-space.
Instead, we fit a wider SwiGLU network \(\mathcal G_K\) with \(K>m\)
neurons to macroscopic FFN input--output pairs from
\eqref{eq:relu-singleton-residual}--\eqref{eq:relu-forward-bridge}
(the extra width avoids poor local minima of exact-width training),
and evaluate its smoothed Hessians analytically from the fitted weights.
Specifically, at each chosen input \(\bw\), we average the Hessians of
the odd part of \(\mathcal G_K\) at nearby inputs \(\bw+\boldsymbol{\zeta}\),
as in~\eqref{eq:swiglu-ffn-smoothed-hessian}. Here
\(\boldsymbol{\zeta}\) is a Gaussian offset added to \(\bw\).
For each neuron, only the part of the offset along its
gate vector changes the gate input. With that part fixed,
opposite offsets perpendicular to the gate contribute
equally and with opposite signs through the linear value
term. These contributions cancel in the average. We
therefore integrate only over the offset along the gate
direction.

We represent each resulting Hessian by the entries of its upper
triangle, with off-diagonal entries multiplied by \(\sqrt{2}\) to
preserve the Frobenius norm~\cite{swigluExtraction}.
Stacking these vectors across the chosen centers and output coordinates
forms a matrix whose leading \(2m\) right singular vectors estimate the
subspace \(\mathcal E\).

\paragraph{Recovering directions from the subspace.}
Following Shi and Fu~\cite{swigluExtraction},
we search for unit row vectors \(\bv\) whose rank-one matrices
\(\bv^\top\bv\) are close to \(\mathcal E\).
In the exact space \(\mathcal E\), the true gate directions
give distance zero. To see when these are the only such
directions, write a general matrix in \(\mathcal E\) as
\begin{equation}
    \mathcal T(\bc)=\sum_{j=1}^{m}
        \bigl(\bc[2j-1]\bA_j+\bc[2j]\bB_j\bigr),
    \qquad \bc\in\mathbb R^{2m}.
    \label{eq:swiglu-ffn-hessian-combination}
\end{equation}
We now ask which nonzero matrices \(\mathcal T(\bc)\) have rank one.
This happens exactly when the determinant of every
\(2\times2\) submatrix of \(\mathcal T(\bc)\) is zero. Since each
entry of \(\mathcal T(\bc)\) is linear in \(\bc\), each determinant
is a sum of terms involving products \(\bc[a]\bc[b]\).
We collect these products in \(\bc^\top\bc\), whose
\((a,b)\)-entry is \(\bc[a]\bc[b]\). To write the determinant
conditions as linear equations, replace each product
\(\bc[a]\bc[b]\) by the entry \(\bW[a,b]\) of a symmetric matrix
\(\bW\in\mathbb R^{2m\times2m}\). Let
\(\Phi_{\mathcal E}(\bW)\) be the vector of the resulting
linear expressions, one for each determinant. Thus a
nonzero \(\mathcal T(\bc)\) has rank one exactly when
\(\bc^\top\bc\in\ker\Phi_{\mathcal E}\).

\begin{proposition}[Gate directions in the exact subspace]
\label{prop:swiglu-ffn-gate-directions}
Suppose the \(2m\) matrices in
\eqref{eq:swiglu-ffn-generators} are linearly independent
and the exact averaged Hessians at the chosen inputs and
output coordinates span \(\mathcal E\).
If \(\dim\ker\Phi_{\mathcal E}=2m\), then the unit row vectors
\(\bv\) satisfying \(\bv^\top\bv\in\mathcal E\) are
precisely \(\pm\bA_g[\cdot,j]^\top/\|\bA_g[\cdot,j]\|_2\), for \(1\leq j\leq m\).
\end{proposition}

\begin{proof}
Let \(\be_k\) denote the \(k\)-th standard coordinate row vector of
\(\mathbb R^{2m}\). For neuron \(j\), put \(k_j=2j-1\).
The matrix \((\bA_g[\cdot,j]+t\bA_u[\cdot,j])(\bA_g[\cdot,j]+t\bA_u[\cdot,j])^\top\)
has rank at most one. Its expansion is
\(\bA_j+t\bB_j+t^2\bA_u[\cdot,j]\bA_u[\cdot,j]^\top\). Consequently, the
constant and linear terms of the minors of
\(\bA_j+t\bB_j\) vanish. Applying
\(\Phi_{\mathcal E}\) to
\((\be_{k_j}+t\be_{k_j+1})^\top(\be_{k_j}+t\be_{k_j+1})\)
and comparing these two terms shows that
\(\be_{k_j}^\top\be_{k_j}\) and
\(\be_{k_j}^\top\be_{k_j+1}+\be_{k_j+1}^\top\be_{k_j}\)
belong to the kernel. Over all \(j\), these give \(2m\)
independent matrices, so they form a basis of the kernel.

If \(\mathcal T(\bc)=\bv^\top\bv\) for a unit vector \(\bv\),
then \(\bc^\top\bc\) belongs to this kernel. The basis
matrices have zero entries on the even diagonal positions
and between distinct odd indices. Thus \(\bc[2j]^2=0\)
for all \(j\), and \(\bc[2j-1]\bc[2k-1]=0\) for \(j\ne k\).
Exactly one coefficient is nonzero, so \(\bv^\top\bv\)
is a multiple of a single \(\bA_j\). Its row space
gives \(\bv=\pm\bA_g[\cdot,j]^\top/\|\bA_g[\cdot,j]\|_2\). Conversely,
\(\bA_j/\|\bA_g[\cdot,j]\|_2^2\) belongs to \(\mathcal E\) for
every \(j\).
\end{proof}

The rank-one search gives an initial estimate of each gate
direction. We then seek a second direction to describe the
part of the value vector perpendicular to the gate.

To see how this second direction appears in the Hessians,
let \(\bu_j\) be a unit row vector along \(\bA_g[\cdot,j]^\top\),
and let \(\boldsymbol{t}_j\) be the unit row direction of the
component of \(\bA_u[\cdot,j]^\top\) perpendicular to \(\bu_j\). We can
then write
\(\bA_u[\cdot,j]^\top=b_{j,1}\bu_j+b_{j,2}\boldsymbol{t}_j\).
The independence of \(\bA_j\) and \(\bB_j\) ensures that
\(b_{j,2}\ne 0\), since otherwise \(\bA_u[\cdot,j]\) would be
parallel to \(\bA_g[\cdot,j]\). Substituting this expression into
\(\bB_j=\bA_g[\cdot,j]\bA_u[\cdot,j]^\top+\bA_u[\cdot,j]\bA_g[\cdot,j]^\top\)
shows that \(\bB_j\) is a linear combination of
\(\bu_j^\top\bu_j\) and
\(\bu_j^\top\boldsymbol{t}_j+\boldsymbol{t}_j^\top\bu_j\),
with a nonzero coefficient on the second matrix.
Thus both matrices belong to \(\mathcal E\).

We start with a candidate gate direction \(\bv\) and
introduce a unit vector \(\boldsymbol{t}\) perpendicular
to \(\bv\). We then refine both vectors, keeping them
at unit length and maintaining
\(\bv\boldsymbol{t}^\top=0\).
We minimize the sum of the squared Frobenius distances
from the two matrices
\begin{equation}
    \bv^\top\bv,\qquad
    \frac{\bv^\top\boldsymbol{t}+\boldsymbol{t}^\top\bv}{\sqrt{2}}
    \label{eq:swiglu-ffn-pair-matrices}
\end{equation}
to \(\mathcal E\).
The first matrix constrains the gate direction \(\bv\),
and the second constrains the perpendicular direction
\(\boldsymbol{t}\) using the gate--value cross term.
A correct pair gives a basis for the plane containing the
gate and value vectors. We use this basis later to fit
\(\bP_j\) from the odd responses.

If the search finds fewer than \(m\) pairs, we project out
the matrix directions associated with the pairs already
found and search again. We check each new pair against
the original space \(\mathcal E\).

\paragraph{Estimating gate lengths.}
Once the gate direction \(\bu_j\) is known, two quantities
remain in \(\bA_g[\cdot,j]^\top=s_j\alpha_j\bu_j\): the length
\(\alpha_j=\|\bA_g[\cdot,j]\|_2>0\) and the sign \(s_j\in\{-1,1\}\).
We estimate the length from the Hessians and determine
the sign later from the even responses. Shi and Fu estimate
the lengths and coupling matrices jointly from odd
responses~\cite{swigluExtraction}.

For each input center \(\bw_p\) and output coordinate \(i\),
we express the averaged Hessian as a linear combination
of the two matrices associated with each recovered pair
\((\bu_j,\boldsymbol{t}_j)\):
\begin{equation}
    \bu_j^\top\bu_j,
    \qquad
    \bu_j^\top\boldsymbol{t}_j+\boldsymbol{t}_j^\top\bu_j.
\end{equation}
We fit all \(2m\) coefficients by linear least squares.
Let \(h_{j,ip}\) be the coefficient of the second matrix
for neuron \(j\). By repeating this fit, we obtain one
such coefficient for each center and output coordinate.

To see how this coefficient depends on the gate length,
substitute \(\bA_g[\cdot,j]^\top=s_j\alpha_j\bu_j\) and
\(\bA_u[\cdot,j]^\top=b_{j,1}\bu_j+b_{j,2}\boldsymbol{t}_j\) into
\(\bB_j=\bA_g[\cdot,j]\bA_u[\cdot,j]^\top+\bA_u[\cdot,j]\bA_g[\cdot,j]^\top\).
The second matrix above appears in \(\bB_j\) with
coefficient \(s_j\alpha_j b_{j,2}\).
In the Hessian, \(\bB_j\) is multiplied by
\(\phi'(s_j\alpha_j\bq\bu_j^\top)\). Since \(\phi'\) is odd,
this supplies another factor of \(s_j\), giving
\(s_j^2=1\). Averaging around \(\bw_p\), with
\(\boldsymbol{\zeta}\bu_j^\top\) distributed as
\(\sigma z\) for \(z\sim\mathcal N(0,1)\), gives
\begin{equation}
    \begin{aligned}
        h_{j,ip}
        &= \bA[j,i]\,\alpha_j b_{j,2}\,
           \bp_j(\alpha_j)[p],\\
        \bp_j(a)[p]
        &= \mathbb E_{z\sim\mathcal N(0,1)}
           \bigl[\phi'(a(\bw_p\bu_j^\top+\sigma z))\bigr].
    \end{aligned}
    \label{eq:swiglu-ffn-magnitude-coefficient}
\end{equation}
The first equality holds for exact directions and
Hessians. The second specifies what we compute at each
center for a trial length \(a>0\).

For a fixed output coordinate \(i\), the factor
\(\bA[j,i]\alpha_j b_{j,2}\) is the same at every center.
Thus every row of the coefficient matrix
\((h_{j,ip})_{i,p}\) is a scaled copy of the row vector
\(\bp_j(\alpha_j)\). We estimate the common vector
up to scale by taking the leading right singular vector
\(\by_j\) of the fitted coefficient matrix. We then
compare \(\by_j\) with the vector predicted by each
trial length \(a\), allowing one scale factor \(c\) for
all centers:
\begin{equation}
    \min_{a\in\mathcal I,\,c\in\mathbb R}
        \|\by_j-c\,\bp_j(a)\|_2^2,
    \qquad \mathcal I\subset(0,\infty).
    \label{eq:swiglu-ffn-magnitude-fit}
\end{equation}
The fit therefore uses the relative changes across
centers. With exact data, it identifies \(\alpha_j\)
provided that the coefficient matrix is nonzero,
\(\alpha_j\in\mathcal I\), and \(\alpha_j\) is the only
length in \(\mathcal I\) whose predicted vector is a
nonzero multiple of \(\bp_j(\alpha_j)\).
\paragraph{Recovering the matrices \(\bP_j\) and the gate signs.}
With the directions and lengths fixed, we return to the
odd responses to recover \(\bP_j\). Define
\(\bc_{j,1}=b_{j,1}\bA[j,\cdot]\) and
\(\bc_{j,2}=b_{j,2}\bA[j,\cdot]\). Since \(\phi\) is even,
substitution into~\eqref{eq:swiglu-ffn-odd-even} gives
\begin{equation}
    \mathcal G_{\mathrm{odd}}(\bw)
    =\sum_{j=1}^{m}\phi(\alpha_j\bw\bu_j^\top)
        \bigl[(\bw\bu_j^\top)\bc_{j,1}
             +(\bw\boldsymbol{t}_j^\top)\bc_{j,2}\bigr].
    \label{eq:swiglu-ffn-coupling-fit}
\end{equation}
Each input gives a row of \(2m\) known scalar weights
multiplying the \(2m\) unknown row vectors. With exact
responses and correct directions and lengths, full column
rank determines these vectors. In particular,
\(\bP_j=\bu_j^\top\bc_{j,1}+
\boldsymbol{t}_j^\top\bc_{j,2}\).
In the numerical fit, we take a rank-one approximation of
the \(2\times d\) matrix with rows \(\bc_{j,1},\bc_{j,2}\)
to obtain the two factors of \(\bP_j\).

Finally, the even responses give linear equations for the
remaining signs \(s_j\)~\cite{swigluExtraction}:
\begin{equation}
    \mathcal G_{\mathrm{even}}(\bw)
    =\frac12\sum_{j=1}^{m}
        s_j\alpha_j(\bw\bu_j^\top)\,\bw\bP_j.
    \label{eq:swiglu-ffn-sign-fit}
\end{equation}
With exact responses and exact directions, lengths, and
matrices \(\bP_j\), full column rank determines the signs
uniquely. In the numerical fit, we solve for real
coefficients by least squares and round each to the
nearest of \(-1\) and \(1\). With the signs fixed, we set
\(\bA_g[\cdot,j]=s_j\alpha_j\bu_j^\top\). These columns and the fitted
factors of \(\bP_j\) initialize a SwiGLU network with \(m\)
hidden neurons. We then use damped Gauss--Newton to refine
all FFN parameters jointly by fitting the network outputs
to the original FFN observations.

\section{Recovering Multi-Head Attention and Full-Block Refinement}
\label{sect:Attention}
\subsection{Obtaining Attention Outputs from Complete-Block Queries}
\label{sec:att-observations}
\label{sec:relu-attention-observations}

The preceding sections recover the aggregate value map and
the FFN for the two model branches. The next task is to obtain
the Attention output of a chosen sequence from the complete
block output. This step is shared by ReLU and SwiGLU.

\paragraph{Using a sequence and its negation.}
For an input \(\bX=[\bx_1,\ldots,\bx_\ell]\), define
\begin{equation}
    \Psi_{\mathcal G}(\bz)=\bz+\mathcal G(\mathcal N_2(\bz)),
    \qquad
    \bz_{\bX}=\bx_\ell+\operatorname{Att}_{\bS,\bC}(\bX).
    \label{eq:relu-known-residual-map}
\end{equation}
Thus \(\mathcal F(\bX)=\Psi_{\mathcal G}(\bz_{\bX})\). Recovering \(\bz_{\bX}\)
gives the Attention output by subtracting \(\bx_\ell\).
A single response need not determine \(\bz_{\bX}\) uniquely,
so we also query the sequence \(-\bX\), obtained by
negating every input token. Both sequences must belong
to the permitted query domain \(\mathcal X\).

Since \(\mathcal N_1\) is odd, negating the tokens leaves every
Attention score unchanged:
\begin{equation}
    (-\bu_\ell)\bS_h(-\bu_i)^\top
    =\bu_\ell\bS_h\bu_i^\top,
    \qquad \bu_i=\mathcal N_1(\bx_i).
    \label{eq:relu-negated-scores}
\end{equation}
The weights are therefore unchanged, whereas every
value vector changes sign. Consequently,
\begin{equation}
    \operatorname{Att}_{\bS,\bC}(-\bX)
    =-\operatorname{Att}_{\bS,\bC}(\bX),
    \label{eq:relu-attention-oddness}
\end{equation}
and the first residual of \(-\bX\) is \(-\bz_{\bX}\).
The two responses give the paired row equation
\begin{equation}
    \mathcal P_{\mathcal G}(\bz_{\bX})=\by_{\bX},
    \qquad
    \mathcal P_{\mathcal G}(\bz)=
    \bigl[\Psi_{\mathcal G}(\bz),\ \Psi_{\mathcal G}(-\bz)\bigr],
    \qquad
    \by_{\bX}=
    \bigl[\mathcal F(\bX),\ \mathcal F(-\bX)\bigr].
    \label{eq:relu-paired-observations}
\end{equation}
Here \(\mathcal P_{\mathcal G}:\mathbb R^d\to\mathbb R^{2d}\)
is known once \(\mathcal G\) is recovered.

\paragraph{Starting from a known hidden state.}
To initialize the inversion, choose a normalized vector
\(\bq\in\mathcal B_d\), set \(\bx_0=\mathcal N_1^{-1}(\bq)\),
and use \(\ell\) identical tokens. Every head then returns
\(\bq\bC_h\), giving the known first residual
\begin{equation}
    \bX_0=[\bx_0,\ldots,\bx_0],
    \qquad
    \bz_0=\bx_0+\bq\bM.
    \label{eq:relu-paired-anchor}
\end{equation}
This construction does not require \(\bM\) to be invertible.
Connect \(\bX_0\) to the target \(\bX\) by a continuous
path \(\bX(\tau)\), \(0\leq\tau\leq1\), for which
both \(\bX(\tau)\) and \(-\bX(\tau)\) belong to
\(\mathcal X\). One such construction interpolates the
normalized tokens and its formula and input constraints
are given in Appendix~\ref{app:relu-paired-continuation}.
Along the path, write
\begin{equation}
    \by(\tau)=
    \bigl[\mathcal F(\bX(\tau)),\ \mathcal F(-\bX(\tau))\bigr]
    =\mathcal P_{\mathcal G}(\bz(\tau)),
    \label{eq:relu-continuation-data}
\end{equation}
where \(\bz(\tau)\) is the true first residual at \(\bX(\tau)\).
At successive path points, the preceding recovered
state provides a starting estimate for the next solve.

\paragraph{Solving the paired equations.}
In the numerical procedure, we evaluate \(\mathcal P_{\mathcal G}\)
with the recovered FFN \(\mathcal G\) and let \(\by_{j+1}\)
contain the two returned outputs at the next path point \(\tau_{j+1}\).
We solve the paired equations for the unknown first residual \(\bz\),
keeping the parameters of \(\mathcal G\) fixed.
Define the equation residual by
\begin{equation}
    \br_j(\bz)
    =\mathcal P_{\mathcal G}(\bz)-\by_{j+1}.
    \label{eq:att-numerical-residual}
\end{equation}
Let \(\boldsymbol{J}(\bz)\in
\mathbb R^{d\times 2d}\) denote the matrix used to
linearize these equations (acting on row perturbations from
the right). At differentiable points,
it equals \(D\mathcal P_{\mathcal G}(\bz)\) and is computed
from the recovered FFN and the known normalization map.
For ReLU, the implementation uses the current activation
pattern, assigning derivative zero at a zero preactivation.
No target-model derivatives are requested.

Given the preceding estimate \(\bz_j\),
a linear least-squares prediction first estimates the
required change:
\begin{equation}
    \bdelta_j
    =-\br_j(\bz_j)\,
       \boldsymbol{J}(\bz_j)^\dagger.
    \label{eq:att-paired-predictor}
\end{equation}
The size of this prediction sets a local search bound
\(b_j>0\), with a positive minimum bound.
Starting from \(\bz_j+\bdelta_j\), clipped
to the interior of the search region, we approximately
solve
\begin{equation}
    \min_{\|\bz-\bz_j\|_\infty\leq b_j}
        \frac12\|\br_j(\bz)\|_2^2.
    \label{eq:att-paired-least-squares}
\end{equation}
The constraint limits the change in each coordinate
relative to the preceding estimate.
The implementation uses a trust-region reflective
least-squares solver, following the bound-constrained
approach of Branch et al.~\cite{branch1999subspace}.
It updates the Jacobian as \(\bz\) changes.
After this solve, additional linear least-squares
corrections
\(-\br_j(\bz)\boldsymbol{J}(\bz)^\dagger\)
are retained only when they stay inside the search
region and decrease the full residual norm.

A finite candidate in the search region is accepted
only when its full residual norm and Jacobian condition
number satisfy the prescribed thresholds. Thus a small
optimization gradient alone is insufficient.
To avoid format-specific branching across FP64, FP32, FP16,
and BF16, the residual threshold is self-calibrated online
before inversion: eight opposite uniform-token anchor pairs
(16 complete-block queries) measure the empirical residual
floor \(e_{\mathrm{floor}}\) between the observed responses
and \(\mathcal P_{\mathcal G}(\bz_0)\), and the
homotopy acceptance tolerance is set to
\(\tau_{\mathrm{hom}}=\max(10^{-7},\,10\,e_{\mathrm{floor}})\).
On failure, the procedure retries with a larger local
bound and, if necessary, inserts an intermediate path
point. The initial value
\(\bz_0=\bx_0+\bq\bM\)
is also checked against the paired responses and locally
corrected when needed. The number of iterations, path
subdivisions, and alternative starting sequences is
bounded; if these attempts fail, no Attention label is
returned. All inner iterations evaluate only the recovered
\(\mathcal G\). Additional target queries are needed
when new path points are introduced.

At the final path point, subtracting \(\bx_\ell\) recovers the
Attention output:
\begin{equation}
    \operatorname{Att}_{\bS,\bC}(\bX)
    =\bz_{\mathrm{end}}-\bx_\ell.
    \label{eq:att-numerical-label}
\end{equation}
Whenever \(\Psi_{\mathcal G}\) is continuously differentiable near \(\pm\bz_{\bX}\) and
the paired Jacobian
\(D\mathcal P_{\mathcal G}(\bz_{\bX})=\bigl[D\Psi_{\mathcal G}(\bz_{\bX}),\,-D\Psi_{\mathcal G}(-\bz_{\bX})\bigr]\in\mathbb R^{d\times 2d}\)
has full row rank \(d\), the inverse function theorem ensures that
\(\mathcal P_{\mathcal G}(\bz)=\by_{\bX}\) has \(\bz_{\bX}\) as its unique local
solution; convergence of continuation along the path from the known
anchor~\eqref{eq:relu-paired-anchor} is analyzed in
Appendix~\ref{app:relu-paired-continuation}.

\subsection{Separating Heads with Symmetric Token Sequences}
\label{sec:att-macro-anchor}

Section~\ref{sec:att-observations} provides Attention outputs
from complete-block queries.
We now use these observations to separate the contributions
of different heads.

For a normalized token sequence
\(\bU=[\bu_1,\ldots,\bu_\ell]\in\mathbb R^{\ell\times d}\), write
\begin{equation}
    \mathcal A(\bU)
    =
    \operatorname{Att}_{\bS,\bC}\bigl(\mathcal N_1^{-1}(\bU)\bigr),
    \label{eq:att-normalized-observations}
\end{equation}
where \(\mathcal N_1^{-1}\) acts tokenwise.
In the numerical procedure, \(\mathcal A(\bU)\) is evaluated via the
paired inversion~\eqref{eq:att-numerical-label}.

\paragraph{Constructing the observations.}
Fix positive numbers \(a_1,\ldots,a_k\), and vary
\(\bq\) in the symmetric sequence
\begin{equation}
    \bU_{\bu_0}(\bq)
    =
    [\bq+a_1\bu_0,\ \bq-a_1\bu_0,\ \ldots,\
     \bq+a_k\bu_0,\ \bq-a_k\bu_0,\ \bq].
    \label{eq:att-centered-probe}
\end{equation}
The last token is \(\bq\), and each preceding pair
has the same midpoint \(\bq\).
The implementation uses three pairs, giving seven
tokens. We choose \(\bq\), \(\bu_0\), and \(a_1,\ldots,a_k\)
so that every normalized token lies in \(\mathcal B_d\)
and the corresponding physical sequence, obtained by
applying \(\mathcal N_1^{-1}\) to each token, belongs to the
query domain \(\mathcal X\).

For a fixed direction \(\bu_0\), form \(\bV,\bT\in\mathbb R^{H\times d}\)
with rows
\begin{equation}
    \bV[h,\cdot]=\bu_0\bC_h,
    \qquad
    \bT[h,\cdot]=\bu_0\bS_h^\top,
    \qquad h=1,\ldots,H,
    \label{eq:att-anchor-factors}
\end{equation}
and set
\begin{equation}
    t_h=\bq\bS_h\bu_0^\top
        =\bq\bT[h,\cdot]^\top.
    \label{eq:att-anchor-score}
\end{equation}
The scores of \(\bq+a_j\bu_0\) and
\(\bq-a_j\bu_0\) are
\(\bq\bS_h\bq^\top+a_jt_h\) and
\(\bq\bS_h\bq^\top-a_jt_h\), respectively.
The final token has score \(\bq\bS_h\bq^\top\).
The common score \(\bq\bS_h\bq^\top\) cancels from
the softmax weights. The weights assigned to
\(\bq+a_j\bu_0\), \(\bq-a_j\bu_0\), and the final
token \(\bq\) are therefore
\begin{equation}
    \frac{e^{a_jt_h}}{z(t_h)},
    \qquad
    \frac{e^{-a_jt_h}}{z(t_h)},
    \qquad
    \frac{1}{z(t_h)},
    \label{eq:att-symmetric-weights}
\end{equation}
respectively, where
\begin{equation}
    z(t)
    =1+\sum_{j=1}^{k}
        \bigl(e^{a_jt}+e^{-a_jt}\bigr)
    =1+2\sum_{j=1}^{k}\cosh(a_jt).
    \label{eq:att-symmetric-denominator}
\end{equation}
Here the term \(1\) corresponds to the final token.
The softmax-weighted average of the normalized tokens
in this head is therefore
\begin{equation}
    \begin{aligned}
        &\frac{
          \bq+
          \sum_{j=1}^{k}
          \bigl[
            e^{a_jt_h}(\bq+a_j\bu_0)
            +e^{-a_jt_h}(\bq-a_j\bu_0)
          \bigr]}
          {z(t_h)}
        \\
        &\qquad=\bq+\psi(t_h)\bu_0,
        \qquad
        \psi(t)=
        \frac{2\sum_{j=1}^{k}a_j\sinh(a_jt)}
             {1+2\sum_{j=1}^{k}\cosh(a_jt)}.
    \end{aligned}
    \label{eq:att-symmetric-head-average}
\end{equation}
Thus each head contributes
\(\bq\bC_h+\psi(\bq\bT[h,\cdot]^\top)\,\bV[h,\cdot]\).
Summing over heads and subtracting the aggregate
term gives
\begin{equation}
    \begin{aligned}
        \mathcal Y_{\bu_0}(\bq)
        &:=
        \mathcal A(\bU_{\bu_0}(\bq))-\bq\bM\\
        &=\sum_{h=1}^{H}
          \psi\bigl(\bq\bT[h,\cdot]^\top\bigr)\,\bV[h,\cdot]
         =\psi(\bq\bT^\top)\bV,
    \end{aligned}
    \label{eq:att-macro-ridge}
\end{equation}
where \(\psi\) acts on each coordinate.
The function \(\psi\) is known from the chosen token
offsets and is odd. In particular,
\begin{equation}
    \psi'(0)
    =
    \frac{2\sum_{j=1}^{k}a_j^2}{1+2k}>0.
    \label{eq:att-psi-origin}
\end{equation}

\paragraph{The common structure across queries.}
The Jacobian of the true response (acting on row perturbations
from the right) is
\begin{equation}
    \boldsymbol{J}(\bq)
    =
    D\mathcal Y_{\bu_0}(\bq)
    =
    \bT^\top\bD(\bq)\bV,
    \qquad
    \bD(\bq)
    =
    \operatorname{diag}
       \bigl(\psi'(\bq\bT^\top)\bigr).
    \label{eq:att-macro-jacobian}
\end{equation}
Changing \(\bq\) therefore changes only the diagonal
factor. The left and right factors remain the same.

To recover the rows of \(\bV\), we first remove
the common left factor \(\bT^\top\).
Since \(\boldsymbol{J}(\bq)\) has rank at most \(H\),
this is done after reducing it to an \(H\times H\)
matrix.

\begin{proposition}[Separating the head value directions]
\label{prop:att-anchor-derivative-separation}
Suppose that \(\bV,\bT\in\mathbb R^{H\times d}\)
have row rank \(H\).
Let \(\bq_0,\ldots,\bq_s\) be points at which the query
construction is defined in a neighborhood, and suppose
that their exact derivative matrices
\(\boldsymbol{J}(\bq_0),\ldots,\boldsymbol{J}(\bq_s)\)
are available.
Use \(\bq_0\) as the reference point and assume that
\(\bD(\bq_0)\) is invertible.

For head \(h\), define the derivative ratio
\begin{equation}
    \lambda_h(\bq)
    =
    \frac{\psi'\bigl(\bq\bT[h,\cdot]^\top\bigr)}
         {\psi'\bigl(\bq_0\bT[h,\cdot]^\top\bigr)}.
    \label{eq:att-head-derivative-profile}
\end{equation}
For every pair of distinct heads \(h\neq g\), assume
that at least one evaluation point \(\bq_j\),
\(1\leq j\leq s\), satisfies
\(\lambda_h(\bq_j)\neq\lambda_g(\bq_j)\).
Then the derivative matrices determine the unordered
set of lines
\(\{\operatorname{span}(\bV[h,\cdot]):1\leq h\leq H\}\).
\end{proposition}

\begin{proof}
The rank assumptions imply that
\(\boldsymbol{J}(\bq_0)\) has rank \(H\).
Moreover, its column space equals the row space of \(\bT\),
and its row space equals the row space of \(\bV\).
An SVD of \(\boldsymbol{J}(\bq_0)\) therefore provides
matrices
\begin{equation}
    \bW,\bU\in\mathbb R^{d\times H},
    \qquad
    \bW^\top\bW=\bU^\top\bU=\bI_H,
    \label{eq:att-compression-bases}
\end{equation}
whose columns form orthonormal bases of these two
spaces, respectively.
In particular, \(\bW\) and \(\bU\) can be computed
from the reference derivative without knowing
\(\bT\) or \(\bV\).

Express the unknown factors in these bases by setting
\begin{equation}
    \bL=\bT\bW,
    \qquad
    \bR=\bV\bU.
    \label{eq:att-compressed-factors}
\end{equation}
Both matrices have size \(H\times H\).
Because the columns of \(\bW\) and \(\bU\) span the row
spaces of \(\bT\) and \(\bV\), respectively,
\begin{equation}
    \bT=\bL\bW^\top,
    \qquad
    \bV=\bR\bU^\top.
    \label{eq:att-factor-coordinates}
\end{equation}
The full row ranks of \(\bT\) and \(\bV\) then
imply that \(\bL\) and \(\bR\) are invertible.

For each evaluation point, form the compressed matrix
\begin{equation}
    \begin{aligned}
        \bB(\bq)
        &:=\bW^\top\boldsymbol{J}(\bq)\bU\\
        &=\bW^\top\bT^\top\bD(\bq)\bV\bU\\
        &=\bL^\top\bD(\bq)\bR.
    \end{aligned}
    \label{eq:att-compressed-derivatives}
\end{equation}
The factors \(\bL\) and \(\bR\) remain unknown,
but \(\bB(\bq)\) is computed from the derivative
and the two known bases.
At the reference point,
\(\bB(\bq_0)=\bL^\top\bD(\bq_0)\bR\) is invertible.
Consequently, left-multiplying by its inverse removes
the common left factor:
\begin{equation}
    \begin{aligned}
        \bA(\bq)
        &:=\bB(\bq_0)^{-1}\bB(\bq)\\
        &=\bigl(\bR^{-1}\bD(\bq_0)^{-1}\bL^{-\top}\bigr)
          \bL^\top\bD(\bq)\bR\\
        &=\bR^{-1}\bD(\bq_0)^{-1}\bD(\bq)\bR.
    \end{aligned}
    \label{eq:att-common-similarity}
\end{equation}
The middle matrix is diagonal, and its \(h\)-th
diagonal entry is \(\lambda_h(\bq)\).

Equation~\eqref{eq:att-common-similarity} gives
\begin{equation}
    \bR[h,\cdot]\,\bA(\bq_j)
    =\lambda_h(\bq_j)\,\bR[h,\cdot],
    \qquad j=1,\ldots,s.
    \label{eq:att-common-head-eigenvectors}
\end{equation}
Thus every row of \(\bR\) is a left eigenvector
of all the matrices \(\bA(\bq_j)\).
The separation assumption ensures that these
common left-eigenvector directions cannot mix different heads.
Indeed, write any nonzero common left eigenvector as
\(\bc\bR\), with eigenvalue \(\mu_j\) for
\(\bA(\bq_j)\).
Then
\(\bc[h]\lambda_h(\bq_j)=\mu_j\bc[h]\)
for every \(h\) and \(j\).
If both \(\bc[h]\) and \(\bc[g]\) were nonzero, this would
force
\(\lambda_h(\bq_j)=\lambda_g(\bq_j)\)
at every evaluation point, contrary to the assumption.
Hence \(\bc\) has exactly one nonzero coordinate.
The common left-eigenvector directions are therefore
precisely the row directions of \(\bR\).

Let \(\bp_1,\ldots,\bp_H\)
be the recovered common left eigenvectors.
Since each common eigenspace is one-dimensional,
there exist a permutation \(\pi\) and nonzero
scalars \(c_1,\ldots,c_H\) such that
\(\bp_h=c_h\bR[\pi(h),\cdot]\).
Right-multiplying by \(\bU^\top\) gives
\begin{equation}
    \bp_h\bU^\top
    =c_h\bR[\pi(h),\cdot]\,\bU^\top
    =c_h\bV[\pi(h),\cdot]\,\bU\bU^\top
    =c_h\bV[\pi(h),\cdot].
    \label{eq:att-lift-head-directions}
\end{equation}
The last equality holds because every true head
vector lies in the row space spanned by \(\bU\).
Thus the lifted vectors \(\bp_h\bU^\top\) have the correct head
directions, with an unknown nonzero multiplier
for each head.
\end{proof}

Collect the lifted row vectors \(\bp_h\bU^\top\) as the rows of
\(\bP\in\mathbb R^{H\times d}\).
To determine these multiples, use the known
aggregate map \(\bM\).
Since \(\bV[h,\cdot]=\bu_0\bC_h\), the rows of \(\bV\) satisfy
\begin{equation}
    \sum_{h=1}^{H}\bV[h,\cdot]
    =
    \bu_0\left(\sum_{h=1}^{H}\bC_h\right)
    =
    \bu_0\bM.
    \label{eq:att-anchor-sum}
\end{equation}
Thus the scaled rows of \(\bP\) must sum to
the known vector \(\bu_0\bM\).
Writing their scaling coefficients as the row vector
\(\bgamma\in\mathbb R^H\), solve
\begin{equation}
    \bgamma\bP
    =\bu_0\bM,
    \qquad
    \bV
    =\operatorname{diag}(\bgamma)\bP.
    \label{eq:att-head-scale-recovery}
\end{equation}
The system is consistent because each true head
vector lies on one recovered line.
The rows of \(\bP\) are linearly
independent, so the scaling coefficients are unique.
Consequently, \(\bV\) is recovered up to the ordering of its heads.

With \(\bV\) fixed under the selected head ordering,
it remains to determine \(\bT\).
At a general input \(\bq\), the exact derivative satisfies
\begin{equation}
    \begin{aligned}
        \boldsymbol{J}(\bq)
        &=\bT^\top\bD(\bq)\bV,\\
        \bD(\bq)
        &=\operatorname{diag}
          \bigl(\psi'(\bq\bT[1,\cdot]^\top),
                \ldots,
                \psi'(\bq\bT[H,\cdot]^\top)\bigr).
    \end{aligned}
    \label{eq:att-score-general-derivative}
\end{equation}
Although \(\bV\) has been recovered, the diagonal
coefficients still depend on the unknown score vectors.
Thus removing \(\bV\) alone does not directly give
\(\bT^\top\).

To make these coefficients known, choose
\(\bq=\boldsymbol0\), while keeping the offset
direction \(\bu_0\) fixed.
For every head,
\(\boldsymbol0\bT[h,\cdot]^\top=0\), so all
diagonal entries become the same known positive number:
\begin{equation}
    \bD(\boldsymbol0)=\psi'(0)\bI_H,
    \qquad
    \psi'(0)
    =\frac{2\sum_{j=1}^{k}a_j^2}{1+2k}>0.
    \label{eq:att-origin-diagonal}
\end{equation}
Consequently, the derivative at this point simplifies to
\begin{equation}
    \boldsymbol{J}(\boldsymbol0)
    =\bT^\top\bigl(\psi'(0)\bI_H\bigr)\bV
    =\psi'(0)\bT^\top\bV.
    \label{eq:att-origin-factorization}
\end{equation}

Given this exact derivative, the score vectors
follow by a linear solve.
Since \(\bV\) has row rank \(H\),
\(\bV\bV^\dagger=\bI_H\).
Right-multiplying by \(\bV^\dagger\) and dividing by
\(\psi'(0)\) therefore gives
\begin{equation}
    \bT^\top
    =
    \frac{1}{\psi'(0)}
    \boldsymbol{J}(\boldsymbol0)\bV^\dagger.
    \label{eq:att-score-from-origin-derivative}
\end{equation}
The columns of this matrix recover the rows of \(\bT\) in the same
head ordering as \(\bV\).

\paragraph{Fitting the head vectors from function values.}
For each sampled input \(\bq_i\), we obtain an
Attention output through Section~\ref{sec:att-observations} and form
\begin{equation}
    \by_i
    =
    \mathcal A
       \bigl(\bU_{\bu_0}(\bq_i)\bigr)
    -\bq_i\bM.
    \label{eq:att-anchor-observed-data}
\end{equation}
To estimate derivatives from these function values,
we fit a polynomial containing linear and cubic
terms in input and output subspaces estimated
from the data.
Only odd-degree terms are used, matching the symmetry
of the response in~\eqref{eq:att-macro-ridge}.
Differentiating the fitted polynomial gives
approximate Jacobian matrices.
Following the construction in
Proposition~\ref{prop:att-anchor-derivative-separation},
we compress these matrices and approximately
diagonalize them in a common basis to obtain
initial head value directions.
Their scales are fitted using the known sum
\(\bu_0\bM\), giving an initial matrix \(\bV_0\).

To initialize the score vectors, we use the linear
behavior of the response near the origin.
Since \(\psi(0)=0\), the approximation
\(\psi(t)\approx\psi'(0)t\) gives
\begin{equation}
    \mathcal Y_{\bu_0}(\bq)
    =\psi(\bq\bT^\top)\bV
    \approx
    \bq
    \underbrace{\psi'(0)\bT^\top\bV}
               _{\boldsymbol{J}(\boldsymbol0)}
    \qquad\text{for \(\bq\) near \(\boldsymbol0\)}.
    \label{eq:att-response-linearization}
\end{equation}
Thus the coefficient of this linear approximation
is the origin derivative needed in
\eqref{eq:att-score-from-origin-derivative}.
We estimate \(\boldsymbol{J}(\boldsymbol0)\) from the
observed input--output pairs by solving
\begin{equation}
    \boldsymbol{J}_0
    =
    \underset{\boldsymbol{J}\in\mathbb R^{d\times d}}
              {\operatorname{argmin}}
    \sum_i
       \|\bq_i\boldsymbol{J}-\by_i\|_2^2.
    \label{eq:att-linear-response-fit}
\end{equation}
Together with \(\bV_0\), this gives the initial score matrix
\begin{equation}
    \bT_0^\top
    =
    \frac{1}{\psi'(0)}
    \boldsymbol{J}_0
    \bV_0^\dagger.
    \label{eq:att-anchor-score-initialization}
\end{equation}

The linear approximation omits the nonlinear
dependence of the response on \(\bq\).
We therefore refine \((\bV_0,\bT_0)\) by
fitting the full model
\(\psi(\bq_i\bT^\top)\bV\) to the original
observations \(\by_i\).
For a fixed trial matrix \(\bT\), every scalar
coefficient \(\psi(\bq_i\bT[h,\cdot]^\top)\)
is known, and the prediction
\(\sum_h\psi(\bq_i\bT[h,\cdot]^\top)\,\bV[h,\cdot]\)
is linear in \(\bV\).
The matrix \(\bV\) is obtained by regularized
least squares, with the constraint
\(\boldsymbol{1}_H\bV=\bu_0\bM\)
(where \(\boldsymbol{1}_H=(1,\ldots,1)\in\mathbb R^{1\times H}\))
enforcing the known sum.

Substituting the fitted \(\bV\) into the
objective leaves only \(\bT\) as the nonlinear
variable. This is variable
projection~\cite{golub1973differentiation}:
whenever \(\bT\) changes, \(\bV\) is
recomputed by the linear solve.
The reduced objective is optimized by damped
Gauss--Newton steps, whose linear least-squares
systems are solved using LSMR~\cite{fong2011lsmr}.
Candidate fits are compared on separate selection
samples and checked for numerical rank.
The resulting matrices \(\bV,\bT\) are used in the
next subsection to reconstruct the value matrices.

\subsection{Recovering the Value Matrices with Two-Token Queries}
\label{sec:att-isolation-stitch}

Section~\ref{sec:att-macro-anchor} recovers
\(\bV[h,\cdot]=\bu_0\bC_h\) and
\(\bT[h,\cdot]=\bu_0\bS_h^\top\)
for a fixed direction \(\bu_0\).
These rows describe the action of each head
along that direction.
To recover the full matrices \(\bC_h\), we now
vary the token displacement and use the recovered
matrices \(\bV,\bT\) to separate the resulting head contributions.
All normalized sequences below are chosen so that
applying \(\mathcal N_1^{-1}\) to each token gives a physical
query in \(\mathcal X\).

\paragraph{Two-token observations.}
For a normalized final token \(\bq\) and a
displacement \(\bw\), define
\begin{equation}
    \bU_2(\bq,\bw)
    =
    [\bq+2\bw,\ \bq].
    \label{eq:att-two-token-input}
\end{equation}
For head \(h\), the first token's score exceeds
the final token's score by
\(2\bq\bS_h\bw^\top\).
Writing \(t_h=\bq\bS_h\bw^\top\), the first
token therefore receives weight
\begin{equation}
    \frac{e^{2t_h}}{1+e^{2t_h}}
    =
    \frac{1+\tanh(t_h)}{2}.
    \label{eq:att-two-token-weight}
\end{equation}
The weighted average of the two normalized tokens is
\begin{equation}
    \begin{aligned}
        &\frac{1+\tanh(t_h)}{2}(\bq+2\bw)
         +\frac{1-\tanh(t_h)}{2}\bq\\
        &\qquad=\bq+\bw+\tanh(t_h)\bw.
    \end{aligned}
    \label{eq:att-two-token-average}
\end{equation}
Right-multiplying by \(\bC_h\), summing over heads,
and subtracting the known term \((\bq+\bw)\bM\)
gives
\begin{equation}
    \begin{aligned}
        \mathcal Z(\bq,\bw)
        &:=
        \mathcal A\bigl(\bU_2(\bq,\bw)\bigr)
        -(\bq+\bw)\bM\\
        &=
        \sum_{g=1}^{H}
        \tanh(\bq\bS_g\bw^\top)\,\bw\bC_g.
    \end{aligned}
    \label{eq:att-two-token-centered}
\end{equation}
This response contains the value vectors
\(\bw\bC_g\), but they are still mixed across heads.
The next construction separates one head at a time.

\paragraph{Separating one head.}
Fix a direction \(\bu_0\) for which
\(\bV,\bT\in\mathbb R^{H\times d}\)
have row rank \(H\), and assume \(H<d\).
Let \(\bU_0\in\mathbb R^{d\times H}\) have
orthonormal columns spanning the row space of \(\bV\).
Define
\begin{equation}
    \boldsymbol{N}_0=\bI_d-\bU_0\bU_0^\top,
    \qquad
    \bu_0\bC_g\boldsymbol{N}_0
    =\boldsymbol0
    \quad(g=1,\ldots,H).
    \label{eq:att-value-complement}
\end{equation}
Thus right-multiplying by \(\boldsymbol{N}_0\) removes
every component in the space spanned by the known vectors
\(\bu_0\bC_g\).

To select head \(h\), let \(\be_h\) be the
\(h\)-th standard basis row vector of \(\mathbb R^H\)
and choose
\begin{equation}
    \bq_h
    =
    \gamma_h\be_h(\bT^\top)^\dagger,
    \qquad \gamma_h>0.
    \label{eq:att-biorthogonal-q}
\end{equation}
Because \(\bT^\top\) has full column rank,
\((\bT^\top)^\dagger\bT^\top=\bI_H\) and
\(\bq_h\bT^\top=\gamma_h\be_h\).
Consequently,
\begin{equation}
    \bq_h\bS_g\bu_0^\top
    =
    \bq_h\bT[g,\cdot]^\top
    =
    \begin{cases}
        \gamma_h,&g=h,\\
        0,&g\neq h.
    \end{cases}
    \label{eq:att-selected-head-scores}
\end{equation}
The fixed displacement \(\bu_0\) therefore produces
a nonzero score difference only for the selected head.

For a further displacement \(\bu\), use the two
sequences
\begin{equation}
    \bU_{h,\pm}(\bu)
    =
    [\bq_h+2(\bu_0\pm\bu),\ \bq_h].
    \label{eq:att-symmetric-two-token-inputs}
\end{equation}
The values of \(\gamma_h\) and \(\bu\) are chosen
so that both physical queries belong to \(\mathcal X\).
Combine their responses as
\begin{equation}
    \begin{aligned}
        \boldsymbol{d}_h(\bu)
        &:=
        \frac12
        \bigl(
            \mathcal Z(\bq_h,\bu_0+\bu)
            -\mathcal Z(\bq_h,\bu_0-\bu)
        \bigr)\boldsymbol{N}_0.
    \end{aligned}
    \label{eq:att-symmetric-diff}
\end{equation}

\begin{proposition}[Cancellation of the other heads]
\label{prop:att-symmetric-isolation}
Under the preceding full-rank and query conditions,
suppose that \(\bV,\bT,\bM\) and the Attention
responses are exact.
For every permitted displacement \(\bu\), let
\(\beta_h=\bq_h\bS_h\bu^\top\).
Then
\begin{equation}
    \boldsymbol{d}_h(\bu)
    =
    \frac{\sinh(2\gamma_h)}
         {\cosh(2\gamma_h)+\cosh(2\beta_h)}
    \bu\bC_h\boldsymbol{N}_0.
    \label{eq:att-symmetric-exact}
\end{equation}
In particular, the contributions of all heads
\(g\neq h\) cancel.
\end{proposition}

\begin{proof}
For each head \(g\), write
\(\alpha_g=\bq_h\bS_g\bu_0^\top\) and
\(\beta_g=\bq_h\bS_g\bu^\top\).
Substituting \(\bw=\bu_0\pm\bu\)
into~\eqref{eq:att-two-token-centered} gives
\begin{equation}
    \mathcal Z(\bq_h,\bu_0\pm\bu)
    =
    \sum_{g=1}^{H}
    \tanh(\alpha_g\pm\beta_g)
    (\bu_0\pm\bu)\bC_g.
    \label{eq:att-symmetric-expanded}
\end{equation}
After right-multiplying by \(\boldsymbol{N}_0\), every term
containing \(\bu_0\bC_g\) vanishes.
Taking the half-difference therefore yields
\begin{equation}
    \boldsymbol{d}_h(\bu)
    =
    \sum_{g=1}^{H}
    \frac{
        \tanh(\alpha_g+\beta_g)
        +\tanh(\alpha_g-\beta_g)
    }{2}
    \bu\bC_g\boldsymbol{N}_0.
    \label{eq:att-symmetric-projected-sum}
\end{equation}
For \(g\neq h\), equation~\eqref{eq:att-selected-head-scores}
gives \(\alpha_g=0\).
The corresponding coefficient is
\((\tanh(\beta_g)+\tanh(-\beta_g))/2=0\).
For \(g=h\), we have \(\alpha_h=\gamma_h\), and
\begin{equation}
    \begin{aligned}
        \frac{
            \tanh(\gamma_h+\beta_h)
            +\tanh(\gamma_h-\beta_h)
        }{2}
        =
        \frac{\sinh(2\gamma_h)}
             {\cosh(2\gamma_h)+\cosh(2\beta_h)}.
    \end{aligned}
    \label{eq:att-symmetric-coefficient}
\end{equation}
Only the selected head remains, proving the claim.
\end{proof}

The remaining coefficient depends on the unknown
score matrix through \(\beta_h\).
Dividing by the known quantity \(\tanh(\gamma_h)\)
gives
\begin{equation}
    \frac{\boldsymbol{d}_h(\bu)}{\tanh(\gamma_h)}
    =
    \kappa_h(\bu)\,\bu\bC_h\boldsymbol{N}_0,
    \qquad
    \kappa_h(\bu)
    :=
    \frac{\cosh(2\gamma_h)+1}
         {\cosh(2\gamma_h)+\cosh(2\beta_h)}.
    \label{eq:att-projected-value-approximation}
\end{equation}
Using \(\cosh(2\beta_h)-1=2\sinh^2(\beta_h)\) and
\(\cosh(2\gamma_h)+\cosh(2\beta_h)\geq\cosh(2\gamma_h)+1=2\cosh^2(\gamma_h)\),
the multiplicative scale defect satisfies
\begin{equation}
    0
    \leq
    1-\kappa_h(\bu)
    =
    \frac{2\sinh^2(\beta_h)}{\cosh(2\gamma_h)+\cosh(2\beta_h)}
    \leq
    \frac{\sinh^2(\beta_h)}{\cosh^2(\gamma_h)}
    =O(\beta_h^2).
    \label{eq:att-projected-scale-defect}
\end{equation}
Because \(\beta_h=\bq_h\bS_h\bu^\top\) scales linearly with
\(\|\bu\|_2\), shrinking the displacement radius controls the bias
quadratically, with exact limit
\(\lim_{\varepsilon\to 0}\boldsymbol{d}_h(\varepsilon\bu)/(\varepsilon\tanh\gamma_h)=\bu\bC_h\boldsymbol{N}_0\).
At the finite displacement norms used to balance precision noise,
the left-hand side of~\eqref{eq:att-projected-value-approximation}
serves as a slightly biased estimate of \(\bu\bC_h\boldsymbol{N}_0\).
The resulting matrices from this stage and the subsequent CUR step
initialize the complete-block refinement in
Section~\ref{sec:att-refinement}, which further reduces their
disagreement with the observed block outputs.

Using the recovered head vectors, aggregate map, and
Attention outputs, we form \(\boldsymbol{d}_h(\bu_i)\)
over a retained set of displacements \(\bu_1,\ldots,\bu_n\)
spanning \(\mathbb R^d\) and fit the projected value matrix
\(\bP_h=\bC_h\boldsymbol{N}_0\) by linear least squares:
\begin{equation}
    \bP_h
    =
    \underset{\bP\in\mathbb R^{d\times d}}
              {\operatorname{argmin}}
    \sum_{i=1}^{n}
    \left\|
        \bu_i\bP
        -
        \frac{\boldsymbol{d}_h(\bu_i)}
             {\tanh(\gamma_h)}
    \right\|_2^2.
    \label{eq:att-projected-value-fit}
\end{equation}

\paragraph{Combining different directions.}
The projection \(\boldsymbol{N}_0\) removes the
component of each output in the row space of \(\bV\).
Consequently, the projected matrix from one direction
does not determine the full \(\bC_h\).
Repeat the construction for \(b\geq2\) directions
\(\bu_{0,1},\ldots,\bu_{0,b}\).
For direction \(a\in\{1,\ldots,b\}\), let \(\bV_a\in\mathbb R^{H\times d}\)
be the recovered value matrix along \(\bu_{0,a}\), and write
\begin{equation}
    \begin{aligned}
        \mathcal V_a
        &=
        \operatorname{span}
        \{\bu_{0,a}\bC_h:1\leq h\leq H\},\\
        \boldsymbol{N}_a
        &=
        \bI_d-\bU_a\bU_a^\top,
        \qquad
        \bP_{h,a}=\bC_h\boldsymbol{N}_a,
    \end{aligned}
    \label{eq:att-multiple-direction-maps}
\end{equation}
where the columns of \(\bU_a\) form an orthonormal
basis of \(\mathcal V_a\).
Different directions remove different output
components, allowing the remaining observations
to be combined.

The head ordering may differ between directions.
Use the first direction to fix the head labels.
For another direction \(a\), the projected image
of its fixed vector satisfies
\begin{equation}
    \bu_{0,a}\bP_{h,1}
    =
    \bu_{0,a}\bC_h\boldsymbol{N}_1.
    \label{eq:att-cross-direction-image}
\end{equation}
The right-hand side can also be obtained by
applying \(\boldsymbol{N}_1\) to the recovered rows of \(\bV_a\).
Thus, if \(\bV_a[k,\cdot]\) denotes the vector with
local label \(k\), its matching head is determined by
\begin{equation}
    \bu_{0,a}\bP_{h,1}
    =
    \bV_a[\pi_a(h),\cdot]\,\boldsymbol{N}_1.
    \label{eq:att-universal-matching}
\end{equation}
This comparison matches the action of a head on
the same input vector across the two constructions.

\begin{proposition}[Combining projected value matrices]
\label{prop:att-value-reconstruction}
Suppose that the head matrices \(\bV_a\) and projected maps \(\bP_{h,a}\)
for all \(b\) directions are known exactly.
For every \(a\geq2\), assume that the rows
\(\bV_a[k,\cdot]\,\boldsymbol{N}_1\), \(1\leq k\leq H\),
are pairwise distinct.
Assume further that
\begin{equation}
    \bigcap_{a=1}^{b}\mathcal V_a
    =\{\boldsymbol0\}.
    \label{eq:att-value-complement-coverage}
\end{equation}
Then~\eqref{eq:att-universal-matching} determines
the head correspondence uniquely.
After aligning the head labels, each value matrix
is uniquely recovered as
\begin{equation}
    \bC_h
    =
    \left(\sum_{a=1}^{b}\bP_{h,a}\right)
    \left(\sum_{a=1}^{b}\boldsymbol{N}_a\right)^{-1}.
    \label{eq:att-stitch-formula}
\end{equation}
The recovered matrices share the head ordering
chosen for the first direction.
\end{proposition}

\begin{proof}
The true correspondence satisfies
\eqref{eq:att-universal-matching}.
The assumed distinctness of the projected head
vectors makes this correspondence unique.

To establish invertibility in
\eqref{eq:att-stitch-formula}, let \(\bx\in\mathbb R^d\).
Each \(\boldsymbol{N}_a\) is an orthogonal projector,
so
\begin{equation}
    \bx
    \left(\sum_{a=1}^{b}\boldsymbol{N}_a\right)\bx^\top
    =
    \sum_{a=1}^{b}\|\bx\boldsymbol{N}_a\|_2^2.
    \label{eq:att-projection-sum-positive}
\end{equation}
The right-hand side vanishes exactly when
\(\bx\) belongs to every \(\mathcal V_a\).
By~\eqref{eq:att-value-complement-coverage},
this occurs only for \(\bx=\boldsymbol0\).
Hence the sum of the projectors is positive
definite and invertible.
Finally,
\begin{equation}
    \sum_{a=1}^{b}\bP_{h,a}
    =
    \bC_h\left(\sum_{a=1}^{b}\boldsymbol{N}_a\right),
    \label{eq:att-value-combination-system}
\end{equation}
which gives the reconstruction formula and its
uniqueness.
\end{proof}

For numerical reconstruction, the head correspondence
is chosen by a one-to-one matching that minimizes
the discrepancies between the vectors
in~\eqref{eq:att-universal-matching}.
The left-hand vectors are estimated from the
two-token observations at the other fixed directions,
or from the fitted projected maps.
After matching, we solve
\eqref{eq:att-value-combination-system}
and truncate each resulting matrix to its
leading \(r\) singular components, giving
rank-at-most-\(r\) value matrices \(\bC_1,\ldots,\bC_H\).
These matrices are used next to separate the
head weights and reconstruct the score matrices.

\subsection{Recovering the Score Matrices}
\label{sec:att-cur}

Once the value matrices have been recovered with
consistent head labels, the two-token response
can be used to separate the score measurements
of all heads.
The rank bound \(\operatorname{rank}(\bS_h)\leq r\)
then allows these measurements to be combined
through a low-rank reconstruction.

\paragraph{Recovering the individual score measurements.}
For a displacement \(\bw\), form the known matrix
\(\bV(\bw)\in\mathbb R^{H\times d}\) with rows
\begin{equation}
    \bV(\bw)[h,\cdot]
    =
    \bw\bC_h,
    \qquad h=1,\ldots,H.
    \label{eq:att-unmixing-value-matrix}
\end{equation}
The two-token identity
in~\eqref{eq:att-two-token-centered} becomes
\begin{equation}
    \mathcal Z(\bq,\bw)
    =
    \boldsymbol{\tau}(\bq,\bw)\bV(\bw),
    \qquad
    \boldsymbol{\tau}(\bq,\bw)[h]
    =
    \tanh(\bq\bS_h\bw^\top).
    \label{eq:att-unmixing-system}
\end{equation}
Here the output vector and the rows of
\(\bV(\bw)\) are known.
Only the \(H\) scalar coefficients in the row vector
\(\boldsymbol{\tau}(\bq,\bw)\) remain unknown.
If \(\bV(\bw)\) has row rank \(H\), the linear
system has the unique solution
\begin{equation}
    \boldsymbol{\tau}(\bq,\bw)
    =
    \mathcal Z(\bq,\bw)\bV(\bw)^\dagger.
    \label{eq:att-unmixed-coefficients}
\end{equation}
Each coefficient lies in \((-1,1)\).
Applying the inverse function
\begin{equation}
    \operatorname{artanh}(t)
    =
    \frac12\log\frac{1+t}{1-t},
    \qquad -1<t<1,
    \label{eq:att-inverse-tanh}
\end{equation}
therefore gives
\begin{equation}
    \bq\bS_h\bw^\top
    =
    \operatorname{artanh}
    \Bigl(
        \bigl(\mathcal Z(\bq,\bw)\bV(\bw)^\dagger\bigr)[h]
    \Bigr),
    \qquad h=1,\ldots,H.
    \label{eq:att-artanh-unmix}
\end{equation}
Thus one two-token Attention observation provides
one bilinear measurement of every score matrix.
The ordering of these measurements is fixed by
the rows of \(\bV(\bw)\), and hence by the
recovered value matrices.

\paragraph{Collecting two sets of matrix measurements.}
Choose an integer \(k\) with \(r\leq k\leq d\),
and two sets of known orthonormal row directions,
stacked as the rows of
\begin{equation}
    \begin{aligned}
        \bQ_{\mathrm{test}}
        &=[\bq_1,\ldots,\bq_k]
          \in\mathbb R^{k\times d},\\
        \bU_{\mathrm{test}}
        &=[\bu_1,\ldots,\bu_k]
          \in\mathbb R^{k\times d}.
    \end{aligned}
    \label{eq:att-score-test-directions}
\end{equation}
Let \(\be_1,\ldots,\be_d\) be the standard basis row vectors
of \(\mathbb R^d\).
Use positive scale factors \(\alpha,\beta\)
to control the sizes of the displacement and
the final token, respectively.
Choose them so that
\(\beta+2\alpha<\sqrt d\) and all resulting
physical queries belong to \(\mathcal X\).

For the first set, use
\((\bq,\bw)=(\beta\bq_a,\alpha\be_j)\),
with \(a=1,\ldots,k\) and \(j=1,\ldots,d\).
After recovering the bilinear measurement
by~\eqref{eq:att-artanh-unmix}, divide by the
known factor \(\alpha\beta\).
This gives
\begin{equation}
    \frac{
        (\beta\bq_a)\bS_h(\alpha\be_j)^\top
    }{\alpha\beta}
    =
    \bq_a\bS_h\be_j^\top.
    \label{eq:att-score-left-entries}
\end{equation}
Collecting these values over \(a,j\) gives
\begin{equation}
    \bL_h
    =
    \bQ_{\mathrm{test}}\bS_h
    \in\mathbb R^{k\times d},
    \label{eq:att-score-left-measurements}
\end{equation}
with rows \(\bL_h[a,\cdot]=\bq_a\bS_h\).

For the second set, use
\((\bq,\bw)=(\beta\be_i,\alpha\bu_b)\),
with \(i=1,\ldots,d\) and \(b=1,\ldots,k\).
The same procedure gives
\begin{equation}
    \frac{
        (\beta\be_i)\bS_h(\alpha\bu_b)^\top
    }{\alpha\beta}
    =
    \be_i\bS_h\bu_b^\top.
    \label{eq:att-score-right-entries}
\end{equation}
These values form
\begin{equation}
    \bR_h
    =
    \bS_h\bU_{\mathrm{test}}^\top
    \in\mathbb R^{d\times k},
    \label{eq:att-score-right-measurements}
\end{equation}
with columns \(\bR_h[\cdot,b]=\bS_h\bu_b^\top\).

The two measured matrices also determine
\begin{equation}
    \bK_h
    =
    \bL_h\bU_{\mathrm{test}}^\top
    =
    \bQ_{\mathrm{test}}\bR_h
    =
    \bQ_{\mathrm{test}}
      \bS_h\bU_{\mathrm{test}}^\top
    \in\mathbb R^{k\times k}.
    \label{eq:att-cur-core}
\end{equation}
Its entry \(\bK_h[a,b]=\bq_a\bS_h\bu_b^\top\) is shared
by the two constructions, so computing \(\bK_h\) requires
no additional queries.

The following reconstruction applies the CUR
identity in these chosen coordinates
\cite{hamm2021cur}.
Its rank condition requires the two sets of
directions to preserve the full rank of
the score matrix.

\begin{proposition}[Low-rank score reconstruction]
\label{prop:att-cur}
Suppose that the value matrices, aggregate map,
and Attention observations are exact, and that
all queries above belong to the permitted domain.
Assume that \(\bV(\bw)\) has row rank \(H\)
for every displacement used in the two query sets.
For each head, let
\(r_h=\operatorname{rank}(\bS_h)\leq r\),
and assume
\(\operatorname{rank}(\bK_h)=r_h\).
Then
\begin{equation}
    \bS_h
    =
    \bR_h\bK_h^\dagger\bL_h,
    \qquad h=1,\ldots,H.
    \label{eq:att-cur-formula}
\end{equation}
The recovered score matrices have the same head
ordering as the recovered value matrices.
The construction uses at most \(2kd\)
two-token Attention evaluations.
\end{proposition}

\begin{proof}
The full-rank condition on \(\bV(\bw)\) makes
each coefficient vector in
\eqref{eq:att-unmixed-coefficients} unique.
Consequently, the two query sets recover
\(\bL_h\) and \(\bR_h\) exactly.

Fix a head \(h\).
Since
\(\bR_h=\bS_h\bU_{\mathrm{test}}^\top\) and
\(\bK_h=\bQ_{\mathrm{test}}\bR_h\),
\begin{equation}
    \operatorname{rank}(\bK_h)
    \leq\operatorname{rank}(\bR_h)
    \leq\operatorname{rank}(\bS_h)
    =r_h.
    \label{eq:att-cur-rank-chain}
\end{equation}
The assumed rank of \(\bK_h\) makes all these
ranks equal.
Thus \(\bR_h\) and \(\bS_h\) have the same column
space, and there exists a matrix \(\bX_h\) such that
\begin{equation}
    \bS_h=\bR_h\bX_h,
    \qquad
    \bL_h
    =\bQ_{\mathrm{test}}\bS_h
    =\bK_h\bX_h.
    \label{eq:att-cur-column-representation}
\end{equation}

Moreover, \(\ker(\bR_h)\subseteq\ker(\bK_h)\).
Their equal ranks imply equal nullspaces.
Since
\(\bK_h(\bI_k-\bK_h^\dagger\bK_h)=\boldsymbol0\),
every column of
\(\bI_k-\bK_h^\dagger\bK_h\) also belongs
to \(\ker(\bR_h)\).
Hence
\(\bR_h\bK_h^\dagger\bK_h=\bR_h\), and
\begin{equation}
    \bR_h\bK_h^\dagger\bL_h
    =
    \bR_h\bK_h^\dagger\bK_h\bX_h
    =
    \bR_h\bX_h
    =
    \bS_h.
    \label{eq:att-cur-reconstruction-proof}
\end{equation}
Finally, each query set contains \(kd\) pairs,
and each pair supplies measurements for all
heads simultaneously.
This gives the stated query count.
\end{proof}

In the implementation, choosing \(k=\min(d,2r)\) limits the
number of Attention evaluations in this score-recovery stage
(given \(\{\bC_h\}_{h=1}^{H}\)) to
\begin{equation}
    2kd\leq 4rd=\frac{4d^2}{H},
    \qquad d=Hr.
    \label{eq:att-score-query-count}
\end{equation}
Each evaluation is synthesized through the paired inversion
procedure of Section~\ref{sec:att-observations}.

\paragraph{Numerical reconstruction.}
In the presence of noise, we clip the unmixed coefficients strictly within \((-1,1)\) before applying \(\operatorname{artanh}\).
To avoid amplifying noise in the pseudoinverse, we truncate \(\bK_h = \bL_h\bU_{\mathrm{test}}^\top\) to its leading \(r\) singular components \(\bK_{h,r}\) and set \(\bS_h = \bR_h \bK_{h,r}^{\dagger} \bL_h\).
The resulting matrices \(\{\bS_h, \bC_h\}\) serve as the initialization for joint refinement.

\subsection{Joint Refinement from Complete-Block Outputs}
\label{sec:att-refinement}

The recovered Attention matrices and FFN parameters
provide an initial estimate of the complete block.
To improve this estimate, we fit its predictions
directly to the observed block outputs, without
converting these observations into Attention labels.

Let
\(\theta=(\{\bS_h,\bC_h\}_{h=1}^{H},\vartheta)\),
where \(\vartheta\) contains the FFN parameters, and let
\(\mathcal F_\theta\) denote the block~\eqref{eq:block}
evaluated at \(\theta\).
For each fitting sequence \(\bX_i\), let
\(\by_i\in\mathbb R^d\) denote the
returned last-token output.
The refinement minimizes
\begin{equation}
    \begin{aligned}
    \mathcal L(\theta)
    =
    \frac{1}{2nd}
    \sum_{i=1}^{n}
       \|\mathcal F_\theta(\bX_i)-\by_i\|_2^2
    +
    \frac{\mu^2}{2}
       \left\|\sum_{h=1}^{H}\bC_h-\bM\right\|_F^2,
    \qquad \mu\geq0,
    \end{aligned}
    \label{eq:att-refinement-objective}
\end{equation}
subject to
\(\operatorname{rank}(\bS_h)\leq r\) and
\(\operatorname{rank}(\bC_h)\leq r\), where \(\bM\) is the
aggregate value map recovered in Stage~1.
The first term compares complete-block predictions
with the returned vectors. The second keeps the
aggregate value map close to its Stage~1 estimate,
without imposing exact equality.

Refinement first adjusts the Attention matrices
while keeping the recovered FFN fixed, and then
jointly adjusts the Attention and FFN parameters.
At each iteration, a damped Gauss--Newton step is
computed using LSMR~\cite{fong2011lsmr}.
The solver uses Jacobian--vector and transposed
Jacobian--vector products computed from the
reconstructed block, without storing the full
Jacobian or requesting target-model derivatives.

Updates are parameterized in low-rank coordinates, and each trial Attention matrix is truncated to rank at most \(r\) via SVD.
At each step, we use backtracking line search to ensure \(\mathcal L\) decreases, reducing the damping after an accepted step and increasing it otherwise.
Both stages run for a bounded number of iterations, and we select the final candidate (including the initial estimate) that minimizes the last-token output error on a separate selection set.
Once the fitting and selection responses are collected, the entire refinement is offline and requires no additional target queries.

\subsection{End-to-End Query Complexity}
\label{sec:query-complexity}

We now summarize the oracle query complexity of the complete four-stage
attack across Sections~\ref{sect:ReLU}--\ref{sect:Attention}, counting
each submission of an input sequence \(\bX\in\mathcal X\) returning
\(\mathcal F(\bX)\in\mathbb R^d\) as one block query.
Table~\ref{tab:query-complexity} lists the query count of each stage
(with each Stage~3 Attention evaluation using two paired block queries
\(\pm\bX\) when no path subdivision is needed), together with the
default sampling sizes used in Section~\ref{sect:experiments}.

\begin{table}[t]
\centering
\caption{End-to-end block query complexity of the four-stage extraction
pipeline (\(d=Hr\)). Stage~3 counts assume direct paired inversion
(\(2\) block queries per Attention evaluation); occasional path
subdivisions add \(2\) queries per inserted intermediate point.}
\label{tab:query-complexity}
\footnotesize
\setlength{\tabcolsep}{4.5pt}
\renewcommand{\arraystretch}{1.18}
\begin{tabularx}{\linewidth}{@{}>{\raggedright\arraybackslash}Xlll@{}}
\toprule
\textbf{Stage} & \textbf{Block queries} & \textbf{Asymptotic} & \textbf{Default setting} \\
\midrule
\multicolumn{4}{@{}l}{\emph{Stage 1: Aggregate value map \(\bM\)}} \\
\hspace{0.8em}ReLU (\S\ref{sec:relu-aggregate})
  & \(2nd\) & \(O(d)\) & \(24d\) (\(n=12\)) \\
\hspace{0.8em}SwiGLU (\S\ref{sec:swiglu-aggregate})
  & \(8+2\nu d\) (FP16: \(8+6\nu d\)) & \(O(d)\) & \(8+128d\)\,/\,\(8+384d\) (\(\nu=64\)) \\
\midrule
\multicolumn{4}{@{}l}{\emph{Stage 2: Coupled FFN \(\mathcal G\)}} \\
\hspace{0.8em}ReLU (\S\ref{sec:relu-ffn-recovery})
  & \(N\) one-token & \(O(m)\) & \(\max(5120,\,160m)\) \\
\hspace{0.8em}SwiGLU (\S\ref{sec:swiglu-ffn-recovery})
  & \(N\) opposite one-token & \(O(m)\) & \(\max(4608,\,36m)\) \\
\midrule
\multicolumn{4}{@{}l}{\emph{Stage 3: Shared multi-head Attention \(\bS_h,\bC_h\)}} \\
\hspace{0.8em}Head separation (\S\ref{sec:att-observations}--\S\ref{sec:att-macro-anchor})
  & \(16 + 2b\cdot O(d)\) & \(O(d)\) & \(16+180d\) (\(b=3\)) \\
\hspace{0.8em}Value stitching (\S\ref{sec:att-isolation-stitch})
  & \(4bH(3d+b)\) & \(O(Hd)\) & \(36H(d+1)\) (\(b=3\)) \\
\hspace{0.8em}CUR score recovery (\S\ref{sec:att-cur})
  & \(4kd \leq 8rd = 8d^2/H\) & \(O(d^2/H)\) & \(4kd\) (\(k=\min(d,2r)\)) \\
\midrule
\multicolumn{4}{@{}l}{\emph{Stage 4: Complete-block refinement (\S\ref{sec:att-refinement})}} \\
\hspace{0.8em}Stage 4a (Attention)
  & fitting + selection & \(O(d)\) & \(\max(960,\,50d)\) \\
\hspace{0.8em}Stage 4b (Joint block)
  & \(0\) (reuses Stage~4a) & \(0\) & \(0\) \\
\midrule
\textbf{Total (scheduled)}
  & \multicolumn{2}{l}{\(Q_{\mathrm{scheduled}}=O(m + Hd + d^2/H)\)}
  & \emph{see Table~\ref{tab:scaling-summary}} \\
\bottomrule
\end{tabularx}
\end{table}

Combining the fixed sampling schedules across all four stages gives
\begin{equation}
    Q_{\mathrm{total}}
    =Q_{\mathrm{scheduled}}+2N_{\mathrm{path}},
    \qquad
    Q_{\mathrm{scheduled}}
    =O\!\bigl(m+(H+r)d\bigr)
    =O\!\left(m+Hd+\frac{d^2}{H}\right),
    \label{eq:total-query-complexity}
\end{equation}
where \(d=Hr\) and \(N_{\mathrm{path}}\) counts any intermediate path
points adaptively inserted during paired inversion in
Section~\ref{sec:att-observations}.
To put~\eqref{eq:total-query-complexity} in perspective, a full
Transformer block contains \(\Theta(d^2+md)\) scalar parameters
(\(4d^2\) Attention weights plus \(2md+m+d\) for ReLU or \(3md\) for
SwiGLU, along with \(2d\) normalization gains), while each oracle query
returns a \(d\)-dimensional vector \(\mathcal F(\bX)\in\mathbb R^d\).
Stages~1, 2, 3a, and 4 scale linearly in \(d\) or \(m\):
Stage~1 queries \(O(d)\) rays or orthonormal frames, Stage~2 collects
\(N=O(m)\) one-token FFN pairs (supplying \(Nd=\Theta(md)\) scalar
observations), and Stage~4 uses \(O(d)\) multi-token sequences.
Within Stage~3, two-token value isolation scales as \(O(Hd)\) across
\(b=3\) anchor directions, and low-rank CUR score recovery uses at most
\(4kd\leq 8d^2/H\) paired queries (Proposition~\ref{prop:att-cur}),
avoiding naive \(O(d^2)\) full-matrix probing whenever \(H>1\).
Consequently, as \(d\) and \(m\) grow, the ratio of block queries to
scalar parameters decreases steadily below one (reaching
\(0.26\text{--}0.47\) queries per parameter at \(d=256,m=1024\) in
Section~\ref{sect:experiments}).

All iterative numerical solvers---including iterative scale refinement
(\(S=6\) steps in Section~\ref{sec:relu-aggregate}), wider-network
fitting, hyperplane merging, smoothed Hessian SVD, Gauss--Newton
updates, and Stage~4b joint refinement---operate entirely offline and
incur zero additional oracle queries.
Across all experiments at \(d\geq 64\) in Section~\ref{sect:experiments},
direct paired inversion succeeds with \(N_{\mathrm{path}}=0\), so the
empirical query counts match the scheduled closed-form totals in
Table~\ref{tab:query-complexity} exactly.

\section{Experiments}
\label{sect:experiments}
We evaluate the attack on Transformer blocks with biased
ReLU or bias-free SwiGLU FFNs, each containing two
pre-RMSNorm layers and two residual connections.
The experiments cover five token dimensions,
\(d\in\{16,32,64,128,256\}\), with four or eight Attention
heads and FFN widths ranging from \(32\) to \(1024\).
Table~\ref{tab:scaling-summary} lists the configurations
and raw block parameter counts (up to \(788{,}224\) for
ReLU and \(1{,}049{,}088\) for SwiGLU).
Target projections are drawn i.i.d.\ with
\(\bQ_h,\bK_h,\bV_h,\bO_h\sim\mathcal N(0,1/d)\) and
non-trivial RMSNorm gains
\(\bgamma_1,\bgamma_2\sim\mathcal U(0.75,1.25)\);
FFN weights use \(\bA_1\sim\mathcal N(0,1/d)\),
\(\bb_1\sim\mathcal N(0,0.2^2)\),
\(\bA_2\sim\mathcal N(0,1/m)\),
\(\bb_2\sim\mathcal N(0,0.02^2)\) for ReLU, and
\(\bA_g,\bA_u\sim\mathcal N(0,1/d)\),
\(\bA\sim\mathcal N(0,1/m)\) for SwiGLU, with
\(\epsilon_1=\epsilon_2=10^{-5}\).
Each target executes the unmerged raw
\(\bQ/\bK/\bV/\bO\) and dual pre-RMSNorm computation
graph entirely in native FP64, FP32, FP16, or BF16
arithmetic, returning only the last-token output vector,
while the extracted effective block is evaluated in FP64.
We evaluate two independent random targets per
configuration for \(d\in\{16,32,64,128\}\) and one target
at \(d=256\) (\(72\) end-to-end runs in total across both
FFN families and all four precisions), with all \(72/72\)
runs completing successfully.
The implementation, benchmark configurations, and saved
results are publicly available at
\url{https://github.com/XXXX}.

Table~\ref{tab:scaling-summary} summarizes extraction
accuracy and query cost.
Parameter errors are measured against the native-quantized
effective parameters after Hungarian head and neuron
matching and FFN scale normalization:
\(\bM\) reports the Stage~1 relative Frobenius error;
FFN, Attention, and Block report the final post-refinement errors, where
FFN is the worst per-neuron relative error across all
\(m\) matched neurons (together with the output bias for
ReLU), Attention is the worst per-head relative
Frobenius error across all \(H\) matched heads, and
Block error is the holdout relative \(L_2\) error against
native target outputs on \(256\) independent standard
Gaussian length-five sequences, and query counts record
all complete-block vector queries across all extraction
stages (including calibration, retries, and selection
sets).

Across all \(72\) runs, the maximum block output errors
are \(2.20\times10^{-15}\) in FP64,
\(5.49\times10^{-7}\) in FP32, \(0.11\%\) in FP16,
and \(0.82\%\) in BF16.
Although larger blocks require more queries, the
vector query count per parameter decreases with scale,
reaching \(0.26\text{--}0.47\) at \(d=256\).

\begin{table}[htbp]
\centering
\caption{End-to-end extraction accuracy and query cost.
Errors and query counts are maxima over the evaluated
targets for each configuration.}
\label{tab:scaling-summary}

\footnotesize
\renewcommand{\arraystretch}{1.0}
\setlength{\tabcolsep}{2.5pt}

\resizebox{\linewidth}{!}{%
\begin{tabular}{@{}llrlrrrrrr@{}}
\toprule

\multicolumn{4}{c}{\textbf{Model}}
&
\multicolumn{2}{c}{\textbf{Query Cost}}
&
\multicolumn{4}{c}{\textbf{Relative Error}}
\\
\cmidrule(lr){1-4}
\cmidrule(lr){5-6}
\cmidrule(l){7-10}

\textbf{Configuration}
&
\textbf{FFN}
&
\textbf{Params.}
&
\textbf{Precision}
&
\textbf{Queries}
&
\shortstack{\textbf{Queries /}\\\textbf{param.}}
&
\(\bM\)
&
\textbf{FFN}
&
\textbf{Attn.}
&
\textbf{Block}
\\
\midrule

\multirow{8}{*}{\shortstack[l]{
\(d=16,\ H=4\)\\
\(r=4,\ m=32\)}}
 & \multirow{4}{*}{ReLU}
 & \multirow{4}{*}{2,128}
 & FP64 & 12,356 & 5.81
 & 1.13e-15 & 4.13e-15 & 3.93e-15 & 7.19e-16 \\
 & & & FP32 & 12,356 & 5.81
 & 3.69e-07 & 2.04e-06 & 1.83e-07 & 1.24e-07 \\
 & & & FP16 & 12,362 & 5.81
 & 5.11e-04 & 9.79e-03 & 8.06e-04 & 6.64e-04 \\
 & & & BF16 & 12,344 & 5.80
 & 4.62e-03 & 8.91e-02 & 7.67e-03 & 5.35e-03 \\
\cmidrule(lr){2-10}
 & \multirow{4}{*}{SwiGLU}
 & \multirow{4}{*}{2,592}
 & FP64 & 13,486 & 5.20
 & 6.98e-16 & 3.05e-15 & 3.89e-15 & 8.06e-16 \\
 & & & FP32 & 13,486 & 5.20
 & 1.85e-08 & 5.12e-07 & 1.75e-07 & 1.67e-07 \\
 & & & FP16 & 17,582 & 6.78
 & 6.47e-04 & 2.14e-03 & 9.02e-04 & 8.61e-04 \\
 & & & BF16 & 13,486 & 5.20
 & 1.36e-03 & 2.75e-02 & 8.27e-03 & 6.49e-03 \\
\midrule

\multirow{8}{*}{\shortstack[l]{
\(d=32,\ H=4\)\\
\(r=8,\ m=32\)}}
 & \multirow{4}{*}{ReLU}
 & \multirow{4}{*}{6,272}
 & FP64 & 20,334 & 3.24
 & 1.28e-15 & 4.74e-15 & 1.79e-15 & 7.92e-16 \\
 & & & FP32 & 20,334 & 3.24
 & 4.27e-07 & 2.75e-06 & 1.99e-07 & 1.60e-07 \\
 & & & FP16 & 20,316 & 3.24
 & 5.17e-04 & 1.21e-02 & 8.76e-04 & 7.04e-04 \\
 & & & BF16 & 20,196 & 3.22
 & 4.37e-03 & 6.04e-02 & 7.33e-03 & 5.72e-03 \\
\cmidrule(lr){2-10}
 & \multirow{4}{*}{SwiGLU}
 & \multirow{4}{*}{7,232}
 & FP64 & 22,888 & 3.16
 & 7.36e-16 & 3.26e-15 & 1.95e-15 & 9.45e-16 \\
 & & & FP32 & 22,888 & 3.16
 & 2.24e-08 & 1.99e-07 & 2.17e-07 & 1.96e-07 \\
 & & & FP16 & 31,080 & 4.30
 & 5.76e-04 & 1.04e-03 & 9.90e-04 & 8.51e-04 \\
 & & & BF16 & 22,888 & 3.16
 & 1.27e-03 & 6.47e-03 & 7.61e-03 & 6.75e-03 \\
\midrule

\multirow{8}{*}{\shortstack[l]{
\(d=64,\ H=8\)\\
\(r=8,\ m=256\)}}
 & \multirow{4}{*}{ReLU}
 & \multirow{4}{*}{49,600}
 & FP64 & 80,048 & 1.61
 & 1.65e-15 & 1.26e-14 & 4.57e-15 & 1.48e-15 \\
 & & & FP32 & 80,048 & 1.61
 & 5.51e-07 & 2.76e-06 & 2.68e-07 & 2.45e-07 \\
 & & & FP16 & 80,048 & 1.61
 & 5.15e-04 & 1.23e-02 & 9.02e-04 & 7.49e-04 \\
 & & & BF16 & 80,048 & 1.61
 & 4.01e-03 & 4.98e-02 & 7.10e-03 & 5.99e-03 \\
\cmidrule(lr){2-10}
 & \multirow{4}{*}{SwiGLU}
 & \multirow{4}{*}{65,664}
 & FP64 & 54,968 & 0.84
 & 1.07e-15 & 9.29e-15 & 4.53e-15 & 1.61e-15 \\
 & & & FP32 & 54,968 & 0.84
 & 2.87e-08 & 5.27e-07 & 2.85e-07 & 3.14e-07 \\
 & & & FP16 & 71,352 & 1.09
 & 5.95e-04 & 1.86e-03 & 9.51e-04 & 9.58e-04 \\
 & & & BF16 & 54,968 & 0.84
 & 1.29e-03 & 1.25e-02 & 7.45e-03 & 7.59e-03 \\
\midrule

\multirow{8}{*}{\shortstack[l]{
\(d=128,\ H=8\)\\
\(r=16,\ m=512\)}}
 & \multirow{4}{*}{ReLU}
 & \multirow{4}{*}{197,504}
 & FP64 & 167,984 & 0.85
 & 2.19e-15 & 1.23e-14 & 4.18e-15 & 1.83e-15 \\
 & & & FP32 & 167,984 & 0.85
 & 6.75e-07 & 2.63e-06 & 3.93e-07 & 3.20e-07 \\
 & & & FP16 & 167,984 & 0.85
 & 5.00e-04 & 1.01e-02 & 1.00e-03 & 7.68e-04 \\
 & & & BF16 & 167,984 & 0.85
 & 4.01e-03 & 3.41e-02 & 8.10e-03 & 6.19e-03 \\
\cmidrule(lr){2-10}
 & \multirow{4}{*}{SwiGLU}
 & \multirow{4}{*}{262,400}
 & FP64 & 117,816 & 0.45
 & 1.16e-15 & 7.99e-14 & 3.71e-15 & 2.16e-15 \\
 & & & FP32 & 117,816 & 0.45
 & 3.81e-08 & 5.49e-07 & 4.10e-07 & 3.96e-07 \\
 & & & FP16 & 150,584 & 0.57
 & 6.06e-04 & 1.70e-03 & 1.10e-03 & 9.84e-04 \\
 & & & BF16 & 117,816 & 0.45
 & 1.31e-03 & 9.72e-03 & 8.73e-03 & 7.87e-03 \\
\midrule

\multirow{8}{*}{\shortstack[l]{
\(d=256,\ H=8\)\\
\(r=32,\ m=1024\)}}
 & \multirow{4}{*}{ReLU}
 & \multirow{4}{*}{788,224}
 & FP64 & 368,432 & 0.47
 & 2.51e-15 & 1.23e-14 & 3.19e-15 & 1.78e-15 \\
 & & & FP32 & 368,432 & 0.47
 & 8.75e-07 & 2.73e-06 & 7.30e-07 & 4.55e-07 \\
 & & & FP16 & 368,432 & 0.47
 & 4.99e-04 & 1.15e-02 & 1.39e-03 & 8.14e-04 \\
 & & & BF16 & 368,432 & 0.47
 & 3.97e-03 & 3.07e-02 & 1.11e-02 & 6.63e-03 \\
\cmidrule(lr){2-10}
 & \multirow{4}{*}{SwiGLU}
 & \multirow{4}{*}{1,049,088}
 & FP64 & 268,088 & 0.26
 & 1.64e-15 & 2.57e-15 & 3.37e-15 & 2.20e-15 \\
 & & & FP32 & 268,088 & 0.26
 & 5.19e-08 & 6.33e-07 & 7.47e-07 & 5.49e-07 \\
 & & & FP16 & 333,624 & 0.32
 & 5.87e-04 & 1.19e-03 & 1.50e-03 & 1.07e-03 \\
 & & & BF16 & 268,088 & 0.26
 & 1.29e-03 & 8.26e-03 & 1.16e-02 & 8.16e-03 \\

\bottomrule
\end{tabular}%
}
\end{table}

\section{Conclusion}
\label{sect:conclusions}
We presented a cryptanalytic extraction method for a complete
Transformer block with two pre-RMSNorm layers, two residual
connections, and either a biased ReLU or a bias-free SwiGLU FFN.
Using only chosen input sequences and their last-token outputs,
the method reconstructs the effective Attention and FFN
parameters. Aggregate value-map recovery provides the
observations needed for FFN extraction, while paired inversion
connects the recovered FFN to a shared multi-head Attention
reconstruction method. Joint refinement then improves agreement
with the complete-block outputs.

Experiments cover five block sizes in native FP64, FP32,
FP16, and BF16 arithmetic, with token dimensions up to
\(d=256\), eight Attention heads, and FFN widths up to
\(m=1024\). The largest SwiGLU block contains over one
million parameters. Across the tested configurations,
block output errors are at most
\(2.20\times10^{-15}\), \(5.49\times10^{-7}\),
\(1.07\times10^{-3}\), and \(8.16\times10^{-3}\)
in the four formats, respectively.
At the largest tested scale, extraction requires
approximately \(0.26\text{--}0.47\) complete-block queries
per model parameter.
These results demonstrate that the combined normalization,
residual, Attention, and FFN structure can be extracted
with small output error across the tested numerical formats.

\paragraph{Future work.}
Future directions include extending the method to multiple
coupled Transformer blocks, incorporating positional
encodings and grouped-query Attention, and evaluating
larger blocks from trained language models.

% \paragraph{Future work.}
% Natural directions for future work include combining our
% precision-agnostic, bounded-input primitives with multilayer decoupling
% techniques~\cite{transformer-lin} to extract stacked Transformer
% architectures across low-precision formats, incorporating rotary or
% learned positional encodings and grouped-query Attention, and scaling
% end-to-end evaluation further to larger widths (such as \(d=512, m=2048\))
% as well as trained language-model checkpoints.

% === IACR Generative AI Policy (2026-11-10) - Section 2.4 Disclosure Details ===
\paragraph{Disclosure of Generative AI Use.}
In accordance with the IACR Generative AI Policy (revised 2026-11-10), this work carries the designation \textbf{Substantial AI contribution}.
Large language model assistants were used during the research workflow to assist with drafting and revising manuscript text and notation, drafting and checking intermediate mathematical derivations and proofs, and implementing, debugging, and optimizing the numerical solver, experimental harness, and verification scripts.
All research directions, algorithmic design choices, mathematical claims, proofs, experimental evaluations, and reported results were directed, audited, independently verified, and approved by the human authors, who take full responsibility for the correctness and integrity of all content.

\bibliography{bib}

\appendix
\section{Proof of Proposition~\ref{prop:relu-scale-refinement}}
\label{app:relu-scale-refinement}

\begin{proof}[Proof of Proposition~\ref{prop:relu-scale-refinement}]
Let \(\boldsymbol{1}_n=(1,\ldots,1)\),
\(\boldsymbol{t}=(t_1,\ldots,t_n)\), and \(\boldsymbol{\phi}(\xi)\in\mathbb R^n\)
with \(\boldsymbol{\phi}(\xi)[i]=\varphi_\xi(t_i)\) be row vectors, form
\(\bZ(\xi)=[\boldsymbol{1}_n^\top,\ \boldsymbol{t}^\top,\ \boldsymbol{\phi}(\xi)^\top]\in\mathbb R^{n\times 3}\),
and stack \(\by(t_1),\ldots,\by(t_n)\) as rows of \(\bY\in\mathbb R^{n\times d}\).
By~\eqref{eq:relu-v11-fixed-pattern}, for any \(\xi>0\),
\begin{equation}
    \bY
    =\bZ(\rho)\begin{bmatrix}\bc_*\\ \bk\\ \bw_*\end{bmatrix}
    =\bZ(\xi)\begin{bmatrix}\bc_*\\ \bk\\ \bw_*\end{bmatrix}
     +\bigl(\boldsymbol{\phi}(\rho)-\boldsymbol{\phi}(\xi)\bigr)^\top\bw_*.
\end{equation}
Since \(\bZ(\rho)\) has full column rank \(3\) and \(\bZ(\xi)\) varies
smoothly with \(\xi>0\), \(\bZ(\xi)\) keeps full column rank \(3\) on an
open neighborhood of \(\rho\), where \(\bZ(\xi)^\dagger\bZ(\xi)=\bI_3\).
Taking the second row of \(\bZ(\xi)^\dagger\bY\) with
\(\be_2=(0,1,0)\) gives the unique solution for \(\bk(\xi)\):
\begin{equation}
    \bk(\xi)
    =\be_2\bZ(\xi)^\dagger\bY
    =\bk+\alpha(\xi)\bw_*,
    \qquad
    \alpha(\xi)
    :=\be_2\bZ(\xi)^\dagger\bigl(\boldsymbol{\phi}(\rho)-\boldsymbol{\phi}(\xi)\bigr)^\top.
\end{equation}
At \(\xi=\rho\), \(\alpha(\rho)=0\), which gives \(\bk(\rho)=\bk\) and
\(\Phi(\rho)=\|\bk\|_2^2/d=\rho\), proving (1).

Near \(\rho>10^{-18}\), the floor in~\eqref{eq:relu-v11-trial-scale} is
inactive, so the update reduces to \(\xi_{s+1}=\Phi(\xi_s)\).
Since \(\partial_\xi\varphi_\xi(t_i)=-\tfrac12\varphi_\xi(t_i)^3\) and
\(\boldsymbol{\phi}(\rho)-\boldsymbol{\phi}(\rho)=\boldsymbol{0}\),
differentiating \(\alpha\) and \(\Phi\) at \(\xi=\rho\) yields
\begin{equation}
    \Phi'(\rho)
    =\frac{2\bk\bw_*^\top}{d}\,\alpha'(\rho)
    =\frac{\bk\bw_*^\top}{d}\,\be_2\bZ(\rho)^\dagger\boldsymbol{\phi}_3(\rho)^\top,
    \qquad \boldsymbol{\phi}_3(\rho)[i]=\varphi_\rho(t_i)^3.
\end{equation}
If \(|\Phi'(\rho)|<1\), continuity of \(\Phi'\) yields \(r_0>0\) for
any \(\eta\in(|\Phi'(\rho)|,1)\) such that \(|\Phi'(\xi)|\leq\eta\)
for all \(\xi\in I_\rho=[\rho-r_0,\rho+r_0]\subset(10^{-18},\infty)\).
By the mean value theorem,
\(|\Phi(\xi)-\rho|=|\Phi(\xi)-\Phi(\rho)|\leq\eta|\xi-\rho|\) on
\(I_\rho\), so \(\Phi(I_\rho)\subseteq I_\rho\) and induction gives
\(|\xi_s-\rho|\leq\eta^s|\xi_0-\rho|\).
Continuity of \(\bk(\xi)\) then gives \(\bk^{(s)}\to\bk\), and setting
\(\eta^S|\xi_0-\rho|\leq\delta\) gives~\eqref{eq:relu-v11-iterations}.
\end{proof}

\paragraph{Asymptotic scaling of \(|\Phi'(\rho)|\).}
Note the exact algebraic identity
\begin{equation}
    \varphi_\rho(t_i)^3
    =3\rho^{-1}\varphi_\rho(t_i)-2\rho^{-3/2}
     +\bigl(\varphi_\rho(t_i)-\rho^{-1/2}\bigr)^2
      \bigl(\varphi_\rho(t_i)+2\rho^{-1/2}\bigr).
\end{equation}
Let \(\br\in\mathbb R^{n}\) be the row vector collecting the last term across
\(i=1,\ldots,n\).
Because \(\boldsymbol{1}_n^\top\) and \(\boldsymbol{\phi}(\rho)^\top\) are the
first and third columns of \(\bZ(\rho)\), we have
\(\be_2\bZ(\rho)^\dagger\boldsymbol{1}_n^\top=0\) and
\(\be_2\bZ(\rho)^\dagger\boldsymbol{\phi}(\rho)^\top=0\), so
\(\be_2\bZ(\rho)^\dagger\boldsymbol{\phi}_3(\rho)^\top=\be_2\bZ(\rho)^\dagger\br^\top\).
Fix \(\bk,\bw_*\) (hence \(\rho>0\)) and distinct sample locations
\(0<t_1<\cdots<t_n\).
Subtracting \(\rho^{-1/2}\boldsymbol{1}_n^\top\) from the third column of
\(\bZ(\rho)\) and rescaling it by \(\rho t_1^2/\epsilon_2\) leaves the
second row \(\be_2\bZ(\rho)^\dagger\) unchanged and converges to a
full-rank matrix as \(\epsilon_2\to 0\); hence
\(\|\be_2\bZ(\rho)^\dagger\|_2=O(1)\), while
\(\|\br\|_2=O\!\left((\epsilon_2/(\rho t_1^2))^2\right)\).
Consequently, with \(\bk,\bw_*,t_1,\ldots,t_n\) fixed,
\(|\Phi'(\rho)|=O\!\left((\epsilon_2/(\rho t_1^2))^2\right)\) as
\(\epsilon_2\to 0\) (with the implicit constant proportional to
\(|\bk\bw_*^\top|/d\)).
\section{Anchored Continuation for Attention Observations}
\label{app:relu-paired-continuation}

This appendix concerns exact responses and the true FFN
function \(\mathcal G\). It applies to either FFN branch.
Let \(\bX(\tau)\) be a continuous input path as in
Section~\ref{sec:att-observations}. Its true first residual
\(\bz(\tau)\) is continuous and satisfies
\(\by(\tau)=\mathcal P_{\mathcal G}(\bz(\tau))\).
The following proposition provides a sufficient condition
for following this solution from a known starting point.

\begin{proposition}[Anchored local continuation]
\label{prop:relu-anchored-continuation}
Let \(\bz_j=\bz(\tau_j)\) be known exactly at the
start of a path segment \([\tau_j,\tau_{j+1}]\).
For \(r_j>0\), let
\(\mathcal K_j=\overline B(\bz_j,r_j)\).
Suppose that \(\bA_j\in\mathbb R^{2d\times d}\) is a
known nonzero matrix and \(0\leq q_j<1\) is a known
constant satisfying
\begin{equation}
    \left\|(\bz-\bz')
      -\bigl(\mathcal P_{\mathcal G}(\bz)-\mathcal P_{\mathcal G}(\bz')\bigr)\bA_j
    \right\|_2
    \leq q_j\|\bz-\bz'\|_2
    \label{eq:relu-continuation-defect}
\end{equation}
for every \(\bz,\bz'\in\mathcal K_j\).
Suppose also that a known \(v_j\geq0\) satisfies
\begin{equation}
    \|\by(\tau)-\by(\sigma)\|_2
    \leq v_j|\tau-\sigma|
    \quad
    (\tau,\sigma\in[\tau_j,\tau_{j+1}]).
    \label{eq:relu-continuation-data-bound}
\end{equation}
Set \(c_j=(1-q_j)/\|\bA_j\|_2\), and assume
\begin{equation}
    v_j(\tau_{j+1}-\tau_j)<c_jr_j/2.
    \label{eq:relu-continuation-step}
\end{equation}
Then \(\bz_{j+1}:=\bz(\tau_{j+1})\) belongs to
\(B(\bz_j,r_j/2)\) and is the unique solution of
\(\mathcal P_{\mathcal G}(\bz)=\by(\tau_{j+1})\) in \(\mathcal K_j\).
Moreover, the iteration
\begin{equation}
    \begin{aligned}
        \bz^{(0)}&=\bz_j,\\
        \bz^{(n+1)}
        &=\bz^{(n)}
          -\bigl(\mathcal P_{\mathcal G}(\bz^{(n)})
                 -\by(\tau_{j+1})\bigr)\bA_j
    \end{aligned}
    \label{eq:relu-continuation-corrector}
\end{equation}
stays in \(\mathcal K_j\) and converges to
\(\bz_{j+1}\), with
\begin{equation}
    \|\bz^{(n)}-\bz_{j+1}\|_2
    \leq q_j^n\|\bz_j-\bz_{j+1}\|_2
    \leq q_j^n r_j/2.
    \label{eq:relu-continuation-error}
\end{equation}
For \(n=0\), the factor \(q_j^0\) is understood to be \(1\).
\end{proposition}

\begin{proof}
By the triangle inequality and
\eqref{eq:relu-continuation-defect},
\begin{equation}
    (1-q_j)\|\bz-\bz'\|_2
    \leq \|\bA_j\|_2
          \|\mathcal P_{\mathcal G}(\bz)-\mathcal P_{\mathcal G}(\bz')\|_2.
    \label{eq:relu-continuation-triangle}
\end{equation}
Thus
\begin{equation}
    \|\mathcal P_{\mathcal G}(\bz)-\mathcal P_{\mathcal G}(\bz')\|_2
    \geq c_j\|\bz-\bz'\|_2
    \qquad(\bz,\bz'\in\mathcal K_j),
    \label{eq:relu-continuation-lower-bound}
\end{equation}
so \(\mathcal P_{\mathcal G}\) is one-to-one on \(\mathcal K_j\).

The true residual path starts at the center \(\bz_j\).
If it left \(\mathcal K_j\), continuity would give
a first boundary point \(\bz(\sigma_0)\), with
\(\sigma_0\in[\tau_j,\tau_{j+1}]\).
Applying~\eqref{eq:relu-continuation-lower-bound} to
this point and the center would give
\begin{equation}
    \|\by(\sigma_0)-\by(\tau_j)\|_2
    \geq c_jr_j.
    \label{eq:relu-continuation-exit-lower}
\end{equation}
However, the response-path bound and the step condition
give
\begin{equation}
    \|\by(\sigma_0)-\by(\tau_j)\|_2
    \leq v_j(\sigma_0-\tau_j)<c_jr_j/2,
    \label{eq:relu-continuation-exit-upper}
\end{equation}
a contradiction. The whole segment therefore stays
inside \(\mathcal K_j\). At its endpoint,
\begin{equation}
    \|\bz_{j+1}-\bz_j\|_2
    \leq
    \frac{\|\by(\tau_{j+1})-\by(\tau_j)\|_2}{c_j}
    <r_j/2.
    \label{eq:relu-continuation-endpoint-distance}
\end{equation}
The endpoint solves the paired equation by definition,
and injectivity proves uniqueness in \(\mathcal K_j\).

Define
\begin{equation}
    \mathcal T_j(\bz)
    =\bz-\bigl(\mathcal P_{\mathcal G}(\bz)
               -\by(\tau_{j+1})\bigr)\bA_j.
    \label{eq:relu-continuation-correction-map}
\end{equation}
It fixes \(\bz_{j+1}\), while
\eqref{eq:relu-continuation-defect} implies
\begin{equation}
    \|\mathcal T_j(\bz)-\mathcal T_j(\bz')\|_2
    \leq q_j\|\bz-\bz'\|_2
    \qquad(\bz,\bz'\in\mathcal K_j).
    \label{eq:relu-continuation-contraction}
\end{equation}
The closed ball
\(\mathcal D_j=\overline B(\bz_{j+1},r_j/2)\)
is contained in \(\mathcal K_j\) and contains \(\bz_j\).
Since \(\mathcal T_j\) fixes its center and contracts
distances by at most \(q_j\), it maps \(\mathcal D_j\)
into itself. The contraction mapping principle
\cite[Theorem~7.6.2]{RolleTheorem}
therefore gives convergence to \(\bz_{j+1}\).
Applying the contraction inequality repeatedly proves
\eqref{eq:relu-continuation-error}.
\end{proof}

If a finite partition of \([0,1]\) satisfies these
conditions at every step, take each successive
residual to be the limit of
\eqref{eq:relu-continuation-corrector}.
Starting with~\eqref{eq:relu-paired-anchor}, induction
recovers the true residual at every partition point.
The final residual minus \(\bx_\ell\) is therefore
the exact Attention output. Under finite-precision
perturbations, the numerical behavior of this continuation
is evaluated empirically in Section~\ref{sect:experiments}.

\paragraph{Verifying the condition in a smooth neighborhood.}
Suppose that \(\mathcal P_{\mathcal G}\) is continuously
differentiable near \(\mathcal K_j\), and that
\begin{equation}
    \begin{aligned}
        \sigma_{\min}\bigl(D\mathcal P_{\mathcal G}(\bz_j)\bigr)
            &\geq \sigma_j>0,\\
        \|D\mathcal P_{\mathcal G}(\bz)-D\mathcal P_{\mathcal G}(\bz_j)\|_2
            &\leq \lambda_j\|\bz-\bz_j\|_2
            \quad(\bz\in\mathcal K_j).
    \end{aligned}
    \label{eq:relu-continuation-jacobian-bound}
\end{equation}
Take \(\bA_j=D\mathcal P_{\mathcal G}(\bz_j)^\dagger\in\mathbb R^{2d\times d}\).
Then
\(D\mathcal P_{\mathcal G}(\bz_j)\bA_j=\bI\) and
\(\|\bA_j\|_2\leq1/\sigma_j\).
For \(\bz,\bz'\in\mathcal K_j\), set
\(\boldsymbol{\xi}(s)=\bz'+s(\bz-\bz')\).
The whole segment lies in \(\mathcal K_j\). Integration
along this segment yields
\begin{equation}
    \begin{aligned}
        &(\bz-\bz')-
          \bigl(\mathcal P_{\mathcal G}(\bz)-\mathcal P_{\mathcal G}(\bz')\bigr)\bA_j\\
        &\quad=\int_0^1
          (\bz-\bz')
          \bigl[\bI-D\mathcal P_{\mathcal G}(\boldsymbol{\xi}(s))\bA_j\bigr]
          \,\mathrm ds.
    \end{aligned}
    \label{eq:att-continuation-integral}
\end{equation}
The integrand is bounded using
\eqref{eq:relu-continuation-jacobian-bound}, giving
\begin{equation}
    \left\|(\bz-\bz')-
       \bigl(\mathcal P_{\mathcal G}(\bz)-\mathcal P_{\mathcal G}(\bz')\bigr)\bA_j
    \right\|_2
    \leq\frac{\lambda_jr_j}{\sigma_j}\|\bz-\bz'\|_2.
    \label{eq:relu-continuation-smooth-bound}
\end{equation}
Thus~\eqref{eq:relu-continuation-defect} holds with
\(q_j=\lambda_jr_j/\sigma_j<1\) whenever
\(\lambda_jr_j<\sigma_j\).
These bounds concern the whole chosen neighborhood.
At a ReLU activation boundary, this smooth argument
need not apply; the proposition remains applicable
if its difference inequality can be established
directly.

\paragraph{Constructing the input path.}
For a target sequence with
\(\bu_i=\mathcal N_1(\bx_i)\), interpolate the normalized
tokens from the common starting vector \(\bq\):
\begin{equation}
    \begin{aligned}
        \bu_i(\tau)&=(1-\tau)\bq+\tau\bu_i,\\
        \bX(\tau)&=
        [\mathcal N_1^{-1}(\bu_1(\tau)),\ldots,
         \mathcal N_1^{-1}(\bu_\ell(\tau))].
    \end{aligned}
    \label{eq:relu-continuation-normalized-path}
\end{equation}
Here the inverse is
\begin{equation}
    \mathcal N_1^{-1}(\bu)
    =\frac{\sqrt{\epsilon_1}\,\bu}
           {\sqrt{1-\|\bu\|_2^2/d}},
    \qquad \|\bu\|_2<\sqrt d.
    \label{eq:att-path-norm-inverse}
\end{equation}
If the physical tokens must satisfy
\(\|\bx_i\|_2\leq R\), let
\(R_1=R/\sqrt{\epsilon_1+R^2/d}\).
When \(\|\bq\|_2\leq R_1\) and
\(\|\bu_i\|_2\leq R_1\), convexity of this ball gives
\(\|\bu_i(\tau)\|_2\leq R_1<\sqrt d\).
Consequently,
\begin{equation}
    \|\mathcal N_1^{-1}(\bu_i(\tau))\|_2^2
    =\frac{\epsilon_1\|\bu_i(\tau)\|_2^2}
            {1-\|\bu_i(\tau)\|_2^2/d}
    \leq R^2.
    \label{eq:att-path-token-bound}
\end{equation}
The negated path satisfies the same norm bound.
Without a token-norm bound, convexity of
\(\mathcal B_d\) suffices to define the path whenever
its endpoints lie in this ball. Any other restrictions
in \(\mathcal X\), including the permitted sequence
length, must also hold along the path.

Proposition~\ref{prop:relu-anchored-continuation} analyzes
fixed-matrix correction under exact observations.
In practice, the numerical solver in
Section~\ref{sec:att-observations} accelerates local
convergence by solving bounded least-squares subproblems
with updated Jacobians and adaptive path subdivision.

\end{document}